%% file: main.tex
\documentclass[
	fontsize=12pt, 
	twoside=true, 
	numbers=noenddot, 
]{kaobook}

\usepackage[latin,greek,english]{babel} 
\usepackage[english=american]{csquotes} 

\usepackage{blindtext}
\usepackage{xcolor}
\usepackage{amsmath}
\usepackage{epigraph}
\usepackage{textgreek}
\usepackage{diagbox}
\usepackage[normalem]{ulem}
\hypersetup{
	colorlinks = true,
	citecolor = RedOrange,
	linkcolor = MidnightBlue,
	urlcolor = ForestGreen,
	hyperindex=true,
	pdfborder={0 0 0} 
}
\usepackage{rotating}
\usepackage{makeidx}
\usepackage{etoolbox}
\usepackage{braket}

\usepackage{booktabs}
\usepackage{multirow}
\usepackage{pifont} 
\newcommand{\gdiamond}{\textcolor{green!60!black}{\ding{117}}}
\newcolumntype{C}[1]{>{\centering\arraybackslash}m{#1}}
\newcolumntype{L}[1]{>{\raggedright\arraybackslash}m{#1}}

\DeclareUnicodeCharacter{0301}{\'{e}}

\newcommand{\dd}{\mathrm{d}}
\newcommand{\bp}{\beta p}
\newcommand{\bpp}{\beta p_{\|}}
\newcommand{\ldb}{\Lambda_{\mathrm{dB}}}

\newcommand{\y}{{y_1}}
\newcommand{\yp}{{y_2}}
\newcommand{\yy}{{y_1,y_2}}

\usepackage{nameref}

\usepackage{tabularx}%
\usepackage[style=numeric-comp,sorting=none,backend=biber,maxnames=50,maxcitenames=4]{biblatex}
\DeclareMathAlphabet{\mathcal}{OMS}{zplm}{m}{n}
\DeclareMathAlphabet{\mathsf}{OT1}{\sfdefault}{m}{n}

\makeglossaries 
\input{chapters/00_Acronyms.tex}

\makeatletter
\renewcommand{\part}{%
  \if@openright
    \cleardoublepage
  \else
    \clearpage
  \fi%
  \thispagestyle{empty}%
  \if@twocolumn
    \onecolumn
    \@tempswatrue
  \else
    \@tempswafalse
  \fi%
  \null\vfil%
  \secdef\@part\@spart%
}
\makeatother

\begin{document}
\title{}
\author{}
\date{}
\thispagestyle{empty}
\begin{center}
\vspace*{-2cm}
{\includegraphics[width=0.9\textwidth]{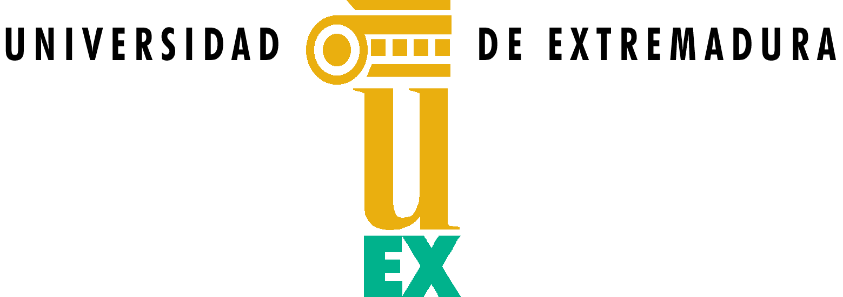}\par}
\vspace{1cm}
{\scshape\large \textbf{TESIS DOCTORAL} \par}
\end{center}
\vspace{0.6cm}
{\large \textbf{PROPIEDADES DE EQUILIBRIO EN FLUIDOS\\FUERTEMENTE CONFINADOS} \par}
\vspace{0.6cm}
{\large \textbf{ANA MAR\'IA MONTERO MART\'INEZ} \par}
\vspace{0.6cm}
{\large \textbf{PROGRAMA DE DOCTORADO EN MODELIZACIÓN Y\\EXPERIMENTACIÓN EN CIENCIA Y TECNOLOGÍA (R007)} \par}
\vspace{0.5cm}
\begin{center}
{Conformidad del director\par}
\vspace{2.5cm}
{Dr. D. Andr\'es Santos Reyes\par}
\end{center}
\vspace{0.2cm}
Esta tesis cuenta con la autorización del director de la misma y de la Comisión Académica del programa. Dichas autorizaciones constan en el Servicio de la Escuela Internacional de Doctorado de la Universidad de Extremadura.
\vspace{0.5cm}
\begin{center}
{\large \textbf{2025} \par}
\end{center}

\frontmatter 

\thispagestyle{empty}
\begin{center}
{\includegraphics[width=0.3\textwidth]{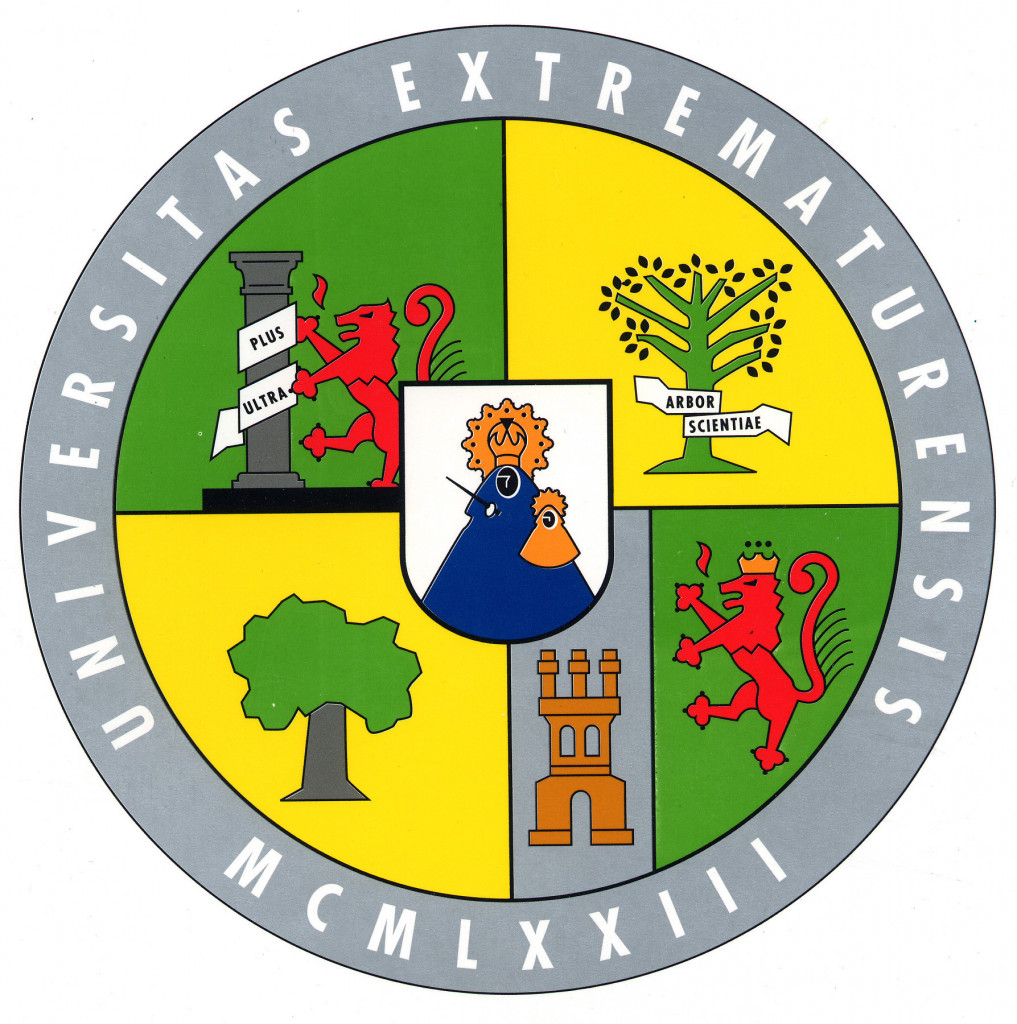}\par}
\vspace{0.2cm}
{\bfseries\LARGE University of Extremadura \par}
\vspace{1.5cm}
{\scshape\Huge Equilibrium properties of strongly confined fluids \par}
\vspace{1cm}
{\itshape A thesis submitted in fulfillment of the requirements for the degree of Doctor of Philosophy \par}
\vspace{3cm}
{\large Author: \par}
{\large Ana Mar\'ia Montero Mart\'inez  \par}
\vspace{0.5cm}
{\large Supervisor: \par}
{\large Dr. Andr\'es Santos Reyes \par}
\vspace{0.5cm}
{\large \textbf{2025} \par}
\end{center}

\thispagestyle{empty}
\dedication{
\flushright {\Large
	\emph{A mi hermana}}	
}

\makeatother

	

\maketitle
	
\input{chapters/01_Resumen.tex}
\input{chapters/01_Abstract.tex}
\input{chapters/01_Agradecimientos.tex}


\setglossarystyle{list}
\printglossary[title=Acronyms, toctitle=Acronyms] 







\tableofcontents


\setchapterstyle{kao}{} 
\pagelayout{wide}

\checkoddpage
\ifoddpage{
	\afterpage{\blankpage}
}\fi

\checkoddpage
\ifoddpage{
	\afterpage{\blankpage}
}\fi

\oldmainmatter 

\setlength{\overflowingheadlen}{\linewidth}
\addtolength{\overflowingheadlen}{\marginparsep}
\addtolength{\overflowingheadlen}{\marginparwidth}

\pagestyle{scrheadings} 

\addpart{Introduction and theory}

\input{chapters/C1_Introduction.tex}

\input{chapters/C2_1_TheoryOf1DLiquids.tex}

\input{chapters/C2_2_TheoryOfQ1DLiquids.tex}

\addpart{Articles}

\input{chapters/C3_OneDimensionalSystems.tex}

\input{chapters/C4_Q1DHardDisks.tex}

\input{chapters/C5_Q1DSWSS.tex}

\input{chapters/C6_Q1DHardSpheres.tex}

\input{chapters/C7_Dumbbells.tex}

\addpart{Results, conclusions, and outlook}

\input{chapters/C8_Results_and_Discussion.tex}

\input{chapters/C9_Conclusions_and_Outlooks.tex}

\addpart{Appendix}

\appendix 

\input{chapters/A1_ProofOfMaxA2.tex}


\backmatter 
\setchapterstyle{plain} 



\printbibliography[heading=bibintoc, title=Bibliography] 

\end{document}

%% file: chapters/00_Acronyms.tex
\newacronym{RDF}{RDF}{Radial distribution function} 
\newacronym{MC}{MC}{Monte Carlo} 
\newacronym{MD}{MD}{Molecular dynamics} 
\newacronym{1D}{1D}{One-dimensional} 
\newacronym{2D}{2D}{Two-dimensional} 
\newacronym{3D}{3D}{Three-dimensional} 
\newacronym{Q1D}{Q1D}{Quasi one-dimensional} 
\newacronym{NN}{NN}{Nearest-neighbor} 
\newacronym{TM}{TM}{Transfer-matrix} 
\newacronym{SW}{SW}{Square-well} 
\newacronym{SS}{SS}{Square-shoulder} 
\newacronym{DOC}{DOC}{Discontinuous oscillatory crossover} 
\newacronym{FW}{FW}{Fisher--Widom} 
\newacronym{PL}{PL}{Parsons--Lee}
\newacronym{RFA}{RFA}{Rational function approximation}
\newacronym{ND}{ND}{Neighbor distribution}

\newacronym{HD}{HD}{Hard-disk}
\newacronym{HR}{HR}{Hard-rod}
\newacronym{HS}{HS}{Hard-sphere}

\newacronym{EOS}{EOS}{Equation of state}

%% file: chapters/01_Resumen.tex
\phantomsection
\addcontentsline{toc}{chapter}{Resumen}
 \thispagestyle{plain}
\begin{center}
\vspace{10pt}
\textcolor{marineblue}{\rule{\textwidth}{1mm}}

\vspace{20pt}
    {\Huge{\textcolor{marineblue}{\textbf{Resumen}}}}
\vspace{10pt}

\textcolor{marineblue}{\rule{\textwidth}{1mm}}
\vspace{10pt}
\end{center}

El estudio mecánico-estadístico de las propiedades de equilibrio de los fluidos, a partir del conocimiento del potencial de interacción entre partículas, es esencial para comprender el papel que la interacción microscópica entre partículas individuales desempeña en las propiedades del fluido. El estudio de estas propiedades desde un punto de vista fundamental es, por tanto, un objetivo central de la física de la materia condensada. Sin embargo, estas propiedades pueden variar enormemente cuando un fluido está confinado.  En esta tesis se investigan fluidos en poros extremadamente estrechos, donde las partículas se ven obligadas a permanecer en formación de \guillemotleft fila~india\guillemotright. Los sistemas resultantes son altamente anisótropos: el movimiento es libre a lo largo del eje del canal, pero está fuertemente restringido transversalmente. Para cuantificar estos efectos, se comparan las propiedades de equilibrio de los fluidos confinados con las de sus homólogos sin confinar, lo que pone de manifiesto el papel de la dimensionalidad. También desarrollamos un novedoso marco teórico basado en una correspondencia entre los fluidos confinados y una mezcla unidimensional equivalente. Este isomorfismo exacto proporciona expresiones cerradas para magnitudes termodinámicas y estructurales, permite calcular el tensor de presión anisótropo y revisa las definiciones de las correlaciones. La teoría se aplica a distintos modelos de núcleo duro, revelando fenómenos como el ordenamiento en zigzag y los cruces estructurales de correlaciones espaciales. Las predicciones analíticas se validan ampliamente con simulaciones de Monte Carlo y de dinámica molecular, mostrando una excelente concordancia en todos los rangos de parámetros estudiados.

\newpage
\blankpage
\newpage

%% file: chapters/01_Abstract.tex
\phantomsection
\addcontentsline{toc}{chapter}{Abstract}
 \thispagestyle{plain}
\begin{center}
\vspace{10pt}
\textcolor{marineblue}{\rule{\textwidth}{1mm}}

\vspace{20pt}
    {\Huge{\textcolor{marineblue}{\textbf{Abstract}}}}
\vspace{10pt}

\textcolor{marineblue}{\rule{\textwidth}{1mm}}
\vspace{10pt}
\end{center}

The statistical-mechanical study of the equilibrium properties of fluids, starting from the knowledge of the interparticle interaction potential, is essential to understand the role that microscopic interaction between individual particles play in the properties of the fluid. The study of these properties from a fundamental point of view is therefore a central goal in condensed matter physics. These properties, however, might vary greatly when a fluid is confined to extremely narrow channels and, therefore, must be examined separately.  This thesis investigates fluids in narrow pores, where particles are forced to stay in single-file formation and cannot pass one another. The resulting systems are highly anisotropic: motion is free along the channel axis but strongly restricted transversely. To quantify these effects, equilibrium properties of the confined fluids are compared with their bulk counterparts, exposing the role of dimensionality. We also develop a novel theoretical framework based on a mapping approach that converts single-file fluids with nearest-neighbor interactions into an equivalent one-dimensional mixture. This exact isomorphism delivers closed expressions for thermodynamic and structural quantities. It allows us to compute the anisotropic pressure tensor and revises definitions of spatial correlations to take into account spatial anisotropy. The theory is applied to hard-core, square-well, square-shoulder, and anisotropic hard-body models, revealing phenomena such as zigzag ordering and structural crossovers of spatial correlations. Analytical predictions are extensively validated against Monte Carlo and molecular dynamic simulations (both original and from the literature), showing excellent agreement across the studied parameter ranges.

\newpage
\blankpage
\newpage

%% file: chapters/01_Agradecimientos.tex
\thispagestyle{empty}

\chapter{Agradecimientos}\label{Agradecimientos}

A lo largo de este proyecto he contado con el apoyo de muchas personas, a quienes agradezco su orientación, sus ánimos y su paciencia. Los avances y logros de esta tesis son el reflejo de la confianza que han depositado en mí, y de la dedicación que han mostrado a lo largo de todo este camino.

En primer lugar, y como no podía ser de otra forma, quiero expresar mi más sincero agradecimiento a mi director de tesis, Andrés Santos Reyes. Este agradecimiento es especial, porque trasciende el marco de esta tesis y se extiende para abarcar también todo su apoyo durante gran parte de mi trayectoria profesional. Durante más de diez años me ha acompañado a lo largo de las distintas etapas, y la confianza que siempre ha depositado en mí ha sido clave para animarme a dar pasos que, de otro modo, quizá no habría dado. Sus palabras y consejos han sido fundamentales en este recorrido, y estoy convencida de que seguirán siendo una guía en los pasos que aún me quedan por dar.

Me siento muy afortunada por haber podido desarrollar esta tesis dentro del grupo SPhinX, y quiero agradecer sinceramente a todos sus integrantes el haber creado un ambiente agradable, donde es fácil trabajar y en el que siempre me he sentido cómoda. Dirijo un agradecimiento especial a los investigadores más veteranos del proyecto en el que se enmarca esta tesis: Vicente, Santos, Juan, Mariano y Enrique, por haberme animado y ofrecido su ayuda cuando la he necesitado. También quiero agradecer a mis compañeros más jóvenes el haber hecho el día a día en la oficina mucho más ameno, especialmente durante los \emph{coffee breaks}: Rubén, Juan, Alberto, Bea y Javi, que me acompañaron en las primeras etapas de esta tesis, y Jesús y Miguel, con quienes además he compartido esta última etapa del camino. 

En el ámbito administrativo, tan indispensable también en el desarrollo de este trabajo, deseo expresar mi agradecimiento a Javier Acero, director del departamento de Física, por estar siempre dispuesto a ayudarme en todo lo posible, así como a Manuel Antón, coordinador del programa de doctorado, por su buena disposición para resolver mis dudas. Mi gratitud también va para Fran y Nuria por su impecable labor como gestores del grupo.

Mis agradecimientos en el plano académico se extienden también a Péter y Sabi, por su cálida hospitalidad durante los tres meses de mi estancia en la Universidad de Pannonia en Hungría. Agradezco igualmente a los revisores externos de este trabajo, quienes, de manera desinteresada, llevaron a cabo esta labor: Achille Giacometti, Eva Gonz\'alez Noya, David Kofke y Gerardo Odriozola.

Gran parte del apoyo necesario durante esta tesis ha venido del ámbito personal. Les estaré siempre agradecida a mis padres, Antonio y Ana, por su amor incondicional y por enseñarme, con su ejemplo, el valor del esfuerzo y la constancia. Mi hermana Ángela ha sido, sin duda, mi mayor punto de apoyo. Nunca podré agradecerle lo suficiente no solo su inagotable confianza en mí, sino también que me haya dado la certeza de que siempre podré contar con ella. También quiero agradecer a mi tía Isabel y a mis padrinos, José Luis y Filo, por haber apoyado mi educación con tanto cariño, y a mi abuela, por estar siempre orgullosa de mis logros y celebrarlos como propios.

Los amigos son la familia que uno elige, y en ese sentido quiero agradecerles a Bea y Patri el estar a mi lado desde los tiempos del instituto. Quiero extender un agradecimiento muy especial a Ainara, Lourdes y Lucía por haberme apoyado siempre, y por haber llevado personalmente ese apoyo a cualquier lugar del mundo en el que me encontrara.

Finalmente, quiero dirigir un agradecimiento muy especial a David, por transmitirme siempre la tranquilidad necesaria, por animarme en todo momento a dar el siguiente paso y, sobre todo, por estar a mi lado cada vez que lo hacía.

Para concluir, quiero expresar mi agradecimiento al Ministerio de Ciencia e Innovación del Gobierno de España por la financiación de esta tesis a través de la ayuda predoctoral PRE2021-097702.

%% file: chapters/C1_Introduction.tex
\thispagestyle{empty}

\setcounter{chapter}{0}
\chapter{Introduction}
\labch{introduction}

Understanding the equilibrium properties of fluids is a central goal in condensed matter physics and physical chemistry. Fluids exhibit a rich array of phases and behaviors (gas, liquids, supercritical fluids, etc.), and the study of their equilibrium properties, such as phase diagrams or correlation functions, is crucial to understanding the principles governing matter.

Equilibrium studies of fluids have their own intrinsic theoretical interest because they reveal how macroscopic behaviors emerge from microscopic interparticle interactions. However, the study of equilibrium properties of fluids also has a broad relevance in more practical applications, where accurate predictions of the behavior of fluids undergoing phase transitions, critical phenomena, or emergent ordered or disordered structures are needed. Information about these processes guides the development of theories applicable in chemical engineering---designing processes involving gases and liquids---and material science---for colloidal assembly or biomolecular solutions~\cite{RWDCVL11,WM13,PPD98}.

In the study of fluids, computer simulations techniques like Monte Carlo (\acrshort{MC}) or molecular dynamics (\acrshort{MD})~\cite{FS02,AT17}, as well as approximate analytical methods~\cite{YS94,AS01}, are widespread and indispensable tools due to the lack of general exact analytical solutions. Because of this, systems that are amenable to analytical or exact results are very important. They provide absolute benchmarks for testing approximations and simulations~\cite{RT95,MS06,S07,SFG08}, often revealing more subtle dependencies that might not be easily recognized by means of simulations or approximations. In this sense, a closed-form expression can yield insight into \emph{why} a system behaves the way it does, whereas numerical simulations might only show \emph{what} the behavior is.

Apart from the conceptual clarity that analytical approaches offer, another clear benefit is the ability to explore extreme conditions and thermodynamic limits with confidence. For example, an exact equation of state allows one to take high/low density and temperature limits to check what the leading behavior is under these conditions~\cite{TS25}, as well as some other limiting cases regarding, for example, the shape of the interaction potential~\cite{MFGS13,FGS15,GVO16,CM00}. This can be challenging to do in simulations due to finite-size or sampling issues.

Within this context, a powerful approach in liquid-state theory is to investigate simple pairwise interaction potentials, that is, pairwise potentials that are relatively straightforward and uncomplicated in form and mathematical representation, involving only basic functional forms. By reducing the complexity of interparticle forces to an idealized form, these simple potentials isolate key factors and allow deeper insight into the causes of fluid behavior~\cite{SBFMS04,YBGDS06,J01b,MU15,M24,BH76,HM13,EL85,BCT07,HS18,PMPSGLVTC22, MCSM22,TL94a,FGMS13}. Because of this, simple potential models usually serve as controllable testbeds for theory and simulation, where underlying mechanisms can be identified without the complexity of real interactions.

The simple potential model \emph{par excellence} is the hard-sphere (\acrshort{HS}) one, where particles interact only though a hard core that creates an excluded volume in the system~\cite{PM86,MU94,WCLSW00,BWQSPP02,RCDRSSV24}. The simplicity of such a model can sometimes be deceptive, as volume exclusion alone can reproduce a lot of phenomena present in real fluids, such as phase transitions~\cite{AW57,AA06,BK11,GA15,KB99,DRE99a,FD06,MLGORFV14}. \acrshort{HS} models have been an important benchmark for understanding matter, supported by a long history of theory and experiments. Colloidal suspensions, for example, often have complex interactions, but they can be tuned to behave like \acrshort{HS} fluids, allowing experimental realization of this simple model.

Many of the more realistic simple potentials also build on the background of the \acrshort{HS} reference, by adding some extra attraction or repulsion along with the hard core. Ramp potentials, like the triangle-well~\cite{BB83c,BS86,MS19} and the Jagla~\cite{J99a,LXASB15,BCT07,RD17,HRYS18,XBAS06} ones, are prime examples of these types of potentials, along with piecewise-constant potentials, like the square-well~(\acrshort{SW})~\cite{BH67, BH67b,YS94,MRR09,RPSP10,TL94b} and square-shoulder~(\acrshort{SS})~\cite{LKLLW99,YSH11,BOO13,GSC11,RS03,RPSP10} potentials. Many of these potentials, though simple, represent real fluids. The one-dimensional (\acrshort{1D}) triangle-well model, for instance, represents the effective Asakura--Oosawa colloid--colloid depletion potential in a colloid--polymer mixture in which the colloids are modeled as hard rods and the polymers are treated as ideal particles excluded from the colloids by a certain distance~\cite{AO54,AO58,V76,BE02}. The \acrshort{SS} potential has also been widely used to study metallic liquids~\cite{SY76} or water anomalies~\cite{J04,BSB11,HU13,HU14,SSGBS01,YBGDS06}. 

While bulk fluids are important, many real-world scenarios involve fluids confined in restricted geometries, for example fluids in porous materials, narrow channels, or between interfaces. These extreme confinements can dramatically alter the properties of the fluid as compared to those of the bulk, and this reduced dimensionality can lead to new phenomena, such as shifts or disappearance of phase transitions and the presence of anisotropic pressure components. In recent decades, advances in nanotechnology have enabled experiments that probe fluids under extreme confinement, driving substantial research interest in this area and making it not only scientifically intriguing but also technologically relevant, with applications in nanofluidics, catalysis, and biological channels~\cite{CLSR02,Ketal11,MCH11,DNHEG08,LG20,LKK14,KHD11}.

In the context of confined fluids, simple interaction potentials---and especially hard spheres---have again proven to be an excellent starting point. They have been investigated in a variety of confining geometries from both theoretical and experimental perspectives. Notable systems include slit pores (two parallel walls forming a narrow gap)~\cite{HSW97,MET06,MET07,TSAHRD18,SFS14,NSHCPBJK16,VMVBO16,MV23,JF23,BGM24}, spherical cavities~\cite{HYS13,WDW21}, and very narrow pores~\cite{BBW89,ALD96,DSBP98,MP00,GSCM04,KKM08,DG09,HKS09,MV19,GVMV18,GVGMV17,SGG20,BGVO21}. Each of these geometries imposes a very different type of confinement, leading to distinct properties of the fluid. For example, in a slit pore, the available space in one dimension is finite and often only a few particle diameters thick, causing the fluid to form well-defined layering between the walls that can undergo two-dimensional (\acrshort{2D}) phase transitions that may have no counterpart in the bulk fluid~\cite{FD06,NSK13,MLGORFV14}. In fluids confined in narrow tubes, movement is restricted along all dimensions but one, which forces particles to arrange in single-file formation and prevents, for example, phase transitions.

\begin{figure}
	\includegraphics[width=0.9\columnwidth]{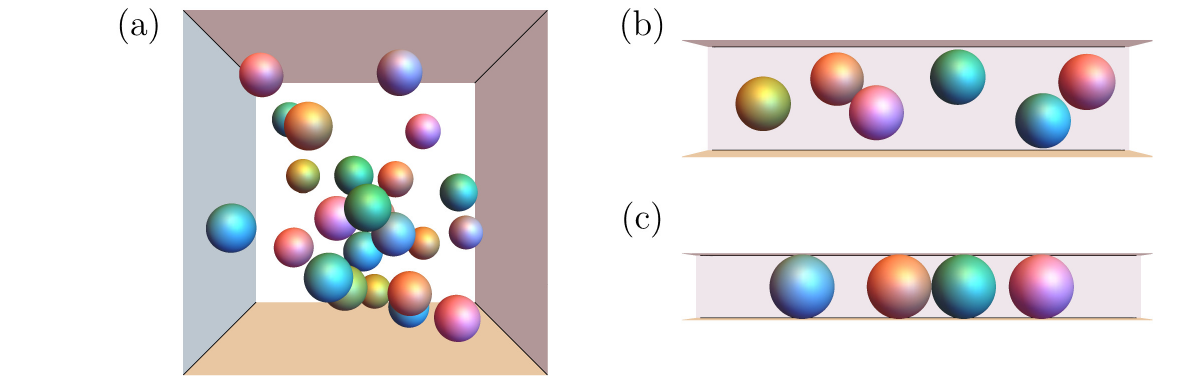}\\
	\caption{Visual representation of particles in different geometries: (a) a bulk system exhibiting isotropic translational invariance in all directions; (b) a confined system with translational invariance only along the longitudinal axis; (c) an ultraconfined system in which particles are restricted to move solely along the longitudinal direction.}
	\label{fig:c1_3dConfinement}
\end{figure}

This thesis focuses on the latter type of confinement, in which particles are free to move along a single spatial dimension while their motion is restricted, to varying degrees, in all other directions. Figure \ref{fig:c1_3dConfinement} schematically illustrates how a bulk three-dimensional (\acrshort{3D}) system can be gradually squeezed in two orthogonal directions until the \acrshort{1D} limit is reached.

Strict \acrshort{1D} fluids, where particles are forced to move on a line with no transverse degree of freedom, are therefore the most extreme limit of this kind of confinement and they have long been studied for their theoretical importance. Many of these \acrshort{1D} systems are amenable to exact analytical solutions~\cite{BNS09,BOP87,HGM34,HC04,VCR03,KT68,K55b,K91,LZ71,R91,N40a,N40b,P76,P82,SZK53,S07,MS19,T42,T36,F16,F17,M94,R71b,TS25}, providing powerful information about how liquids behave in confined dimensions and also offering insight into some behaviors observed in bulk fluids. A famous example of an exactly solvable model is the \acrshort{1D} hard-rod (\acrshort{HR}) fluid. This system is equivalent to the \acrshort{HS} one, but confined to move only in one dimension [see Fig.~\ref{fig:c1_3dConfinement}(c)]. An exact solution for its equation of state was found by~\textcite{T36} in 1936, a solution so well known that this system is now commonly referred to as \emph{Tonks gas}. In 1953,~\textcite{SZK53} proved that it was possible to derive exact expressions for the distribution functions in a \acrshort{1D} system whose particles interact only with their nearest neighbors by means of pairwise interaction potentials. These distribution functions were then used to compute the thermodynamic properties of the fluid. In 1971,~\textcite{LZ71} extended this procedure to \acrshort{1D} mixtures.

Comparing a bulk fluid with its \acrshort{1D} counterpart---especially when the \acrshort{1D} model is exactly solvable---is valuable for two complementary reasons. First, the side-by-side analysis exposes phenomena that arise only in three dimensions or only in one, revealing which features are governed by dimensionality. As an example, \acrshort{1D} fluids with short-range interactions do not exhibit phase transitions because long-range order cannot be sustained in one dimension~\cite{vH50,F10b}, in contrast to the bulk case. Second, when a given effect appears in both one and three dimensions, the exact \acrshort{1D} solution serves as a theoretical background that provides a transparent description that helps clarify the microscopic mechanisms that also operate in the more complex bulk system~\cite{CMT02,FGMS10,CM97,MC99}.

A way to bridge the gap between purely \acrshort{1D} systems and the bulk phase is to study quasi one-dimensional (\acrshort{Q1D}) fluids, in which particles have some degree of freedom to move along the confined (transverse) direction---unlike in the \acrshort{1D} version---but are still forced to stay in single-file formation. The study of fluids restricted to single-file configurations constitutes an active field of study for both equilibrium~\cite{B62,B64b,WPM82,PK92,KP93,P02,KMP04,FMP04,VBG11,GV13,GM14,GM15,HFC18,M14b,M15,M20,HBPT20,P20,JF22,PBT23} and nonequilibrium properties~\cite{GM14,FMP04,KMS14,RGM16,TFCM17,WLB20,LG20,HBPT21,RS23,MBGM22,RGIB23,MGB22, PGIB21,M24}, as well as for jamming effects~\cite{GM14,ZGM20,I20,LM20,LM22}. From a fundamental perspective, studying \acrshort{Q1D} fluids allows us to interpolate between purely \acrshort{1D} and higher-dimensional behavior. As confinement tightens, the system approaches a \acrshort{1D} limit; as it loosens, bulk-like behavior should be recovered. Investigating this crossover enhances our understanding of how dimensionality affects fluid properties.

Examining \acrshort{Q1D} models across the full range of confinement, from extremely tight to relatively loose, reveals how precursors to bulk phenomena emerge. Tracking these changes clarifies the mechanisms that govern the appearance or suppression of certain effects as the geometry moves from strictly \acrshort{1D} systems to fully \acrshort{3D} fluids. As an example, coming back to phase transitions, \acrshort{Q1D} models behave like the \acrshort{1D} counterpart in the sense that they do not present real phase transitions. However, they do present a certain type of structural transitions at certain densities when moving from a disordered fluid to an ordered zigzag arrangement along the channel, an effect that cannot happen in \acrshort{1D} geometries and that brings \acrshort{Q1D} systems closer to pure bulk effects.

Another key reason to study \acrshort{Q1D} models is that they can often be realized in a laboratory~\cite{HP18}, allowing a direct comparison between theory and experiment. The relevance of confined fluids in real-life situations spans across many scales: from nanometers, for example ion channels in biology or gas storage in nanoporous solids, to microns when working with colloidal particles between walls. This makes the study of \acrshort{Q1D} systems relevant for a wide variety of interdisciplinary topics, as it addresses practical questions about how confinement alters fluid structure and thermodynamics, which can inform experimental design for nanoconfined fluids.

From a theoretical point of view, the effectively reduced dimensionality of \acrshort{Q1D} systems can render them mathematically tractable under suitable conditions, as occurred with strictly  \acrshort{1D} models. More specifically, if interactions are limited to nearest neighbors and the pair potential is sufficiently simple, mathematical treatments become feasible~\cite{KP93,VBG11,GV13,GM14,HFC18,ZGM20,HBPT20,HC21,TPHB21,F23,FS24,MS23,MS23b,MS24}. This means that \acrshort{Q1D} geometries occupy a valuable middle ground: confinement is strong enough to permit rigorous analysis, while still exhibiting nontrivial behavior that provides valuable physical insight.

An important characteristic in the study of \acrshort{Q1D} systems is their \emph{anisotropy}, induced by geometrical confinement and which is absent in both \acrshort{1D} and \acrshort{3D} systems with the same interaction potential. This high anisotropy paves the way for studying \acrshort{Q1D} systems, and the thermodynamic formalism used to describe them must be modified accordingly. Many of the usual techniques in \acrshort{3D} systems that assume rotational and translational invariance are no longer valid. Reference quantities, such as the radial distribution function (\acrshort{RDF}), acquire a different meaning in restricted geometries. A careful treatment of the different pressure components~\cite{VBG11,MS24b,FS24}, for example, is also needed.

In fact, due to the high confinement of \acrshort{Q1D} systems, many of the studies effectively treat the fluid's degrees of freedom as \acrshort{1D} when developing an exact solution, and their main focus is on studying their longitudinal properties (i.e., those along the unconfined direction). In this regard~\textcite{B62} presented a very general result in 1962 to obtain the exact solution of almost \acrshort{1D} systems, where particles can be ordered serially and have a well-defined range of interaction. In 1993,~\textcite{KP93} used this result to derive an exact transfer-matrix (\acrshort{TM}) solution for a \acrshort{Q1D} system of hard particles, which could be solved numerically for the thermodynamic properties of the fluid, such as the equation of state, and for some structural properties, such as the transverse density profile across the confined directions. This \acrshort{TM} method has since been widely used to study \acrshort{Q1D} systems of different kinds of particles and interaction potentials~\cite{VBG11,GV13,HFC18,MS23}.

One of the main results of this thesis is the development of a novel method to study systems confined in \acrshort{Q1D} geometries and obtain all its thermodynamic and structural properties~\cite{MS23,MS23b,MS24,MS24b}. The general idea of this method is to establish an exact mapping between the \acrshort{Q1D} model and another \acrshort{1D} mixture, which means that solving the \acrshort{1D} mixture is enough to get all information about the \acrshort{Q1D} counterpart. We then use the probability distribution formalism to derive all properties of the system including the \acrshort{RDF}, which had remained elusive until now.

\section{List of publications}

The thesis is presented as a compilation of articles, all of them related to the study of the equilibrium properties of highly confined fluids and their characteristic differences with respect to the bulk phase. The articles are, in order of appearance in the thesis:

\begin{itemize}
	\item \cite{MRYSH24} \textbf{Article 1}: A. M. Montero, A. Rodr\'iguez-Rivas, S. B. Yuste, A. Santos, and M. {L\'opez de Haro}. \enquote{On a conjecture concerning the {F}isher--{W}idom line and the line of vanishing excess isothermal compressibility in simple fluids}. \textit{Mol. Phys.} \textbf{122} (2024) e2357270. DOI: \href{https://doi.org/10.1080/00268976.2024.2357270}{10.1080/00268976.2024.2357270}
	\vspace{0.35cm}
	\item \cite{MYSL25} \textbf{Article 2}: A. M. Montero, S. B. Yuste, A. Santos, and M. {L\'opez de Haro}. \enquote{Discontinuous Structural Transitions in Fluids with Competing Interactions}. \textit{Entropy} \textbf{27} (2025) 95. DOI: \href{https://doi.org/10.3390/e27010095}{10.3390/e27010095}
	\vspace{0.35cm}
	\item \cite{MS23} \textbf{Article 3}: A. M. Montero and A. Santos. \enquote{Equation of state of hard-disk fluids under single-file confinement}. \textit{J. Chem. Phys.} \textbf{158} (2023) 154501. DOI: \href{https://doi.org/10.1063/5.0139116}{10.1063/5.0139116}
	\vspace{0.35cm}
	\item \cite{MS23b} \textbf{Article 4}: A. M. Montero and A. Santos. \enquote{Structural properties of hard-disk fluids under single-file confinement}. \textit{J. Chem. Phys.} \textbf{159} (2023) 034503. DOI: \href{https://doi.org/10.1063/5.0156228}{10.1063/5.0156228}
	\vspace{0.35cm}
	\item \cite{MS24b} \textbf{Article 5}:  A. M. Montero and A. Santos. \enquote{Exploring anisotropic pressure and spatial correlations in strongly confined hard-disk fluids. {E}xact results}. \textit{Phys. Rev. E} \textbf{110} (2024) L022601. DOI: \href{https://doi.org/10.1103/PhysRevE.110.L022601}{10.1103/PhysRevE.110.L022601}
	\vspace{0.35cm}
	\item \cite{MS24} \textbf{Article 6}: A. M. Montero and A. Santos. \enquote{Exact equilibrium properties of square-well and square-shoulder disks in single-file confinement}. \textit{Phys. Rev. E} \textbf{110} (2024) 024601. DOI: \href{https://doi.org/10.1103/PhysRevE.110.024601}{10.1103/PhysRevE.110.024601}
	\vspace{0.35cm}
	\item \cite{MS25} \textbf{Article 7}:  A. M. Montero and A. Santos. \enquote{Exact anisotropic properties of hard spheres in narrow cylindrical confinement}. \textit{J. Chem. Phys.} \textbf{163} (2025) 024506. DOI: \href{https://doi.org/10.1063/5.0273930}{10.1063/5.0273930}
	\vspace{0.35cm}
	\item \cite{MSGV23} \textbf{Article 8}: A. M. Montero, A. Santos, P. Gurin and S. Varga. \enquote{Ordering properties of anisotropic hard bodies in one-dimensional channels}. \textit{J. Chem. Phys.} \textbf{159} (2023) 154507. DOI: \href{https://doi.org/10.1063/5.0169605}{10.1063/5.0169605}
\end{itemize}

\section{Structure of the thesis}

The first part of the thesis is formed by Chapters~\ref{1dTheory} and~\ref{1DMixtureTheory}, which are devoted to presenting the theoretical results employed throughout the thesis.

In particular, Chapter~\ref{1dTheory} presents the exact solution for the thermodynamic and structural properties of a \acrshort{1D} fluid with nearest-neighbor (\acrshort{NN}) interactions. Following the approach in~\textcite{SZK53}, and working in the isothermal--isobaric ensemble, we first derive the \acrshort{NN} probability distribution and show that this quantity alone is enough to obtain all information about the system and to make a full connection with thermodynamics through the Gibbs free energy. This method is used in three different situations: in a monocomponent system, in a discrete mixture, and in the limit where the mixture is fully polydisperse. A method for analyzing correlation lengths using Laplace-transform techniques is also provided. Although most of the information in this chapter is not a novelty derived during this thesis, and many authors have previously worked in this field, here we present a self-contained and concise summary of the method with a unified notation that is enough to understand all techniques used to study confined systems that were used throughout the thesis.

Chapter~\ref{1DMixtureTheory} is dedicated to present the theoretical framework for the exact solution of \acrshort{Q1D} systems with pairwise interactions between nearest neighbors, which is one of the main results of this work. The material consolidates results published in Articles~3–6, but is organized here in a comprehensive manner, trying to enhance clarity rather than presenting them in the chronological order of their appearance. The chapter first introduces the mapping strategy that translates a \acrshort{Q1D} problem into an exactly solvable \acrshort{1D} mixture---both from a mathematical and an intuitive physical perspective---and it then details the derivation of the exact solution. This entire chapter therefore provides a self-contained and coherent account of this central result.

The following chapters present the results of the articles that constitute this thesis, arranging them into four thematic chapters according to their particular objectives.

Chapter~\ref{1Dand3Dcomparison} encompasses the first set of articles of this thesis, namely, those devoted to analyzing the similarities and differences between the behaviors of \acrshort{1D} and \acrshort{3D} fluids when they interact through a pairwise potential with competing interactions, that is, one featuring both attractive and a repulsive part. This chapter consists of~\nameref{a1} and~\nameref{a2} where, in both cases, we use the exact methods developed in Chapter~\ref{1dTheory} for the analysis of the \acrshort{1D} system, and approximate and simulation methods for the \acrshort{3D} version. Although both articles are devoted to studying the impact of the attractive and repulsive forces of the interparticle potential, each one uses a slightly different approach.  In~\nameref{a1}, the focus of the study is to test, for a standard fixed Jagla potential, how these competing interactions manifest in the structural and thermodynamic properties of the fluid, and whether the different signatures of this competition---observed in, for example, spatial correlations and response functions---have any kind of correlation with one another. In~\nameref{a2}, we investigate a versatile two-step interaction model (a hard core followed by two step-wise potential levels, which can represent either two wells, two shoulders, or one of each) to explore how competing interactions affect structural transitions in both the \acrshort{1D} and \acrshort{3D} versions of the model. By analyzing different variations of the same two-step potential, a complex and intricate pattern of equilibrium structures is revealed, in which \acrshort{1D} confinement can enhance or suppress certain transitions.

Chapter \ref{Q1D_Disks} synthesizes the findings of~\nameref{a3},~\nameref{a4}, and~\nameref{a5}, which together provide an exact description of a \acrshort{Q1D} fluid of hard disks. Although this system is formally a \acrshort{2D} system confined to a \acrshort{Q1D} geometry, it can also be seen as a \acrshort{3D} system with one unconfined direction, one confined direction along which the particles have a limited freedom of movement and one fully confined one, where particles do not have any range of motion. Throughout all these studies the channel width---the length of the confined direction accessible to the centers of the particles---is restricted to be less than or equal to $\sqrt{3}/2$ times the hard-disk (\acrshort{HD}) diameter in order to prevent second \acrshort{NN} interactions.

\nameref{a3} concentrates on the thermodynamic properties using the result from the \acrshort{TM} formalism. Although exact, the \acrshort{TM} method does not provide any closed-form solution for the equation of state, so the limiting low- and high- pressure behaviors are worked out analytically as functions of the pore width. The low-pressure limit is studied by means of the virial expansion~\cite{HM13,S16}, for which the exact third and fourth virial coefficients are obtained---the second virial coefficient was already known exactly~\cite{KMP04,M18}. Deviating from the exact solution, a couple of simple approximations based on the exact low- and high- pressure approximations are proposed for the entire range of the equation of state. 

\nameref{a4} moves away from thermodynamics and focuses on structure. It develops the theory behind the mapping technique that recasts the \acrshort{Q1D} system as an exactly solvable \acrshort{1D} mixture, and employs it to calculate longitudinal properties such as the structure factor, spatial correlations, and the correlation length of the longitudinal \acrshort{RDF}. Comparison with previous \acrshort{MC} and \acrshort{MD} simulations shows an excellent agreement that validates the theoretical framework. 

\nameref{a3} and \nameref{a4} focus exclusively on longitudinal quantities, that is, those defined along the unconfined direction which includes, for example, the longitudinal pressure component and the longitudinal correlation functions. However, a full description of a \acrshort{Q1D} system is not complete without the study of its transverse properties. \nameref{a5} therefore extends the exact formalism to encompass all anisotropic properties such as the transverse pressure component or the complete \acrshort{RDF}. Within this framework, we compute the transverse equation of state and also obtain its behavior at low and high densities. A new definition of the \acrshort{RDF}, specifically tailored to the highly anisotropic geometry of the system, is also formulated.

The next step forward in the development of the exact solution for \acrshort{Q1D} systems is to include repulsive or attractive forces beyond the pure hard-core interactions of the hard disks. In this regard, Chapter~\ref{1DSWSS} presents the results of \nameref{a6}, which investigates the longitudinal thermodynamic and structural properties of single-file confined \acrshort{SW} and \acrshort{SS} disks. In this article, the mapping technique is adapted to account for the extra attractive well or repulsive step and exact results are then derived for key properties, including the equation of state, the internal energy (absent in the \acrshort{HD} case), and the longitudinal \acrshort{RDF}. The asymptotic behavior and the correlation length for both potentials are also derived, and an \enquote{asymptotic behavior} phase diagram in the temperature-density plane is constructed. Theoretical predictions presented here are confirmed by our own \acrshort{MC} simulations, performed both in the canonical and the isothermal--isobaric ensemble.

In Chapter~\ref{Q1D_HardSpheres}, we build again on the \acrshort{Q1D} \acrshort{HD} model of Chapter~\ref{Q1D_Disks} and consider now a \acrshort{3D} fluid of hard spheres confined in a cylindrical pore. In this geometry, the axial direction remains unconfined, whereas the other two directions are restricted, allowing only limited particle movement. This scenario contrasts with the system studied in Chapter~\ref{Q1D_Disks}, where disks could shift only slightly along a single confined direction.

Results are presented in \nameref{a7} where, using again the mapping technique that translates the confined system into an equivalent \acrshort{1D} mixture, we derive exact expressions for both thermodynamic and structural properties. Although these results are formally exact, they cannot be written in closed analytic form for general state points. Therefore we work out analytical expressions in key limiting regimes---narrow pore widths and the low- and high-pressure limits. On the structural side, we analyze fluctuations in the radial position of the particles and different spatial correlations.

Before moving on to the final chapter of results, Table~\ref{tab:confined-properties} offers a concise overview of the different models presented in Chapters \ref{Q1D_Disks},~\ref{1DSWSS}, and~\ref{Q1D_HardSpheres}. For each article, the table lists the specific \acrshort{Q1D} system examined and indicates the quantities addressed. In this way, Table~\ref{tab:confined-properties} highlights how the various contributions of this thesis collectively build a systematic understanding of fluids in spatially confined geometries.

\begin{table}[htpb]
	\centering
	\caption[Confined‐system properties]{Summary of spatially confined systems studied during this thesis, indicating the Chapter (Ch.) in which they appear, the article (Art.) number, the pairwise potential studied, the number (\#) of confined directions along which particles have some freedom of movement and, finally, which properties of the system are studied in each article, specifying whether longitudinal (Long.) or transverse (Trans.) properties were calculated.}
	\label{tab:confined-properties}
	\renewcommand{\arraystretch}{1.3}
	\begin{tabular}{@{}
			C{0.9cm}          
			C{0.9cm}          
			C{2.5cm}        
			C{2.5cm}          
			C{1.5cm}C{1.5cm}
			C{1.5cm}C{1.5cm}
			@{}}
		\toprule
		\textbf{Ch.} 
		& \textbf{Art.} 
		& \textbf{Potential} 
		& \textbf{\# confined dirs.} 
		& \multicolumn{2}{c}{\textbf{Thermodynamics}} 
		& \multicolumn{2}{c}{\textbf{Structural}} \\
		\cmidrule(lr){5-6} \cmidrule(lr){7-8}
		& & & 
		& Long. & Trans. 
		& Long. & Trans. \\
		\midrule
		4 & 3 & \acrshort{HD}                              & 1 & \gdiamond &           &           &           \\
		4 & 4 & \acrshort{HD}                               & 1 &           &           & \gdiamond &           \\
		4 & 5 & \acrshort{HD}                               & 1 & \gdiamond & \gdiamond & \gdiamond & \gdiamond \\
		5 & 6 & \acrshort{SW},~\acrshort{SS} 				& 1 & \gdiamond &           & \gdiamond &           \\
		6 & 7 & \acrshort{HS}                            & 2 & \gdiamond & \gdiamond & \gdiamond & \gdiamond \\
		\bottomrule
	\end{tabular}
\end{table}

The final chapter of results, Chapter~\ref{Dumbbells}, presents the findings from \nameref{a8}, which explores a special class of \acrshort{Q1D} systems representing the minimal model of systems that form necklace-like structures. In this models, particles are restricted to move along a single \acrshort{1D} axis, but they are hard, anisotropic \acrshort{3D} bodies that rotate while their centers remain fixed along the \acrshort{1D} axis. More specifically, the study focuses on prisms and dumbbells that are allowed to adopt two or three discrete rotational orientations. Each orientation corresponds to a different effective length along the longitudinal direction, resulting in a system where particles switch dynamically between multiple longitudinal sizes, depending on their orientation.

For these models, we compare the performance of different theories under additive (prisms) or nonadditive (dumbbells) interactions. Among these theories, we also test the mapping technique previously developed for spatially confined systems, which is shown to yield exact results for this special class of \acrshort{Q1D} models. The equation of state, the \acrshort{RDF}, and the spatial and orientational correlation lengths are also computed.

After the individual publications that constitute this compilation have been fully described, the thesis closes with Chapters~\ref{results} and~\ref{conclusions}, which offer a concise discussion of the reported findings and the principal conclusions drawn from them, along with an outlook on the broader implications of this work.

%% file: chapters/C2_1_TheoryOf1DLiquids.tex
\thispagestyle{empty}

\chapter{Exact solution of one-dimensional liquids}\label{1dTheory}

\section{Monocomponent systems}\label{sec:1dtheory}

Consider a \acrshort{1D} system of $N$ particles in a box of length $L$, where $\lambda=N/L$ represents the number density. The particles interact through a potential $\psi(r)$, which only depends on the distance, $r$, between the particles and has the following properties:

\begin{itemize}
	\item $\lim_{r \to 0} \psi(r) = \infty$, which sets a hard core and ensures that the order of particles remains fixed.
	\item $\lim_{r \to \infty} \psi(r) = 0$, so that the interaction range is finite.
\end{itemize}
Additionally, each particle interacts only with its two nearest neighbors. The total potential energy is then
\begin{equation}\label{eq:1d_potential}
	\Psi_N(\mathbf{x}^N) = \sum_{i=1}^{N} \psi(x_{i+1} - x_i),
\end{equation}
where $x_i$ represents the position of particle $i$ and we have applied periodic boundary conditions, so that $x_{N+1}=x_1+L$. Under these circumstances, one can derive exact expressions for the thermodynamic and structural properties of the system~\cite{T36,SZK53,BB83a,MS19,MS20}. Although several other \acrshort{1D} systems with different interaction potentials are also amenable to an exact solution~\cite{K59,G63,R71b}, these other cases will not be considered here.

Throughout the discussion that follows, all results are derived within the isothermal–isobaric ensemble, which characterizes a system of $N$ particles at a fixed temperature $T$ and an external pressure $p$. In such an ensemble, the volume is allowed to fluctuate under the constraint of the constant pressure. The relevant thermodynamic potential for the $(N, p, T)$ ensemble is the Gibbs free energy $\mathcal{G}$~\cite{C85,GNS12}, defined as
\begin{equation}
\mathcal{G}(N,p,T) = U - TS + pL,
\end{equation}
where $U$ is the internal energy, $S$ is the entropy, and the length $L$ plays the role of the volume $V$ for a \acrshort{1D} system. Its total differential can be written as
\begin{equation}
	\dd \mathcal{G} = - S \,\dd T + L \,\dd p + \mu \,\dd N,
\end{equation}
from where it follows that 
\begin{subequations}
	\begin{equation}\label{eq:1dgibbsL}
		L = \left(\frac{\partial \mathcal{G}}{\partial p}\right)_{T,N},
	\end{equation}
\begin{equation}\label{eq:1dgibbsMu}
	\mu = \left(\frac{\partial \mathcal{G}}{\partial N}\right)_{T,p},
\end{equation}
\end{subequations} 
where $\mu$ is the chemical potential.

\subsection{Exact solution}

\begin{figure}[htpb]
		\includegraphics[width=0.7\columnwidth]{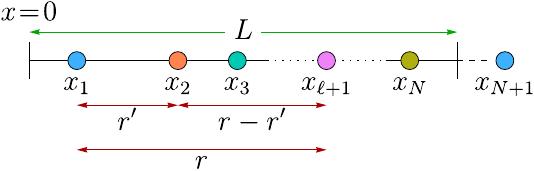}\\
		\caption{Schematic representation of the convolution property from Eq.~\eqref{eq:convolution} for a system of length $L$ where periodic boundary conditions have been applied, enforcing $x_{N+1}=x_1+L$.}
		\label{fig:c2_1dConvolution}
\end{figure}

To derive an exact solution for a \acrshort{1D} system as described in Sec.~\ref{sec:1dtheory} from a statistical-mechanical point of view, we follow the approach derived by~\textcite{SZK53}, who demonstrated that all the structural and thermodynamic properties of a \acrshort{1D} system can be calculated from the knowledge of the \acrshort{NN} probability distribution, as an alternative to the well-known method of computing the full partition function. In order to do this, let us first define $p^{(\ell)}(r)\dd r$ as the probability that the (right) $\ell$-th nearest neighbor of a certain particle is located within the interval $[r,r+\dd r]$. Thanks to the sequential arrangement of the particles (single-file condition), this probability can be recursively determined from the \emph{first} \acrshort{NN} probability distribution $p^{(1)}(r)$ as
\begin{equation}\label{eq:convolution}
	p^{(\ell)}(r) = \int_0^r \dd r' p^{(1)}(r') p^{(\ell-1)}(r-r').
\end{equation}
This integral equation is illustrated in Fig.~\ref{fig:c2_1dConvolution}. It expresses $p^{(\ell)}(r)$ as a convolution of \acrshort{NN} distributions where the probability distribution for successive neighbors can be systematically constructed by iterating over all possible intermediate positions.

\begin{figure}[htbp]
	\includegraphics[width=0.7\columnwidth]{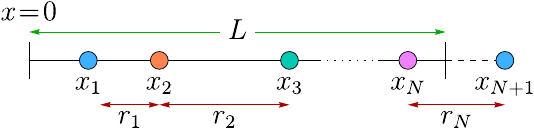}\\
	\caption{Schematic representation a \acrshort{1D} system of $N$ particles, including a visual description of the variables necessary to evaluate the configurational partition function $\Delta_C$. Periodic boundary conditions enforce $x_{N+1}=x_1+L$.}
	\label{fig:c2_1dComputeZ}
\end{figure}

Equation~\eqref{eq:convolution} shows that the key quantity to determine is $p^{(1)}(r)$. One way to compute it is to begin by evaluating the configurational part of the partition function in the isothermal--isobaric ensemble:
\begin{align}
	\Delta_C &= \int_0^\infty \dd L\, e^{-\bp L} \int_{0}^L \dd x_1 \int_{x_1}^L \dd x_2 \cdots \int_{x_{N-1}}^L \dd x_N e^{-\beta \sum_{i=1}^{N} \psi(x_{i+1} - x_i)} \nonumber \\
	&=\int_0^\infty \dd L\, e^{-\bp L} \int_{0}^L \dd r_1 e^{-\beta\psi(r_1)}  \int_{0}^{L-r_1} \dd r_2 e^{-\beta\psi(r_2)} \cdots \int_{0}^{L-r_1-\cdots-r_{N-1}} \dd r_N e^{-\beta\psi(r_N)},
\end{align}
where $\beta=1/k_BT$ is the inverse temperature ($k_B$ being the Boltzmann constant), and the same periodic boundary conditions as in Eq.~\eqref{eq:1d_potential} have been applied. In the second step, the change of variable $r_i = x_{i+1} - x_{i}, (i=1,\cdots, N)$ has been introduced. A representation of this system can be found in Fig.~\ref{fig:c2_1dComputeZ}.

For reasons that will become apparent later, we now swap the order of integration between $L$ and $r_1$ to obtain
\begin{align}
	\Delta_C &= \int_0^\infty \dd r_1\, e^{-\beta\psi(r_1)} \int_{r_1}^\infty \dd L e^{-\bp L}   \int_{0}^{L-r_1} \dd r_2 e^{-\beta\psi(r_2)} \cdots \int_{0}^{L-r_1-\cdots-r_{N-1}} \dd r_N e^{-\beta\psi(r_N)} \nonumber \\
	&=\int_0^\infty \dd r_1\, e^{-\beta\psi(r_1)}  e^{-\bp r_1} \int_{0}^\infty \dd L' e^{-\bp L'}   \int_{0}^{L'} \dd r_2 e^{-\beta\psi(r_2)} \cdots \int_{0}^{L'-\cdots-r_{N-1}} \dd r_N e^{-\beta\psi(r_N)},
\end{align}
where, in the last step, the change of variable $L'= L-r_1$ has been carried out. Defining now
\begin{equation}\label{eq:zeta}
	\zeta =  \int_{0}^\infty \dd L' e^{-\bp L'}   \int_{0}^{L'} \dd r_2 e^{-\beta \psi(r_2)} \cdots \int_{0}^{L'-\cdots-r_{N-1}} \dd r_N e^{-\beta\psi(r_N)},
\end{equation}
which is independent of $r_1$, one can rewrite $\Delta_C$ as
\begin{equation}\label{eq:Z3}
	\Delta_C = \zeta \int_0^\infty \dd r_1\, e^{-\beta\psi(r_1)}  e^{-\bp r_1}.
\end{equation}
Without loss of generality we can now take the particles at $x_1$ and $x_2$ as the representative \acrshort{NN} pair. In this case, the \acrshort{NN} probability distribution $p^{(1)}(r)$ represents the probability density of finding these two particles at a distance $r$ and can therefore be computed as
\begin{align}\label{eq:P11}
	 p^{(1)}(r) &=\frac{1}{\Delta_C}\int_0^\infty \dd L\, e^{-\bp L} e^{-\beta\psi(r)}  \int_{0}^{L-r} \dd r_2 e^{-\beta\psi(r_2)} \cdots \int_{0}^{L-r-\cdots-r_{N-1}} \dd r_N e^{-\beta\psi(r_N)}, \nonumber \\
	&=\frac{1}{\Delta_C}e^{-\beta \psi(r)}e^{-\bp r} \int_0^\infty \dd L'\, e^{-\bp L'}  \int_{0}^{L'} \dd r_2 e^{-\beta \psi(r_2)} \cdots \int_{r}^{L'-\cdots-r_{N-1}} \dd r_N e^{-\beta \psi(r_N)}\nonumber \\
	&=\frac{\zeta}{\Delta_C}e^{-\beta \psi(r)}e^{-\bp r},
\end{align}
where, again, the change of variable $L'= L-r$ has been made in the second line of Eq.~\eqref{eq:P11}. Finally, taking into account Eq.~\eqref{eq:Z3}, $p^{(1)}(r)$ becomes
\begin{equation}\label{eq:P12}
	 p^{(1)}(r) = \frac{e^{-\beta \psi(r)}e^{-\bp r}}{\int_0^\infty \dd r'\, e^{-\beta \psi(r')}  e^{-\bp r'}}.
\end{equation}
Note that, as expected, $p^{(1)}(r)$ is normalized as
\begin{equation}
	\int_0^\infty \dd r\, p^{(1)}(r) = 1,
\end{equation}
because the first (right) neighbor must be found somewhere within the system. This normalization is also extendable to any $p^{(\ell)}(r)$ through Eq.~\eqref{eq:convolution}.

The convolution structure of Eq.~\eqref{eq:convolution} and the denominator of Eq.~\eqref{eq:P12} suggest the introduction of the Laplace transform
\begin{equation}\label{eq:P1s1}
	\hat{P}^{(\ell)}(s) \equiv \int_0^\infty \dd r e^{-rs}p^{(\ell)}(r),
\end{equation}
which allows Eq.~\eqref{eq:convolution} to be rewritten as
\begin{equation}
	\hat{P}^{(\ell)}(s)=\hat{P}^{(1)}(s)\hat{P}^{(\ell-1)}(s)=\left[\hat{P}^{(1)}(s) \right]^{\ell}.
\end{equation}
The determination of $\hat{P}^{(1)}(s)$ from Eqs.~\eqref{eq:P12} and~\eqref{eq:P1s1} is done by defining the Laplace transform of the pair Boltzmann factor $e^{-\beta \psi(r)}$,
\begin{equation}\label{eq:1dOmega}
	\hat{\Omega}(s, \beta) \equiv \int_0^\infty \dd r e^{-rs}e^{-\beta \psi(r)}.
\end{equation}
Using this definition, $\hat{P}^{(1)}(s)$ becomes
\begin{equation}\label{eq:P1s2}
	\hat{P}^{(1)}(s) = \frac{\hat{\Omega}(s+\bp, \beta)}{\hat{\Omega}(\bp, \beta)}.
\end{equation}

Note that the dependence of $\hat{P}^{(\ell)}(s)$ on $\bp$ and $\beta$ has been omitted for clarity. This convention is used throughout this entire chapter and extend to the rest of quantities directly derived from these.

\subsection{The radial distribution function}

The \acrshort{RDF} $g(r)$ is one of the most fundamental quantities to analyze correlations between particles. Its physical meaning is that $\lambda g(r) \dd r$ gives the total number of particles located within a region of thickness $\dd r$ at a distance $r$ from a reference particle. In this definition, all possible neighbors of the reference particle must be taken into account, so that
\begin{equation}
	g(r) = \frac{1}{\lambda}\sum_{\ell=1}^{\infty} p^{(\ell)}(r),
\end{equation}
whose Laplace transform is
\begin{equation}\label{eq:1dGs1}
\hat{G}(s) = \frac{1}{\lambda} \sum_{\ell=1}^{\infty}\left[\hat{P}^{(1)}(s) \right]^{\ell} = \frac{1}{\lambda} \frac{\hat{P}^{(1)}(s)}{1-\hat{P}^{(1)}(s)} = \frac{1}{\lambda} \frac{\hat{\Omega}(s+\bp,\beta)}{\hat{\Omega}(\bp,\beta)-\hat{\Omega}(s+\bp,\beta)} ,
\end{equation}
where, in the last step, we have made use of Eq.~\eqref{eq:P1s2}. The final form of Eq.~\eqref{eq:1dGs1} means that the Laplace transform of the \acrshort{RDF} is fully determined by Eq.~\eqref{eq:1dOmega}, apart from the prefactor $1/\lambda$ involving the number density. This latter quantity is unknown \emph{a priori} since the control variables are $\bp$ and $\beta$. We will address this remaining factor later in Sec.~\ref{sec:1deos}.

The inverse Laplace transform of $\hat{G}(s)$, which allows us to recover $g(r)$ in real space, can be done analytically for the simplest potentials~\cite{S16, MS19}, whereas numerical algorithms~\cite{AW92,EulerILT} can be employed for more complex cases.

Other common correlation functions can be directly obtained from the \acrshort{RDF}. The total correlation function
\begin{equation}
h(r) = g(r) - 1
\end{equation}
and its Laplace transform
\begin{equation}
	\hat{H}(s) = \int_0^\infty \dd r e^{-rs}h(r) = \hat{G}(s) - s^{-1},
\end{equation}
are a clear example. Its Fourier transform,
\begin{equation}
	\tilde{h}(k) = \left[ \hat{H}(s)+\hat{H}(-s)\right]_{s=\imath k},
\end{equation}
where $\imath$ is the imaginary unit, allows us to obtain the structure factor as
\begin{equation}
	\tilde{S}(k) = 1 + \lambda \tilde{h}(k),
\end{equation}
 which is a key quantity in scattering experiments.

\subsection{Equation of state}\label{sec:1deos}

Because we are working in the isothermal--isobaric ensemble, the length $L$ of the system and, consequently, the density $\lambda$ are not fixed variables. This means that to fully close the method of determining the \acrshort{RDF} from Eq.~\eqref{eq:1dGs1}, one needs to compute $\lambda \equiv \lambda(\beta p,\beta)$, i.e., the equation of state.
This can be easily done from a physically consistency condition, where we impose that
\begin{equation}\label{eq:1dFVT}
	\lim_{r\to \infty}g(r) = 1 \implies \lim_{s \to 0} s \hat{G}(s) = 1,
\end{equation}
where the final value theorem for the Laplace transform has been applied. Expanding now $\hat{\Omega}(s+\bp,\beta)$ in powers of $s$ and imposing Eq.~\eqref{eq:1dFVT}, the equation of state becomes
\begin{equation}\label{eq:1deos1}
\lambda = - \frac{ \hat{\Omega}(\bp,\beta) }{ \hat{\Omega}_s(\bp,\beta) },
\end{equation}
with
\begin{equation}
	\hat{\Omega}_s(s,\beta) \equiv \frac{\partial \hat{\Omega}(s,\beta) }{\partial s} =- \int_0^\infty \dd r \, r e^{-rs}e^{-\beta \psi(r)}.
\end{equation}

\subsection{Connection to thermodynamics}\label{sec:c2_Gibbs}

Once the equation of state has been determined, the last step to fill in the gap and obtain all thermodynamic quantities is to derive the Gibbs free energy $\mathcal{G}$, which can be easily obtained by taking into account Eqs.~\eqref{eq:1dgibbsL} and~\eqref{eq:1deos1},
\begin{equation}\label{eq:1d_gibbs}
	\frac{\beta \mathcal{G}}{N} = \ln \frac{\ldb(\beta)}{\hat{\Omega}(\bp, \beta)}, 
\end{equation}
where $\ldb(\beta) = h/\sqrt{2 \pi m / \beta}$ is the de Broglie wavelength ($h$ being Planck's constant and $m$ the mass of the particles). The integration constant has been determined by imposing the ideal gas limit 
\begin{equation}
	\lim _{\bp \to 0}\frac{\beta \mathcal{G}}{N} = \frac{\beta \mathcal{G}^{\mathrm{ideal}}}{N} = \ln (\bp \ldb).
\end{equation}
The determination of the Gibbs free energy shows how the neighbor distribution can provide all structural and thermodynamic properties of the system. Once we have the Gibbs free energy, using standard thermodynamic relations one can obtain
\begin{subequations}
	\begin{align}
		\mu &= \left(\frac{\partial \mathcal{G}}{\partial N}\right)_{T,p} = \frac{\mathcal{G}}{N}, \\[7pt]
		\frac{S}{N k_\mathrm{B}} &= \left(\frac{\partial \mathcal{G}}{\partial T}\right)_{p,\mu} 
		= \frac{1}{2} - \ln \frac{\ldb(\beta)}{\hat{\Omega}(\bp,\beta)} - \frac{\bp \Omega_s(\bp, \beta) + \beta \Omega_\beta(\bp, \beta)}{\Omega(\bp, \beta)}, \\[8pt]
		\frac{\beta U}{N} &= \frac{1}{2} - \frac{\beta \Omega_\beta(\bp,\beta)}{\Omega(\bp, \beta)},
	\end{align}
\end{subequations}
where
\begin{equation}
	\hat{\Omega}_\beta(s,\beta) \equiv \frac{\partial \hat{\Omega}(s,\beta) }{\partial \beta} =- \int_0^\infty \dd r \, \psi(r) e^{-rs}e^{-\beta \psi(r)}.
\end{equation}

In many applications in theory of liquids, the equation of state is usually studied in terms of the compressibility factor
\begin{equation}
	Z(\lambda, \beta) \equiv \frac{\bp}{\lambda} = -\bp \frac{ \hat{\Omega}_s(\bp,\beta) }{ \hat{\Omega}(\bp,\beta) } ,
\end{equation}
which is equal to 1 for all densities in the ideal-gas case and can be either smaller or larger than 1 in real liquids, depending on the prevalence of attractive or repulsive interactions, respectively.
Other thermodynamic quantities that are usually studied are response functions, because they quantify how a thermodynamic property changes in reaction to a variation of another control variable, while specific constraints are held fixed. In the context of this thesis, we highlight here the (reduced) isothermal susceptibility, defined as
\begin{equation}
	\chi_T = \frac{1}{\beta}\left(\frac{\partial \lambda}{\partial p}\right)_{\beta}.
\end{equation}

\subsection{Correlation length}\label{sec:c2_CorrleationLength}

The correlation length $\xi$ of a system measures the characteristic distance over which particle correlations persist. It essentially tells us how far the mutual influence of two particles extends before they become uncorrelated. It can be measured by fitting the total correlation function $h(r) \equiv g(r) - 1$ to an exponential decay, $h(r) \sim A e^{-r/\xi}$, at sufficiently large values of $r$. Equivalently, one could fit the decay to $g(r) \sim 1+A e^{-r/\xi}$.

It is worth noting that, while some \acrshort{1D} liquids can present algebraic decay, $h(r) \sim A r^{-\eta}$, this decay is only characteristic of systems with long-range particle interactions. For the short-range interaction potentials we are working with, correlations cannot propagate indefinitely  and, once local constraints are resolved, the system becomes effectively uncorrelated.

To obtain the correlation length $\xi$, the exponential decay of $h(r)$ can be directly accessed through the knowledge of the poles of the Laplace transform $\hat{G}(s)$. We start from the Bromwich inversion formula
\begin{equation}\label{eq:Bromwich}
	g(r) = \frac{1}{2 \pi \imath} \int_{\gamma- \imath \infty}^{\gamma + \imath \infty} \dd s e^{s x} \hat{G}(s),
\end{equation}
where the integration is done along a vertical line $\mathrm{Re} \left[ s\right] = \gamma $ in the complex plane such that all singularities of $\hat{G}(s)$ lie to the left. Taking into account the residue theorem, Eq.~\eqref{eq:Bromwich} becomes
\begin{equation}\label{eq:1dgrpoles}
	g(r) = \sum_j \mathrm{Res} \left[ e^{s r} \hat{G}(s)\right]_{s=s_j},
\end{equation}
where $\{s_j\}$ is the set of all the poles of $\hat{G}(s)$. The poles can be calculated as solutions to the equation [see Eq.~\eqref{eq:1dGs1}]
\begin{equation}\label{eq:1dDet}
	\hat{\Omega}(\bp,\beta)-\hat{\Omega}(s+\bp,\beta)=0,
\end{equation}
which, in principle, can have infinitely many isolated solutions. From  Eq.~\eqref{eq:1dDet} we notice that there is always a pole at $s=0$, which does not contribute to the exponential decay in Eq.~\eqref{eq:1dgrpoles} but rather to the constant term $\lim_{r \to \infty}g(r)=1$. The rest of the nonzero poles have negative real parts and they all come as either a real pole or a pair of complex conjugates since $\hat{G}(s)$ is the Laplace transform of a real function. The asymptotic exponential decay of $h(r)$ is then given by the nonzero pole $s_0$ with the largest real part,
\begin{equation}\label{eq:1dgrpole}
	\lim_{r \to \infty} h(r) = \mathrm{Res} \left[e^{sr}  \hat{G}(s)\right]_{s=s_0}.
\end{equation}
This pole is always simple for the straightforward, short-range potentials considered here, and therefore its residue is given by
\begin{equation}
	\mathrm{Res} \left[  e^{sr}\hat{G}(s)\right]_{s=s_0} = \lim_{s \to s_0}\left[ (s-s_0)e^{sr}\hat{G}(s) \right] = \mathcal{A} e^{s_0r}.
\end{equation}
We now need to take into account whether this leading pole is real or part of a complex conjugate pair.
\begin{enumerate}
	\item For a real pole: $s_0 = -\kappa$, the contribution of the pole is given by
	\begin{equation}
	\mathrm{Res} \left[  e^{sr}\hat{G}(s)\right]_{s=s_0} = \mathcal{A} e^{- \kappa r},
	\end{equation}
	where $\mathcal{A}$ is a real number.

\item For a complex conjugate pair of the form $s_0=-\kappa + \imath \omega$ and $s_0^* = -\kappa - \imath \omega$, the contributions of the poles are
\begin{equation}
		\mathrm{Res} \left[  e^{sr}\hat{G}(s)\right]_{s=s_0} = \mathcal{A} e^{-(\kappa-\imath\omega)r}, \qquad 	\mathrm{Res} \left[  e^{sr}\hat{G}(s)\right]_{s=s_0^*} = \mathcal{A}^* e^{-(\kappa+\imath\omega)r},
\end{equation}
where $\mathcal{A}=|\mathcal{A}|e^{\imath \delta}$ and $\mathcal{A}^* =|\mathcal{A}|e^{-\imath \delta} $ are complex conjugates. Taking into account both contributions, we obtain
	\begin{equation}
	\lim_{r \to \infty} h(r) = 2 |\mathcal{A}| e^{-\kappa r} \cos(\omega r + \delta).
\end{equation}
\end{enumerate}
This analysis shows that, even though the decay of $g(r)$ is always exponential with a correlation length given by
\begin{equation}
	\xi = \kappa^{-1},
\end{equation}
the decay can be either monotonic or oscillatory, depending on whether $s_0$ is real or part of a complex conjugate pair, respectively. Note also that from Eq.~\eqref{eq:1dDet} it is clear that the values of the set $\{s_j\}$, and consequently the leading pole, depend on the value of $\bp$ and $\beta$. Furthermore, it is possible for a crossing to occur between the leading and subleading poles at a given pressure or temperature, resulting in an interchange of their positions. Although the correlation length changes continuously (even if its derivative may not), the oscillation frequency exhibits a discontinuous jump if two pairs of complex conjugate poles cross.
If this crossing occurs between a real pole and a pair of complex conjugates, the asymptotic behavior of the \acrshort{RDF} transitions from monotonic to oscillatory (or vice versa). The locus of points in the $(p,T)$ plane for which this transition occurs is called the Fisher--Widom~(\acrshort{FW}) line~\cite{FW69} and has been studied extensively in \acrshort{1D} systems~\cite{PF72,RC90,FS17,MS19,FGMS09}.


\section{Discrete mixtures}\label{sec:1dmixture}

\begin{figure}[htpb]
	\includegraphics[width=0.7\columnwidth]{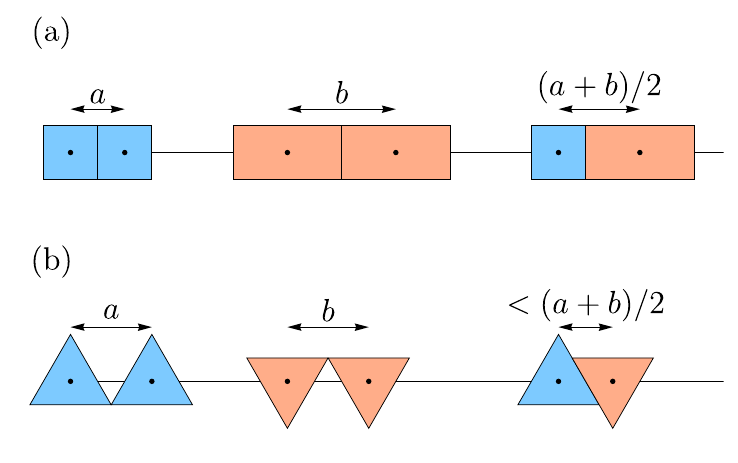}\\
	\caption{Visual example of mixtures where particles interact only through hard-core volume exclusion: (a) a binary mixture of additive particles and (b) a binary mixture of negative nonadditive particles.}
	\label{fig:c2_DiffAddNonAdd}
\end{figure}

In this section, we study the properties of \acrshort{1D} mixtures of $M$ different components. The theoretical framework developed in this section applies to both additive and nonadditive mixtures. In additive mixtures, the interaction distance between particles of different species is equal to the arithmetic mean of their individual interaction distances. In contrast, nonadditive mixtures deviate from this rule: the interaction distance between two particles from different species can be either greater than (positive nonadditivity) or less than (negative nonadditivity) the sum of their respective individual interaction distances. Figure~\ref{fig:c2_DiffAddNonAdd} illustrates this distinction with a visual comparison of binary additive and nonadditive mixtures of \acrshort{1D} particles.

As before, we work within the isothermal–isobaric ensemble, in which the temperature $T$, the pressure $p$, and the particle numbers $N_i$ of each species are held fixed. The relevant thermodynamic potential is again the Gibbs free energy of a mixture, its total differential being
\begin{equation}\label{eq:1dmgibbs}
	\dd \mathcal{G} = - S \,\dd T + L \,\dd p + \sum_i \mu_i \,\dd N_i,
\end{equation}
from where one obtains the length of the system and the chemical potential of each species as
\begin{subequations}
	\begin{equation}\label{eq:1dmgibbsL}
		L = \left(\frac{\partial \mathcal{G}}{\partial p}\right)_{T,\{N_i\}},
	\end{equation}
	\begin{equation}\label{eq:1dmgibbsMu}
		\mu_i = \left(\frac{\partial \mathcal{G}}{\partial N_i}\right)_{T,p,\{N_{j \neq i}\}}.
	\end{equation}
\end{subequations}

\subsection{Exact solution}

 Following a similar approach as for the monocomponent case, we define the \acrshort{NN} probability distribution, $p^{(1)}_{ij}(r)$, as the probability of finding the first (right) neighbor of a reference particle of species $i$ at a distance $r$ \emph{and} that it belongs to species $j$. If we define $\phi_i$ such that $\phi_i^2$ represents the mole fraction of species $i$ in the mixture, the normalization condition is
\begin{equation}\label{eq:1dm_norm}
	\sum_j \phi^2_j = 1.
\end{equation}
The probability distribution in the isothermal--isobaric ensemble, $p^{(1)}_{ij}(r)$ is given by~\cite{LZ71}
\begin{equation}\label{eq:1dm_pij}
	p^{(1)}_{ij}(r) =\frac{\phi_j}{\phi_i} A_i A_j e^{-\beta \psi_{ij}(r) - \bp r},
\end{equation}
where $\psi_{ij}(r)$ is the interaction potential between particles of species $i$ and $j$ and the parameters $\{A_i\}$ are solutions to the nonlinear set of equations,
\begin{equation}\label{eq:1dm_nlin}
	A_i \sum_j \hat{\Omega}_{ij}(\bp,\beta)\phi_j A_j = \phi_i,
\end{equation}
with
\begin{equation}
	\hat{\Omega}_{ij}(\bp,\beta) = \int_0^{\infty}\dd r e^{-s r} e^{-\beta \psi_{ij}(r)}
\end{equation}
being the Laplace transform of the Boltzmann factor. Note that Eq.~\eqref{eq:1dm_nlin} constitutes a nonlinear set of equations with, in principle, $M$ different solutions for parameters $\{A_i\}$, given a fixed composition $\{\phi_i\}$. Among all possible solutions, the physical one is identified as the one that exhibits physically meaningful behavior in the ideal gas limit, $\bp \to 0$.

\subsection{Structural properties}

\begin{figure}[htpb]
	\includegraphics[width=0.7\columnwidth]{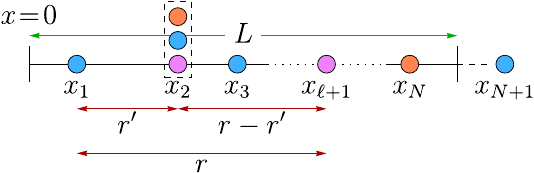}\\
	\caption{Schematic illustration of the convolution property [Eq.~\eqref{eq:convolutionmixture}] for a ternary mixture ($M=3$). The summation over species in Eq.~\eqref{eq:convolutionmixture} is shown explicitly, indicating that the particle at position $x_2$ might belong to any of the three species. Periodic boundary conditions impose $x_{N+1}=x_1+L$.}
	\label{fig:c2_1dConvolutionMixture}
\end{figure}

Starting from the definition of \acrshort{NN} probability distribution in Eq.~\eqref{eq:1dm_pij}, the single-file structure of the \acrshort{1D} system allows for the definition of any $\ell$-th \acrshort{NN} distribution as a convolution of the form
\begin{equation}\label{eq:convolutionmixture}
	p_{ij}^{(\ell)}(r)=\sum_k\int_0^r \dd r'\,p_{ik}^{(1)}(r')p_{kj}^{(\ell-1)}(r-r'),
\end{equation}
where the summation runs over all species because it is necessary to consider that, although the species of particles $i$ and $j$ are fixed, intermediate particles can belong to any species, as depicted in Fig.~\ref{fig:c2_1dConvolutionMixture}. The normalization condition is therefore
\begin{equation}
	\sum_j \int_0^\infty \dd r p_{ij}^{(\ell)}(r) = 1.
\end{equation}

The total $\ell$-th neighbor probability distribution, defined as the probability of finding the $\ell$-th neighbor from a reference particle at a distance $r$ (independently of the species of both particles) is defined as
\begin{equation}\label{eq:convolutionmixture2}
	p^{(\ell)}(r) = \sum_{i,j} \phi^2_i  p^{(\ell)}_{ij}(r).
\end{equation}
From Eqs.~\eqref{eq:convolutionmixture} and~\eqref{eq:convolutionmixture2} one can now define the longitudinal partial and total \acrshort{RDF} as 
\begin{subequations}\label{eq:1dm_g}
	\begin{equation}
		\label{eq:q1dgij}
		g_{ij}(r)=\frac{1}{\lambda \phi^2_j}\sum_{\ell=1}^\infty p_{ij}^{(\ell)}(r),
	\end{equation}
	\begin{equation}
		\label{eq:q1dgx}
		g(r)=\sum_{i,j}\phi^2_i \phi^2_j g_{ij}(r)=\frac{1}{\lambda}\sum_{\ell=1}^\infty p^{(\ell)}(r),
	\end{equation}
\end{subequations}
respectively. The partial \acrshort{RDF} is related to spatial correlations between particles of species $i$ and $j$ and the total \acrshort{RDF} is related to correlations of the overall system. As in the purely monocomponent case, the convolution structure of Eq.~\eqref{eq:convolutionmixture} suggests the introduction of the Laplace transform of Eqs.~\eqref{eq:1dm_pij} and~\eqref{eq:convolutionmixture},
\begin{subequations}
	\begin{equation}
		\hat{P}_{ij}^{(1)}(s)=\frac{\phi_j}{\phi_i} A_i A_j \hat{\Omega}_{ij}(s+\bp,\beta),
	\end{equation}
	\begin{equation}
		\hat{P}_{ij}^{(\ell)}(s)=\left(\left[\hat{\mathsf{P}}^{(1)}(s)\right]^\ell\right)_{ij},
	\end{equation}
\end{subequations}
where $\hat{\mathsf{P}}^{(1)}(s)$ is the $M \times M$ matrix of elements $\hat{P}_{ij}^{(1)}(s)$. The Laplace transform of Eqs.~\eqref{eq:1dm_g} then yields
\begin{subequations}\label{eq:1dm_Gs}
	\begin{align}\label{eq:1dm_Gsa}
		\hat{G}_{ij}(s)=&\frac{1}{\lambda \phi^2_j}\left(\sum_{\ell=1}^\infty\left[\hat{\mathsf{P}}^{(1)}(s)\right]^\ell\right)_{ij}=\frac{1}{\lambda \phi^2_j}\left(\hat{\mathsf{P}}^{(1)}(s)\cdot\left[\mathsf{I}-\hat{\mathsf{P}}^{(1)}(s)\right]^{-1}\right)_{ij},
	\end{align}
	\begin{equation}
		\hat{G}(s)=\sum_{i,j} \phi^2_i \phi^2_j \hat{G}_{ij}(s).
	\end{equation}
\end{subequations}
As in the pure monocomponent \acrshort{1D} case, the recovery of the longitudinal \acrshort{RDF} in real space, $g(r)$, can be done by performing the inverse Laplace transform of Eqs.~\eqref{eq:1dm_Gs} analytically~\cite{S16} or numerically~\cite{AW92,EulerILT}.
The longitudinal structure factor is then
\begin{equation}
	\tilde{S}(k) = 1 + \lambda \left[ \hat{G}(s)+\hat{G}(-s)\right]_{s=\imath k}.
\end{equation}

\subsection{Thermodynamic quantities}\label{sec:1dm_thermo}

As in Sec.~\ref{sec:1deos}, where the equation of state for a monocomponent \acrshort{1D} fluid was obtained by enforcing physical constraints on its structural properties, one can similarly derive the equation of state for a \acrshort{1D} mixture. The algebra, however, is considerably more complicated. For brevity, we summarize only the key results here. A detailed derivation is provided in Sec. III~B of~\nameref{a4}.

From the physical condition $\lim_{r\to\infty}g_{ij}(r)=1$ one finds the equation of state to be
\begin{equation}\label{eq:1dm_density}
	\frac{\beta}{\lambda}=-\sum_{i,j}\phi_i \phi_j A_i A_j\partial_p\Omega_{ij}(\bp,\beta).
\end{equation}
Connections with other thermodynamic properties can be done by first deriving the Gibbs free energy of the system. Using Eq.~\eqref{eq:1dm_nlin} and the fact that $\partial_p \sum_{i,j}\phi_i \phi_j A_i A_j \hat{\Omega}_{ij}(\bp,\beta)=\partial_p\sum_i \phi_i^2=0$, Eq.~\eqref{eq:1dm_density} can be rewritten as
\begin{equation}\label{eq:1dm_density2}
	\frac{\beta}{\lambda}= \sum_i \phi_i^2 \partial_p \ln A_i^2.
\end{equation}
Although Eqs.~\eqref{eq:1dm_density} and~\eqref{eq:1dm_density2} are equivalent, the former is more convenient for practical purposes, whereas the latter is primarily of theoretical interest. The key distinction lies in the fact that the dependence of $A_i$ on the pressure $p$ is unknown, preventing the direct differentiation required by Eq.~\eqref{eq:1dm_density2}, in contrast with the derivative in Eq.~\eqref{eq:1dm_density}, which can be directly performed.
The Gibbs free energy can be computed now by taking into account the thermodynamic relation in Eq.~\eqref{eq:1dmgibbsL} along with Eq.~\eqref{eq:1dm_density2}, which yields
\begin{equation}\label{eq:1dm_gibbs2}
	\frac{\beta \mathcal{G}}{N} = \sum_i \phi^2_i \ln (A_i^2 \ldb),
\end{equation}
where the integration constant has been determined by the ideal-gas condition
\begin{equation}
	\lim_{p \to 0}	\frac{\beta \mathcal{G}}{N} = \frac{\beta \mathcal{G}^{\mathrm{id}}}{N} = \sum_i \phi_i^2 \ln(\phi_i^2 \bp \ldb).
\end{equation}
In certain applications, it is also convenient to derive the excess Gibbs free energy per particle,
\begin{equation}
	\beta g^{\mathrm{ex}} =\frac{\beta(\mathcal{G}-\mathcal{G}^{\mathrm{id}})}{N} = \sum_i \phi_i^2 \ln \frac{A_i^2}{\phi_i^2 \bp}.
\end{equation}
Equation~\eqref{eq:1dm_gibbs2} provides the Gibbs free energy of the \acrshort{1D} mixture. Unlike the monocomponent case in Eq.~\eqref{eq:1d_gibbs}, the explicit dependence of $\mathcal{G}$ on its natural variables $(T, p, \{N_i\})$ remains unknown, primarily because the parameters $\{A_i\}$ are not explicitly determined as function of these variables. Nevertheless, it is still possible to derive other thermodynamic properties including the chemical potential [see Eq.~\eqref{eq:1dmgibbsMu}], which can be shown to take the form
\begin{equation}\label{eq:1d_mmu}
	\beta \mu_i= \ln (A_i^2 \ldb),
\end{equation}
where it has been assumed that all species have the same mass so that $\ldb$ is common to all of them. Equation~\eqref{eq:1d_mmu} provides a physical interpretation of the parameters $\{A_i\}$, thereby directly relating them to the chemical potential of each species as
\begin{equation}
	A_i ^2 = \frac{e^{\beta \mu_i}}{\ldb}.
\end{equation}
The internal energy can also be computed as
\begin{equation}
	U=\left(\frac{\partial \beta \mathcal{G}}{\partial \beta}\right)_{\bp,\{N_i\}}=	\frac{N}{\beta} \left[ \frac{1}{2}-\beta\sum_{i,j}\phi_i \phi_j A_i A_j \left(\frac{\partial \hat{\Omega}_{ij}(\bp,\beta)}{\partial\beta}\right)_{\bp} \right].
\end{equation}

\subsection{Correlation length}\label{sec:1dm_corrlength}
Following the same reasoning as in the monocomponent case, each partial correlation function $h_{ij}(r)\equiv g_{ij}(r)-1$ of a \acrshort{1D} mixture also decays exponentially at large distances,
\begin{equation}
	h_{ij}(r) \simeq A_{ij}\,e^{-r/\xi},
\end{equation}
for sufficiently large $r$. Since the correlation length $\xi$ is determined by the poles of the Laplace transform $\hat{G}_{ij}(s)$, it follows from Eq.~\eqref{eq:1dm_Gsa} that all $\hat{G}_{ij}(s)$ share the same set of poles. Specifically, these poles are given by the zeros of the determinant of $\mathsf{I} - \hat{\mathsf{P}}^{(1)}(s)$ [see Eq.~\eqref{eq:1dm_Gsa}], which is common to all pairs $\{i,j\}$. To recover the partial correlation functions from the poles, we can write
\begin{subequations}\label{eq:1dmgrpoles}
	\begin{equation}\label{eq:1dmgrpolesa}
		h_{ij}(r) = \sum_{k}\mathrm{Res}[e^{sr}\hat{G}_{ij}(s)]_{s=s_k},
	\end{equation}
\end{subequations}
where the sum is over all nonzero poles $s_k$. The asymptotic behavior of $h_{ij}(r)$ then depends on whether the leading pole $s_0$---the one with the real part closest to zero---is real or part of a complex conjugate pair. If $s_0 = -\kappa$ is real, its contribution to $h_{ij}(r)$ is
\begin{equation}
\mathrm{Res} [e^{sr}\,\hat{G}_{ij}(s)]_{s=s_0}= \mathcal{A}_{ij} e^{- \kappa r},
\end{equation}
where $\mathcal{A}_{ij}$ is real. The asymptotic behavior is then
\begin{equation}\label{eq:1dm_real}
	\lim_{r \to \infty} h_{ij}(r) = \mathcal{A}_{ij} e^{-\kappa r}.
\end{equation}
If the leading poles are complex conjugates $s_0=-\kappa + \imath \omega$ and $s_0^* = -\kappa - \imath \omega$, then
\begin{equation}
	\mathrm{Res} \left[  e^{sr}\hat{G}_{ij}(s)\right]_{s=s_0} = \mathcal{A}_{ij} e^{-(\kappa-\imath\omega) r}, \qquad 	\mathrm{Res} \left[  e^{sr}\hat{G}_{ij}(s)\right]_{s=s_0^*} = \mathcal{A}_{ij}^* e^{-(\kappa+\imath\omega)r},
\end{equation}
where $\mathcal{A}_{ij} = \lvert\mathcal{A}_{ij}\rvert\,e^{i\,\delta_{ij}}$ and $\mathcal{A}_{ij}^* = \lvert\mathcal{A}_{ij}\rvert\,e^{-i\,\delta_{ij}}$ are also complex conjugates.\footnote{In these expressions, $\delta_{ij}$ denotes the phase of the complex coefficient and should not be confused with the Kronecker delta symbol.} The sum of both contributions yields the asymptotic form
\begin{equation}\label{eq:1dm_complex}
	\lim_{r \to \infty} h_{ij}(r) = 2 |\mathcal{A}_{ij}| e^{-\kappa r} \cos(\omega r + \delta_{ij}).
\end{equation}
In both scenarios, the \acrshort{RDF} presents an exponential decay with a correlation length
\begin{equation}
	\xi = \kappa^{-1}.
\end{equation}

An important subtlety arises from the fact that, although all $\hat{G}_{ij}(s)$ share the same poles, the corresponding residues can differ across pairs $\{i,j\}$. As Eqs.~\eqref{eq:1dm_real} and \eqref{eq:1dm_complex} show, the leading pole’s contribution to $h_{ij}(r)$ depends on the residue $\mathcal{A}_{ij}$, which depends on the species' indices. It is then possible that $\mathcal{A}_{ij}=0$ for certain combinations of $\{i,j\}$. In this situation, the leading pole does not contribute to the decay of that specific pair, and the correlation function instead decays according to the first subleading pole with a nonzero residue.
A similar effect can occur in the total correlation function,
\begin{equation}\label{eq:1dmgrpolesb}
	h(r) =  \sum_k \sum_{i,j} \phi^2_i \phi^2_j  \mathrm{Res} \left[ e^{s r} \hat{G}_{ij}(s)\right]_{s=s_k},
\end{equation}
where symmetry considerations or cancellations in the sum $\sum_{i,j}$ may again result in the leading pole’s contribution vanishing, so that the subleading pole dominates the large-$r$ decay of $h(r)$, even if this contribution does not vanish for each individual $h_{ij}(r)$.


\section{The polydisperse limit}\label{sec:1dpolydisperse}

In Sec.~\ref{sec:1dmixture}, calculations were conducted for a mixture with a discrete number of species. However, it is also possible to work with a fully polydisperse mixture, where the particle species are continuously distributed over a property, here denoted by $y$, that characterizes them (particle size, magnetization, etc.). Besides their intrinsic interest~\cite{GKM82,PF04,P87b,SYHO17,SYHOO14,W98}, polydisperse mixtures will play a key role later on when dealing with confined systems.
In this polydisperse limit, the Gibbs free energy from Eq.~\eqref{eq:1dmgibbs} becomes
\begin{equation}\label{eq:1dpgibbs}
	\dd \mathcal{G} = - S \,\dd T + L \,\dd p + \int_\epsilon \dd y  N(y) \mu(y),
\end{equation}
where now the chemical potential $\mu(y)$ is a function of the polydisperse variable $y$, $\int_\epsilon$ represents the integral over the domain of $y$, which spans the entire range of the continuous property, and $N(y)dy$ is the number of particles with a value of the polydisperse variable between $y$ and $y+dy$. From Eq.~\eqref{eq:1dpgibbs} we obtain the length of the system and the chemical potential of each species as
\begin{subequations}
	\begin{equation}\label{eq:1dpgibbsL}
		L = \left(\frac{\partial \mathcal{G}}{\partial p}\right)_{T,N(y)},
	\end{equation}
	\begin{equation}\label{eq:1dpgibbsMu}
		\mu(y) = \left(\frac{\delta \mathcal{G}}{\delta N(y)}\right)_{T,p}.
	\end{equation}
\end{subequations}
The normalization condition for the composition distribution function, analogous to Eq.~\eqref{eq:1dm_norm}, becomes
\begin{equation}\label{eq:compositiondistribution}
	\int_\epsilon  \dd y \phi^2(y) = 1,
\end{equation}
where and $\phi^2(y)=N(y)/N$. Equation~\eqref{eq:compositiondistribution} serves as the polydisperse counterpart of the normalization condition for discrete mixtures in Eq.~\eqref{eq:1dm_norm}, where the sum over all species is now replaced by an integral over the domain of $y$. Most of the results derived in Sec.~\ref{sec:1dmixture} for discrete mixtures can be straightforwardly extended to the polydisperse limit using this transformation. Nevertheless, we explicitly present the results here to provide a complete and transparent overview. With this in mind, the resulting \acrshort{NN} probability distribution is given by
\begin{equation}\label{eq:1dp_pij}
	p^{(1)}_{\yy}(r) =\frac{\phi_\yp}{\phi_\y} A_\y A_\yp e^{-\beta \psi_\yy(r) - \bp r},
\end{equation}
 where, for conciseness and closer analogy to the discrete mixture case, the notation $f_y$ is used instead of the more conventional $f(y)$ to denote the dependence of any function $f$ on the polydisperse variables. Note also that the dependence of $A_y$ and $\phi_y$  on $\bp$ and $\beta$ has been omitted for brevity. For a given composition distribution $\phi_y^2$, the function $A_y$ is the solution to the integral equation
 \begin{equation}\label{eq:1dp_nlin}
 	A_\y \int_\epsilon \dd \yp \hat{\Omega}_\yy(\bp,\beta)\phi_\yp A_\yp = \phi_\y,
 \end{equation}
where $\hat{\Omega}_\yy(s,\beta)$ is the Laplace transform of the Boltzmann factor
\begin{equation}
	\hat{\Omega}_\yy(s,\beta) = \int_0^{\infty}\dd r e^{-s r} e^{-\beta \psi_\yy(r)}.
\end{equation}
The function $A_y$ can again be related to the chemical potential using Eq.~\eqref{eq:1dpgibbsMu} to obtain
\begin{equation}\label{eq:1dp_mmu}
	\beta \mu_y=\left(\frac{\delta \beta \mathcal{G}}{\delta{N_y}}\right)_{\beta,p,\{N_{y'\neq y}\}}=\frac{1}{N}\left(\frac{\delta \beta \mathcal{G}}{\delta{\phi^2_y}}\right)_{\beta,p,\{N_{y'\neq y}\}} = \ln (A_y^2 \ldb),
\end{equation}
where we have assumed that all species share the same mass so that $\ldb$ is common across the entire polydisperse variable.

\subsection{Structural properties}
The structural properties of the polydisperse mixture can be obtained as in the discrete case by taking into account the convolution property of the $\ell$-th \acrshort{NN} probability distribution
\begin{equation}\label{eq:1dp_conv}
		p_\yy^{(\ell)}(r)=\int_\epsilon \dd y_3 \int_0^r \dd r'\,p_{y_1,y_3}^{(1)}(r')p_{y_3,y_2}^{(\ell-1)}(r-r'),
\end{equation}
from which the partial and total \acrshort{RDF}s can be derived. The polydisperse counterparts of Eqs.~\eqref{eq:1dm_g} are
\begin{subequations}\label{eq:1dp_g}
	\begin{equation}
		\label{eq:1dp_gij}
		g_\yy(r)=\frac{1}{\lambda \phi^2_\yp}\sum_{\ell=1}^\infty p_\yy^{(\ell)}(r),
	\end{equation}
	\begin{equation}
		\label{eq:1dp_gx}
		g(r)=\int_\epsilon \dd \y \int_\epsilon \dd y_2 \phi^2_\y \phi^2_\yp g_\yy(r).
	\end{equation}
\end{subequations}
Because in the polydisperse limit the convolution structure of $p_\yy^{(\ell)}(r)$ is maintained [see Eq.~\eqref{eq:1dp_conv}], we introduce the Laplace transform of Eqs.~\eqref{eq:1dp_pij} and~\eqref{eq:1dp_conv} as
\begin{subequations}
\begin{equation}\label{eq:1dp_pijs}
		\hat{P}_\yy^{(1)}(s)=\frac{\phi_\yp}{\phi_\y} A_\y A_\yp \hat{\Omega}_\yy(s+\bp,\beta),
\end{equation}
\begin{equation}\label{eq:1dp_pl}
	\hat{P}_\yy^{(\ell)}(s)=\int_\epsilon \dd y_3 \hat{P}_{y_1,y_3}^{(1)}(s)\hat{P}_{y_3,y_1}^{(\ell-1)}(s)=\left(\left[\hat{\mathsf{P}}^{(1)}(s)\right]^\ell\right)_\yy,
\end{equation}
\end{subequations}
where, in the second step of Eq.~\eqref{eq:1dp_pl}, the standard definition for matrix multiplication of infinite-dimensional matrices (analogous to the finite case) has been applied. The Laplace transform of Eqs.~\eqref{eq:1dp_g} is then
\begin{subequations}\label{eq:1dp_Gs}
	\begin{align}\label{eq:1dp_Gsa}
		\hat{G}_\yy(s)=&\frac{1}{\lambda \phi^2_\yp}\left(\sum_{\ell=1}^\infty\left[\hat{\mathsf{P}}^{(1)}(s)\right]^\ell\right)_\yy=\frac{1}{\lambda \phi^2_\yp}\left(\hat{\mathsf{P}}^{(1)}(s)\cdot\left[\mathsf{I}-\hat{\mathsf{P}}^{(1)}(s)\right]^{-1}\right)_\yy,
	\end{align}
	\begin{equation}\label{1dp_Gsb}
		\hat{G}(s)=\int_\epsilon \dd \y \int_\epsilon \dd y_2  \phi^2_\y \phi^2_\yp \hat{G}_\yy(s),
	\end{equation}
\end{subequations}
where the ($\yy$) element of the unit matrix $\mathsf{I}$ is $\delta(\y-\yp)$. Equation \eqref{eq:1dp_Gsa} is simply the formal solution to the integral equation
\begin{equation}\label{eq:1dp_Gs2}
\frac{\phi_{y_2}}{ A_{y_2}}\hat{G}_{y_1 y_2}(s)=\int_{\epsilon} {d}y_3\,\phi_{y_3}\hat{G}_{y_1,y_3}(s)A_{y_3}\hat{\Omega}_{y_2,y_3}(s+\bp)+\frac{A_{y_1}}{\lambda\phi_{y_1}}\hat{\Omega}_{y_3,y_2}(s+\bp).
\end{equation}

\subsection{Thermodynamic quantities}
The derivation of the thermodynamic properties can be done following analogous steps as in Sec.~\ref{sec:1dm_thermo}. Again, we highlight here the main results, but a more detailed derivation can be found in Sec. III~D of \nameref{a4}. The equation of state is found to be
\begin{equation}\label{eq:1dp_density}
	\frac{\beta}{\lambda}=-\int_\epsilon \dd \y \phi_\y  A_\y \int_\epsilon \dd \yp \phi_\yp A_\yp \, \partial_p\hat{\Omega}_\yy(\bp,\beta),
\end{equation}
from where, using again Eqs.~\eqref{eq:1dpgibbsL} and~\eqref{eq:1dp_density}, we can obtain the Gibbs free energy as
\begin{equation}\label{eq:1dp_gibbs}
	\frac{\beta \mathcal{G}}{N} = \int_\epsilon \dd y \phi^2_y \ln (A_y^2 \ldb).
\end{equation}
The integration constant in Eq.~\eqref{eq:1dp_gibbs} has been determined by the ideal-gas condition for a polydisperse mixture
\begin{equation}
	\lim_{p \to 0}	\frac{\beta \mathcal{G}}{N} = \frac{\beta \mathcal{G}^{\mathrm{id}}}{N} = \int_\epsilon \dd y \phi_y^2 \ln(\phi_y^2 \bp \ldb),
\end{equation}
where, again, we have assumed that $\ldb$ is common to all species.The excess Gibbs free energy per particle then becomes
\begin{equation}\label{eq:1dp_gibbs2}
	\beta g^{\mathrm{ex}} = \int_\epsilon \dd y \phi_y^2 \ln \frac{A_y^2}{\phi_y^2 \bp}.
\end{equation}

\subsection{Correlation length}
The calculation of the correlation length is analogous to the one made for the discrete mixture in Sec.~\ref{sec:1dm_corrlength}. By using the same logic, one arrives at the fact that the decay of the correlation function is either an exponential monotonic decay
\begin{equation}\label{eq:1dp_real}
		\lim_{r \to \infty} h_\yy(r) = \mathcal{A}_\yy e^{-\kappa r},
\end{equation}
if the leading pole is real $s_0 = -\kappa$, or an exponentially damped oscillatory decay
\begin{equation}\label{eq:1dp_complex}
	\lim_{r \to \infty} h_\yy(r) =2 |\mathcal{A}_\yy| e^{-\kappa r} \cos(\omega r + \delta_\yy),
\end{equation}
in case the leading poles are a pair of complex conjugates.
The decay of the total correlation function, which defines the large-$r$ correlations of the overall system by taking into account all species, is then given by the double integral
\begin{equation}\label{eq:1dp_grpolesb}
	h(r) = \sum_k \int_\epsilon \dd \y \int_\epsilon \dd \yp \phi^2_\y \phi^2_\yp  \mathrm{Res} \left[ e^{s r} \hat{G}_\yy(s)\right]_{s=s_k}.
\end{equation}
Once again, situations may arise in which, due to special symmetries in the function $\mathrm{Res} \left[ e^{s r} \hat{G}_\yy(s)\right]_{s=s_k}$, the integrals $\int_\epsilon \dd \y \int_\epsilon \dd \yp$ might vanish for a specific pole $s_k$. If this is the case for the leading pole $s_0$, then it is possible that the leading pole contributes to the partials \acrshort{RDF} $g_{\yy}(r)$ for all or specific pairs of species, but that this pole's contribution is absent in the total \acrshort{RDF} $g(r)$.

%% file: chapters/C2_2_TheoryOfQ1DLiquids.tex
\thispagestyle{empty}

\chapter{Quasi one-dimensional liquids}\label{1DMixtureTheory}

Quasi-one-dimensional liquids are systems confined within geometries where the available space in one dimension is significantly larger than in the remaining directions, forcing particles to remain in single-file formation. This description typically refers to particles confined in nanopores or very narrow channels. The higher-dimensional nature of the system makes it harder to obtain exact results than in the pure \acrshort{1D} case studied in Chapter~\ref{1dTheory}, and the pronounced anisotropy invalidates many techniques developed for isotropic bulk systems.

These limitations highlight the need for a dedicated theoretical framework to study these systems. The \acrshort{TM} technique is a powerful method that yields the exact equation of state and some structural properties. However, key quantities such as the \acrshort{RDF} remain inaccessible. In this chapter we present a novel mapping method developed in this thesis that overcomes those gaps. The approach reproduces all \acrshort{TM} results and, crucially, delivers exact expressions for every thermodynamic and structural property of a \acrshort{Q1D} fluid, including the full \acrshort{RDF} that had previously eluded exact theoretical treatment.

\section{The mapping approach}\label{th_mapping}

The key to determining structural and thermodynamic properties of \acrshort{Q1D} systems lies in the fact that they are isomorphic to a polydisperse mixture of \acrshort{1D} nonadditive rods, in which all species of the mixture share the same chemical potential. In this context, \emph{isomorphic} refers to the existence of a one-to-one correspondence between the physical properties of the \acrshort{Q1D} system and those of its equivalent \acrshort{1D} mixture. Although the equivalence of both statistical ensembles is more easily shown using the grand canonical ensemble (see Appendix~A of~\nameref{a6}), the equivalence of statistical ensembles in the thermodynamic limit ensures that the correspondence holds in any ensemble of choice.

Following the same approach used for general \acrshort{1D} systems, we work in the $(\{N_i\},\bpp, \beta)$ ensemble to derive the exact solution of the mapped \acrshort{1D} mixture in order to obtain the corresponding thermodynamic potential: the Gibbs free energy $\mathcal{G}(\{N_i\},\bpp, \beta)$. Note the change in notation $\bp \to \bpp$ with respect to Chapter~\ref{1dTheory} to emphasize that in a \acrshort{Q1D} geometry the pressure has several components and $\bpp$ represents the \emph{longitudinal} component: the one that represents the single pressure $\bp$ when the \acrshort{Q1D} system reduces to a purely \acrshort{1D} one. Adopting $\bpp$ therefore helps distinguish the longitudinal pressure component (along the nonconfined direction) from the pressures arising in other, confined directions.

Once the thermodynamic potential is obtained, ensemble equivalence with the \acrshort{Q1D} counterpart allows it to be reinterpreted as the Gibbs–Helmholtz free energy of the \acrshort{Q1D} system, $\mathcal{G}(\bpp,L_\perp, \beta)$, where $L_\perp$ represents the size of the system along the confined directions. It is important to clarify that the term \emph{confined} directions encompasses not only spatial confinement---i.e., directions with limited available volume---but also internal degrees of freedom that are similarly restricted. For example, orientational degrees of freedom can also be considered as part of the confined dimensions we are referring to.

This equivalence is only valid as long as the condition of equal chemical potential across all species in the \acrshort{1D} mixture is imposed. From a physical perspective, although each species in a general mixture would typically have its own chemical potential, our theoretical \acrshort{1D} mixture treats each species as the same particle from the \acrshort{Q1D} system, differing only by its transverse position, orientation, or other confined degrees of freedom. Such a transformation does not change the Gibbs free energy. Therefore, each species must have the same chemical potential for the equivalence to remain valid.

\begin{figure}[htpb]
	\includegraphics[width=0.7\columnwidth]{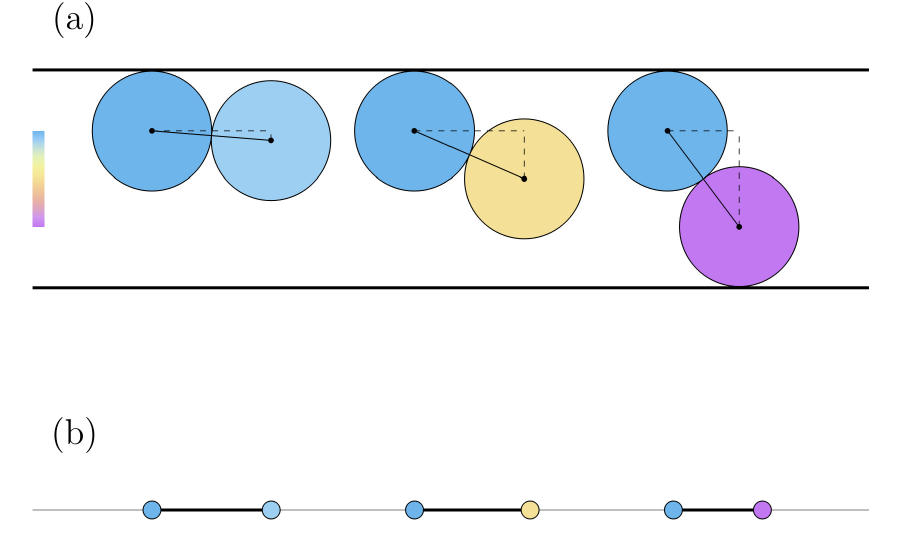}\\
	\caption{(a) Q1D system of hard disks confined within a channel that allows a single transverse degree of freedom. Each disk is colored according to its transverse coordinate. Both the transverse and longitudinal components of the contact distance between disks are indicated. (b) Equivalent 1D mixture obtained by mapping each disk to a  particle on a line. Each circle, colored according to species, represents the center of a 1D particle. The contact distance between a pair of particles (illustrated by a thick, solid line) corresponds to the longitudinal contact distance shown in panel (a).}
	\label{fig:c2_q1dDisk}
\end{figure}

Apart from the mathematical methods, the ensemble equivalence between the \acrshort{Q1D} system and its \acrshort{1D} mixture counterpart is often more intuitively understood through visual representations. As an illustrative example, Fig.~\ref{fig:c2_q1dDisk} depicts the mapping for a \acrshort{Q1D} system of hard disks. In panel~(a), the \acrshort{Q1D} configuration is shown, with disks colored according to their transverse $y$-coordinate. Although the true distance between two particles at contact is always the same, the longitudinal component of that distance varies depending on the transverse positions of both particles. In the corresponding \acrshort{1D} mixture shown in panel~(b), the transverse positional information (i.e., the $y$-coordinate) is encoded in the particle species. As a result, the longitudinal contact distance, now the only physically meaningful one in one dimension, becomes species-dependent, capturing the geometric constraints of the original \acrshort{Q1D} system.

\begin{figure}[htpb]
	\includegraphics[width=0.7\columnwidth]{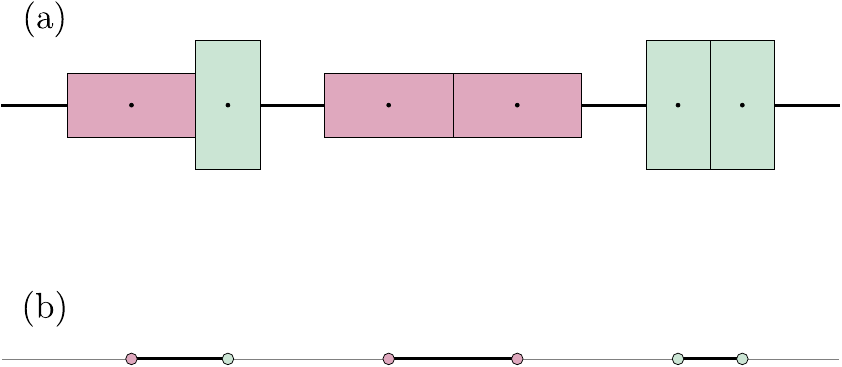}\\
	\caption{(a) \acrshort{Q1D} system of hard rectangles confined to move along a single axis, with two states of rotational freedom. Rectangles are colored by their rotation angle. (b) Equivalent \acrshort{1D} mixture obtained by mapping each rectangle to \acrshort{1D} particles. In this representation, the contact distance is the same as the contact distance from panel (a) and points, which represent the position of each particle's center, are colored by species.}
	\label{fig:c2_q1dRectangles}
\end{figure}

While Fig.~\ref{fig:c2_q1dDisk} shows a system with a single spatially confined dimension, the same rationale can be straightforwardly extended to geometries where two spatial dimensions are confined. Furthermore, as discussed previously, confined directions encompass more than just spatial limitations. Figure~\ref{fig:c2_q1dRectangles} provides an example where the confined coordinate is the orientational degree of freedom, restricted to only two possible orientations.

Beyond the physical distinction between spatial and orientational confinement, there is also a fundamental difference in how the two cases illustrated in Figs.~\ref{fig:c2_q1dDisk} and \ref{fig:c2_q1dRectangles} map to \acrshort{1D} mixtures. In the first example, the mapping variable---the $y$-coordinate---varies continuously, leading to a polydisperse mixture upon mapping into one dimension. On the other hand, in the second example, the orientational coordinate is discrete, producing a mixture with a finite number of components.

It is important to emphasize that, while particles in the \acrshort{Q1D} system can be disks, rectangles, etc., where any interaction potential is valid, provided it complies with restrictions imposed in  Sec.~\ref{sec:1dtheory}, the \acrshort{1D} representation treats them as rod-like particles constrained to move along a single axis. The interaction potential or contact distance between these effective rods, determined by their species, does not necessarily correspond to any physical interaction in a real \acrshort{1D} system. It merely serves as a mathematical construct designed to reproduce the structural and thermodynamic behavior of the original \acrshort{Q1D} system.

\section{Exact solution}\label{th_q1dEQC}

By means of the mapping between \acrshort{Q1D} systems and \acrshort{1D} mixtures explained in Sec.~\ref{th_mapping}, the exact solution for \acrshort{Q1D} systems can be directly obtained from the exact solution for \acrshort{1D} mixtures derived in Secs.~\ref{sec:1dmixture} and~\ref{sec:1dpolydisperse}. In order for this solution to be applicable to a \acrshort{Q1D} system, the equal chemical potential condition must be applied to the \acrshort{1D} discrete mixture, in the case of discrete degrees of freedom, or to a polydisperse mixture in the case of continuous degrees of freedom. Throughout this section, we take the unconfined (longitudinal) axis to be oriented along the $x$-direction and the vector $\mathbf{r}$ to be the higher-dimensional position vector.

\subsection{Discrete degrees of freedom}

In the discrete case, the condition of equal chemical potential is imposed through parameters $\{A_i\}$, which are related to the chemical potential by Eq.~\eqref{eq:1d_mmu}, which means
\begin{equation}
	\mu_i = \mu \implies A_i = A.
\end{equation}
The nonlinear set of equations in Eq.~\eqref{eq:1dm_nlin} becomes
\begin{equation}\label{eq:q1d_eig}
 \sum_j \hat{\Omega}_{ij}(\bpp,\beta)\phi_j = \frac{1}{A^2}\phi_i,
\end{equation}
where Eq.~\eqref{eq:q1d_eig} is now an eigenvalue equation in which the relevant eigenpair is the one corresponding to the maximum eigenvalue, as argued in Appendix~\ref{ProofOfMaxA2}. Note that the maximum eigenvalue of Eq.~\eqref{eq:q1d_eig} corresponds to the minimum value of $A$.
The first step to obtain the exact solution for the \acrshort{Q1D} system is then to solve Eq.~\eqref{eq:q1d_eig} and to keep the largest eigenvalue and its corresponding eigenvector. After normalization according to Eq.~\eqref{eq:1dm_norm}, we can also derive the mole fraction of each species, i.e. the set $\{\phi^2_i\}$.

It is important to keep in mind that the information about the transverse position, orientation, or any other possible degree of freedom of the particle is now encoded in the species information via the mapping approach. Throughout the remainder of this section, we refer to this dependence as the dependence on the species of the particle, but one should keep in mind that each \emph{species} has its corresponding physical meaning in the original \acrshort{Q1D} system under study. As an example, the mole fractions $\{\phi^2_i\}$ of the species in the \acrshort{1D} mixture correspond to the number fractions of particles at a specific transverse position (spatial confinement) or at a specific orientation (rotational freedom).

The structural and thermodynamic properties can then be obtained from the \acrshort{NN} probability distribution in Eq.~\eqref{eq:1dm_pij}, which now becomes\footnote{To better reflect the \acrshort{Q1D} nature of the system, throughout this section the distance along the unconfined axis is now denoted by $x$.}
\begin{equation}\label{eq:q1d_pij}
	p^{(1)}_{ij}(x) =\frac{\phi_j}{\phi_i} A^2 e^{-\beta \psi_{ij}(x) - \bpp x},
\end{equation}
its Laplace transform being
\begin{equation}
	\hat{P}_{ij}^{(1)}(s)=\frac{\phi_j}{\phi_i} A^2 \hat{\Omega}_{ij}(s+\bpp,\beta).
\end{equation}
The equation of state from Eq.~\eqref{eq:1dm_density} is
\begin{equation}\label{eq:q1d_density}
	\frac{\beta}{\lambda}=-A^2 \sum_{i,j}\phi_i \phi_j\partial_p\hat{\Omega}_{ij}(\bpp,\beta),
\end{equation}
and the Gibbs free energy can be now directly obtained from Eq.~\eqref{eq:1dm_gibbs2} as
\begin{equation}\label{eq:q1d_gibbs}
	\frac{\beta \mathcal{G}}{N} = \ln (A^2 \ldb).
\end{equation}
The \acrshort{RDF} is still defined through the knowledge of $p^{(1)}_{ij}(x)$ as in Eqs.~\eqref{eq:1dm_g}. The only difference is that the Laplace transform of the partial \acrshort{RDF} $\hat{G}_{ij}(s)$ from Eq.~\eqref{eq:1dm_Gsa}, can be rewritten as
	\begin{align}\label{eq:q1d_Gsa}
	\hat{G}_{ij}(s)=&\frac{1}{\lambda \phi^2_j}\left(\hat{\mathsf{P}}^{(1)}(s)\cdot\left[\mathsf{I}-\hat{\mathsf{P}}^{(1)}(s)\right]^{-1}\right)_{ij} \nonumber \\
	=& \frac{A^2}{\lambda \phi_i \phi_j} \left(\hat{\mathsf{\Omega}} (s+\bpp,\beta)\cdot\left[\mathsf{I}-A^2\hat{\mathsf{\Omega}}(s+\bpp, \beta)\right]^{-1}\right)_{ij}.
\end{align} 

\subsection{Continuous degrees of freedom}\label{sec:q1d_continuous}

If the mapped variable of the \acrshort{Q1D} system is a continuous one (continuous transverse position, continuous rotation, etc.), the mapping must be done to a polydisperse mixture, whose exact solution was derived in Sec.~\ref{sec:1dpolydisperse}. Imposing the equal chemical potential condition now implies that, according to Eq.~\eqref{eq:1dp_mmu}, $A_y = A$ is a constant on the variable $y$ (note that it can still depend on $\bpp$ and $\beta$).

The integral equation from Eq.~\eqref{eq:1dp_nlin} now becomes a homogeneous Fredholm integral equation of the second kind:
 \begin{equation}\label{eq:q1dp_eig}
	\int_\epsilon \dd \yp \hat{\Omega}_\yy(\bpp,\beta)\phi_\yp = \frac{\ell}{\bpp} \phi_\y, \qquad \ell = \frac{\bpp}{A^2},
\end{equation}
which recovers the \acrshort{TM} results~\cite{KP93}. The \acrshort{NN} probability distribution in real and Laplace spaces are directly obtained from Eqs.~\eqref{eq:1dp_pij} and~\eqref{eq:1dp_pijs} as
\begin{equation}\label{eq:q1dc_pij}
	p^{(1)}_{\yy}(x) =\frac{\phi_\yp}{\phi_\y} A^2 e^{-\beta \psi_\yy(x) - \bpp x},
\end{equation}
and
\begin{equation}\label{eq:q1dc_pijs}
	\hat{P}_\yy^{(1)}(s)=\frac{\phi_\yp}{\phi_\y} A^2 \hat{\Omega}_\yy(s+\bpp,\beta),
\end{equation}
respectively.
The equation of state and the Gibbs free energy become
\begin{equation}\label{eq:q1dc_density}
	\frac{\beta}{\lambda}=- A^2 \int_\epsilon \dd \y \phi_\y \int_\epsilon \dd \yp \phi_\yp \, \partial_p\hat{\Omega}_\yy(\bpp,\beta),
\end{equation}
\begin{equation}\label{eq:q1dc_gibbs}
	\frac{\beta \mathcal{G}}{N} = \ln (A^2 \ldb).
\end{equation}
Note that Eq.~\eqref{eq:q1dc_gibbs} is formally equivalent to its discrete counterpart in Eq.~\eqref{eq:q1d_gibbs}, but parameter $A$ must be calculated differently in both cases. The RDF corresponds to the infinite-dimensional, continuous analogue of Eq.~\eqref{eq:q1d_Gsa},
\begin{equation}\label{eq:q1dp_Gsa}
	\hat{G}_\yy(s)= \frac{A^2}{\lambda \phi_\y \phi_\yp} \left(\hat{\mathsf{\Omega}}(s+\bpp,\beta)\cdot\left[\mathsf{I}-A^2\hat{\mathsf{\Omega}}(s+\bpp, \beta)\right]^{-1}\right)_\yy,
\end{equation}
where now the $(\yy)$ element of the unit matrix $\mathsf{I}$ is $\delta(\y - \yp)$.

\section{Spatially confined Q1D systems}

Although the mapping approach presented in this chapter is broadly applicable, the majority of the models examined in this thesis are \acrshort{Q1D} fluids confined by narrow geometries. Accordingly, this section focuses on refining the treatment of such models: we highlight their distinctive features and derive selected thermodynamic and structural properties by applying the proposed theoretical framework.

In this class of systems, the variable $y$, as defined for polydisperse systems in Sec.~\ref{sec:1dpolydisperse}, now denotes the transverse coordinate of each particle, which can be either \acrshort{1D} or \acrshort{2D}, depending on the number of confined directions in which particles are allowed to move. The integral $\int_\epsilon$, initially introduced in Eq.~\eqref{eq:compositiondistribution}, now represents the integral over the volume of the transverse direction(s), which is again dependent on the dimensionality. Consequently, the parameter $\epsilon$ represents either the width of the channel---for only one confined direction---or the area of the transverse section---for two confined directions.

\subsection{Thermodynamic quantities}

All thermodynamic properties follow from the Gibbs free energy in Eq.~\eqref{eq:q1dc_gibbs}. However, for some applications, it may be more suitable to work with the \emph{excess} Gibbs free energy per particle. In order to derive it, we first need to calculate the Gibbs free energy of the corresponding ideal system. For an ideal gas confined in the pore, every point of the transverse cross-section is equally likely, so the transverse mole-fraction density is uniform: $\phi_y = c$. Taking into account the normalization condition in Eq.~\eqref{eq:compositiondistribution}, that constant is found to be $\phi_y = 1/\sqrt{\mathcal{V}_\epsilon}$, where
\begin{equation}
	\mathcal{V}_\epsilon = \int_\epsilon \dd y
\end{equation}
is the volume of the transverse cross section of the confining channel. This means that the ideal-gas Gibbs free energy for this mixture becomes
\begin{equation}
	\frac{\beta \mathcal{G}^{\mathrm{id}}}{N} = \int_\epsilon \dd y \phi^2_y \ln\left(\phi_y^2 \bpp \ldb\right) =\ln\frac{\bpp \ldb}{\mathcal{V}_\epsilon}.
\end{equation}
The excess Gibbs free energy per particle is then easily derived as
\begin{equation}\label{eq:spa_gibbs2}
	\beta g^{\mathrm{ex}} = \ln \left(\mathcal{V}_\epsilon \frac{A^2}{\bpp}\right) \equiv - \ln \frac{\ell}{\mathcal{V}_\epsilon},
\end{equation}
which recovers the \acrshort{TM} results.

The excess Gibbs free energy derived in Eq.~\eqref{eq:spa_gibbs2} for a purely \acrshort{1D} polydisperse mixture can now be viewed as the Gibbs--Helmholtz thermodynamic potential for the confined system. Both components of the compressibility factor, the longitudinal $Z_\|$ and transverse one $Z_\perp$, are then computed using standard thermodynamic relations as
\begin{subequations}
\begin{align}
	Z_\|= &1+\bpp \left(\frac{\partial \beta g^{\mathrm{ex}}}{\partial \bpp}\right)_\epsilon, \\[8pt]
	Z_\perp= &1-\mathcal{V}_\epsilon \left(\frac{\partial \beta g^{\mathrm{ex}}}{\partial \mathcal{V}_\epsilon}\right)_{\bpp}.
\end{align}
\end{subequations}

\subsection{Structural properties}\label{th_rdf3D}

In Sec.~\ref{sec:q1d_continuous}, we introduced the partial longitudinal \acrshort{RDF} $g_\yy(x)$, which quantifies correlations between two particles whose transverse coordinates lie at $\y$ and $\yp$, respectively, with a longitudinal separation $x$.  By integrating over all transverse positions we obtain the total longitudinal \acrshort{RDF}, $g(x)$, which captures the overall pair correlations in the system.
Both $g_\yy(x)$ and $g(x)$ refer exclusively to \emph{longitudinal} correlations, i.e., correlations measured along the unconfined axis. Because that direction retains translational invariance, these longitudinal \acrshort{RDF}s are well defined and depend only on the longitudinal distance $x$.

\begin{figure}[htpb]
	\includegraphics[width=0.9\columnwidth]{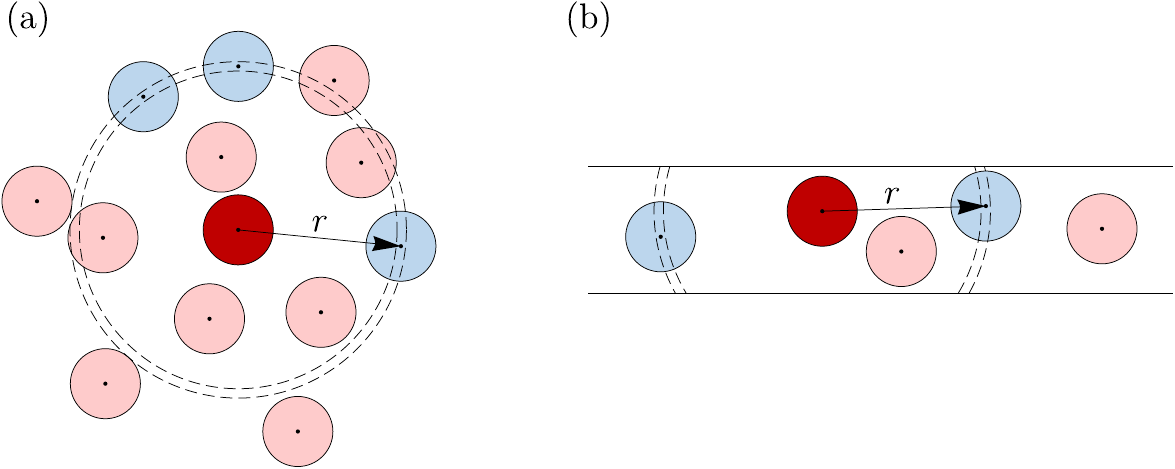}\\
	\caption{Schematic representation of the definition of the usual \acrshort{RDF} in (a) a bulk system, where spatial isotropy is conserved, versus (b) its counterpart under confinement, where the definition of a global \acrshort{RDF} is not as straightforward due to the loss of translation invariance.}
	\label{fig:c2_DefineRDF}
\end{figure}

Defining a global \acrshort{RDF} $g(r)$ that measures correlations between particles at a distance $r$ along any given direction is therefore not as straightforward as it was for its longitudinal counterpart, due to the loss of translational invariance along the transverse direction. Firstly, the geometry of the system is inherently anisotropic, and only certain interparticle distances $r$ are geometrically allowed. This leads to a nonuniform sampling of distances, as illustrated in Fig.~\ref{fig:c2_DefineRDF}. Secondly, the density profile along the confined direction is not constant, which makes the system inhomogeneous. As a result, the probability of finding a particle at a given position depends strongly on the local environment.

These two issues introduce key complications. Any attempt to define a global, scalar $g(r)$ necessarily involves averaging over positions, thereby discarding the position-dependent correlations that are intrinsic to inhomogeneous systems and resulting in an inherent loss of spatial information. Additionally, normalization becomes ambiguous: in homogeneous systems, $g(r)$ is typically normalized by dividing the two-body correlation function by the square of the average density, assuming uncorrelated, ideal-gas-like, particle distributions. In inhomogeneous confined systems, this assumption fails, and alternative normalization schemes become necessary. For instance, one could normalize using an ideal-gas reference of noncorrelated but confined system, or construct a reference distribution of non-interacting particles that still respects the actual system's spatial inhomogeneity along the transverse directions.

Despite these challenges, it is still possible to define a nominal radial distribution function, which we denote $\hat{g}(r)$, that remains meaningful in a confined geometry. Specifically, one can define $\hat{g}(r)$ such that:
$2 \lambda\, \hat{g}(r)\, \dd r $ denotes the average number of particles located at a distance between $r$ and $r + \dd r$ from a reference particle.

The position of any particle in the \acrshort{Q1D} system is given by $\mathbf{r} = x \hat{\mathbf{x}} + \mathbf{r}_\perp$, where $\hat{\mathbf{x}}$ is the unit vector along the unconfined direction and $\mathbf{r}_\perp$ represents the position vector in the transverse sub-space, whose dimensionality can be 1 or 2, depending on the geometry of the system. We can now define $n_1(\mathbf{r}) = \lambda \phi_y^2$ as the local number density. The two-body configurational distribution function $n_2(\mathbf{r_1},\mathbf{r_2})$, that measures the number of pairs of particles such that one of the particles is inside the region $[\mathbf{r_1},\mathbf{r_1}+\dd \mathbf{r_1}]$ and the other one sits inside $[\mathbf{r_2},\mathbf{r_2}+\dd \mathbf{r_2}]$, is given by
\begin{equation}
n_2(\mathbf{r_1},\mathbf{r_2}) = n_1(\mathbf{r_1})n_1(\mathbf{r_2}) g(\mathbf{r_1},\mathbf{r_2}) = \lambda^2 \phi_\y^2 \phi_\yp^2 g_\yy(x_{12}),
\end{equation} 
where $x_{12} = |x_2 - x_1|$. The normalization condition is
\begin{equation}
\int \dd \mathbf{r_1} \int \dd \mathbf{r_2} n_2(\mathbf{r_1},\mathbf{r_2}) = N^2.
\end{equation}
Let us now define $\hat{n}(r)$ such that $\hat{n}(r)\dd r$ is the average number of particles at a distance between $r$ and $r+\dd r$ from a reference particle. As a marginal distribution, it can be derived from $n_2(\mathbf{r_1},\mathbf{r_2})$ as
\begin{equation}
	\hat{n}(r)=\frac{1}{N} \int \dd \mathbf{r_1} \int \dd \mathbf{r_2}n_2(\mathbf{r_1},\mathbf{r_2}) \delta(r - r_{12}),
\end{equation}
where $r_{12}=|\mathbf{r}_1-\mathbf{r}_2|$. The \acrshort{RDF} $\hat{g}(r)$ can now be computed from $\hat{n}(r)$ as
\begin{equation}\label{eq:q1d_gr0}
	\hat{g}(r) = \frac{\hat{n}(r)}{2 \lambda}.
\end{equation}
As an example, in a \acrshort{Q1D} system of confined hard disks, it can be shown that 
\begin{equation}\label{eq:q1d_gr}
	\hat{g}(r)=\int_{\epsilon}^\dagger \dd \y \int_{\epsilon}^\dagger \dd \yp\, \phi^2_{\y}\phi^2_{\yp} \hat{g}_\yy(r),
\end{equation}
where the dagger symbolizes the geometric constraint $y_{12}^2<r^2$ imposed on the integrals and
\begin{equation}
	\hat{g}_\yy(r) = \frac{r}{\sqrt{r^2-y_{12}^2}} g_{\yy}\left(\sqrt{r^2-y_{12}^2}\right).
\end{equation}
A full derivation of Eq.~\eqref{eq:q1d_gr} can be found in \nameref{a5}. In addition, the specific expression for Eq.~\eqref{eq:q1d_gr0} for the case of the cylindrical confinement of hard spheres is derived in Sec. III~C of \nameref{a7}.

%% file: chapters/C3_OneDimensionalSystems.tex
\thispagestyle{empty}

\chapter{Structural transitions in one- and three-dimensional systems}\label{1Dand3Dcomparison}

\section{Summary}

One-dimensional fluids with short-range interactions cannot undergo thermodynamic phase transitions at finite temperature~\cite{vH50}. However, it is not uncommon for them to experience structural transitions, where at certain values of pressure and temperature, the correlation length presents a kink and the oscillatory asymptotic decay of the \acrshort{RDF} has a discontinuous jump. If this jump occurs between two different oscillatory decays, a discontinuous oscillatory crossover~(\acrshort{DOC}) emerges. If the jump occurs between an oscillatory and monotonic decay, we find a \acrshort{FW} line.

\begin{figure}
	\includegraphics[width=0.6\columnwidth]{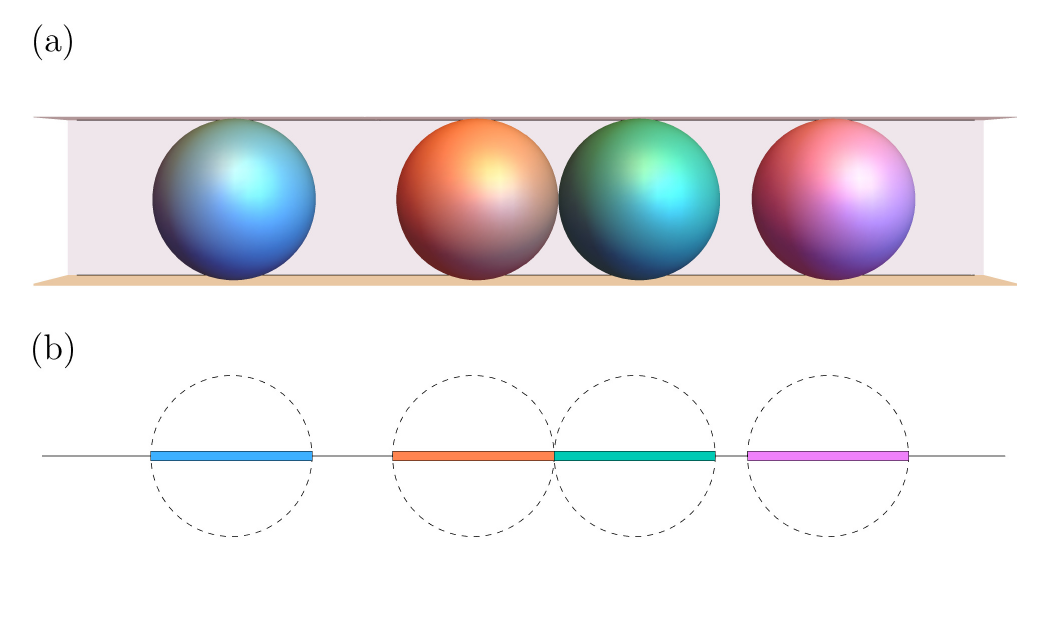}\\
	\caption{Schematic representation of a (a) \acrshort{3D} system of particles confined in a channel and (b) its representation as a purely \acrshort{1D} system, where \acrshort{3D} particles are now viewed as rods in a \acrshort{1D} space.}
	\label{fig:c3_3dUltraConfinement}
\end{figure}

These kinds of structural transitions can also be found in bulk \acrshort{3D} systems. Because purely \acrshort{1D} models can be seen as \acrshort{3D} ultraconfined systems, where particles are so confined that they are forced to move along a single spatial dimension (see Fig.~\ref{fig:c3_3dUltraConfinement}), the study of these structural transitions in bulk \acrshort{3D} systems and their \acrshort{1D} counterparts can shed light on the impact of dimensionality and confinement on the properties of fluids.

In this chapter, we analyze certain aspects of these structural transitions for \acrshort{1D} and \acrshort{3D} systems of hard-core particles with short-range interaction potentials, and compare how the dimensionality affects them. In particular, we focus on how \acrshort{DOC} and \acrshort{FW} structural transitions are affected by the competition between the repulsive and attractive nature of the interparticle potential.

\begin{figure}
	\includegraphics[width=0.95\columnwidth]{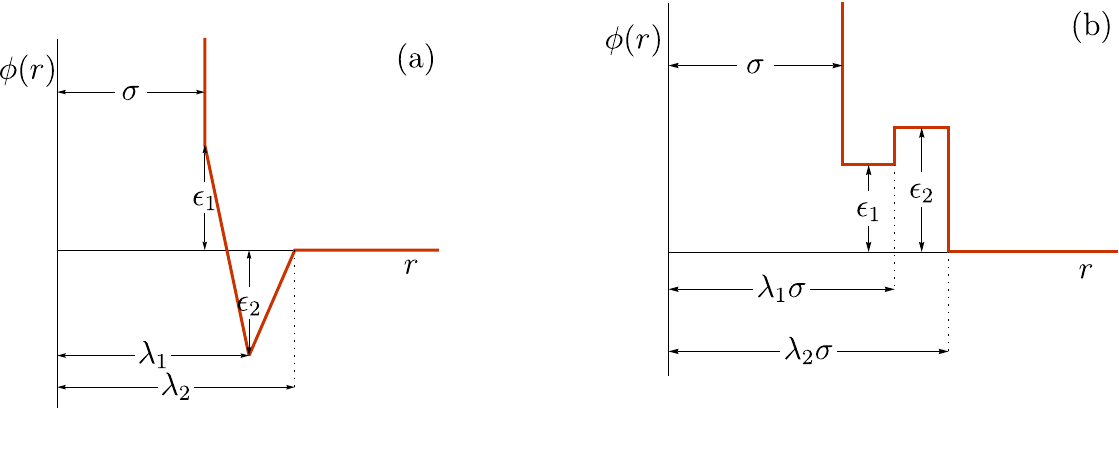}\\
	\caption{Representation of (a) the Jagla potential studied in \nameref{a1} and (b) the piecewise two-step potential used to study competing interactions in \nameref{a2}. Both potentials show the parameters used in the articles.}
	\label{fig:c3_potentials}
\end{figure}

In \nameref{a1}, the Jagla potential~\cite{J99a}---consisting of a hard-core repulsive part plus a triangle well attractive part---is used to investigate this competition in both the \acrshort{1D} and the \acrshort{3D} system, which allows for a detailed examination of the interplay between attractive and repulsive forces, and their impact on the thermodynamic and structural properties. A representation of this potential is shown in Fig.~\ref{fig:c3_potentials}(a).

From a thermodynamic point of view, the outcome of this competition can be determined by quantities such as the compressibility factor, $Z(\rho,T) = p/\rho k_B T$, or the isothermal susceptibility $\chi_T(\rho,T) = k_B T (\partial \rho/\partial p)_T$, where $\rho$ is the number density.\footnote{Note that the number density is denoted by $\rho$ in this chapter, instead of the usual $\lambda$ from Chapters~\ref{1dTheory} and~\ref{1DMixtureTheory} to follow the notation of~\nameref{a1} and~\nameref{a2}} In systems with interaction potentials with attractive and repulsive parts, when the attractive interactions dominate, then $Z$ tends to be smaller than one, while if the repulsive part dominates, $Z$ tends to be higher than one. The opposite trend will occur with the isothermal susceptibility. The pair of values $(\rho, T)$ for which $Z(\rho, T) = 1$ is referred to as the Zeno line~\cite{BH90,PBL24}. Similarly, the line for which $\chi_T = 1$ is called the Seno line in \nameref{a1}, although it is also commonly referred to in the literature as the line of vanishing excess isothermal compressibility~\cite{SHRE19}.

From a purely structural perspective, this competition can be assessed by the asymptotic decay of the total correlation function, since the effect of a dominant attractive component manifests itself in a monotonic asymptotic decay. This is measured by the \acrshort{FW} line~\cite{J99a}, discussed in Sec.~\ref{sec:c2_CorrleationLength}, which does not exist in the absence of attractive interactions and is formed by the points in the phase diagram $(\rho, T)$ for which a crossing of the poles of $g(r)$ occurs, such that its asymptotic decay goes from monotonic to damped oscillatory.

While the Zeno, Seno, and \acrshort{FW} lines qualitatively measure the same phenomenon, their quantitative behaviors do not necessarily coincide. A recent conjecture proposed by \textcite{SHRE19} suggests that the Seno line approximates the \acrshort{FW} line in simple fluids. In \nameref{a1}, approximate theoretical results using the rational function approximation (\acrshort{RFA}) approach~\cite{SYH12,SYHBO13,SYHOO14} and \acrshort{MC} simulations show that this conjecture is satisfied reasonably well, particularly at intermediate densities (see Fig.~4 of \nameref{a1}). However, when we reduce the dimensionality of the system and move to its \acrshort{1D} counterpart, for which exact results can be derived, the conjecture is not satisfied for any density range (see Fig.~1 of \nameref{a1}). This is not very surprising, as the criterion of using the ideal-gas-like isothermal compressibility to estimate the FW line~\cite{SHRE19} is essentially a mean-field approach, which tends to be more accurate in higher dimensions.

While \nameref{a1} focuses on structural transitions and their connection to thermodynamic properties, \nameref{a2} offers a more detailed examination of how these transitions are influenced by variations in the attractive or repulsive components of a potential with competing interactions. The potential studied in \nameref{a2} consists of a hard core plus two steps of height $\epsilon_1$ and $\epsilon_2$, as depicted in Fig.~\ref{fig:c3_potentials}(b). The sign of these two parameters determines whether the corresponding step is a shoulder (repulsive) or a well (attractive). We vary these parameters and compute how the \acrshort{DOC} and \acrshort{FW} lines evolve.

Results for both \acrshort{1D} and \acrshort{3D} systems indicate that, while they share many common characteristics (for instance, the \acrshort{FW} line in Figs. 4 and 8 of \nameref{a2}), the dimensionality introduces significant differences, such as shifts in temperature ranges where transitions occur. Of particular relevance is the intricate behavior of the \acrshort{DOC} lines reported in the \acrshort{1D} case, where the solution is exact. \acrshort{DOC} lines present in the phase diagram exhibit a very complex behavior, including reentrant properties, lines forming lobes, and triple points (see Fig.~3 of \nameref{a2}). Many of these features might vanish completely in the \acrshort{3D} case, although some of them are still present in a slightly different shape, such as the lobes, which do exist but occupy a smaller region in phase space.

\newpage
\section{Article 1}\label{a1}
\underline{\textbf{Title:}}\\ On a conjecture concerning the Fisher–Widom line and the line of vanishing excess isothermal compressibility in simple fluids. \\
\underline{\textbf{Authors:}}\\ Ana M. Montero$^a$, \'Alvaro Rodr\'iguez-Rivas$^b$, Santos B. Yuste$^{a,c}$, Andr\'es Santos$^{a,c}$ and Mariano L\'opez de Haro$^{a,d}$.\\
\noindent \underline{\textbf{Affiliations:}}\\
$^a$ Departamento de Física, Universidad de Extremadura, Badajoz, Spain\\
$^b$ Departamento de Matem\'atica Aplicada II, Estuela T\'ecnica Superior de Ingenier\'ia, Universidad de Sevilla, Seville, Spain\\
$^c$ Instituto de Computación Científica Avanzada (ICCAEx), Universidad de Extremadura, Badajoz, Spain\\\vspace{-0.6cm} \\
$^d$ On sabbatical leave from Instituto de Energ\'ias Renovables, Universidad Nacional Aut\'onoma de M\'exico (U.N.A.M.), Temixco, M\'exico\\\vspace{-1.1cm}
\begin{flushleft}
	\begin{tabular}{@{}ll@{}}
		\underline{\textbf{Journal:}} & Molecular Physics \\[5pt]  
		\underline{\textbf{Volume:}}  & 122 \\[5pt] 
		\underline{\textbf{Number:}}  & 21-22 \\[5pt]   
		\underline{\textbf{Pages:}}   & e2357270 \\[5pt]   
		\underline{\textbf{Year:}}    & 2025 \\[5pt]   
		\underline{\textbf{DOI:}}     & \href{https://doi.org/10.1080/00268976.2024.2357270}{10.1080/00268976.2024.2357270} \\[5pt]   
	\end{tabular}
\end{flushleft}

\includepdf[pages=-, fitpaper=false,pagecommand={}, width=\textwidth, frame, offset=3.8mm -3mm,trim=18mm 20mm 17mm 16mm,clip]{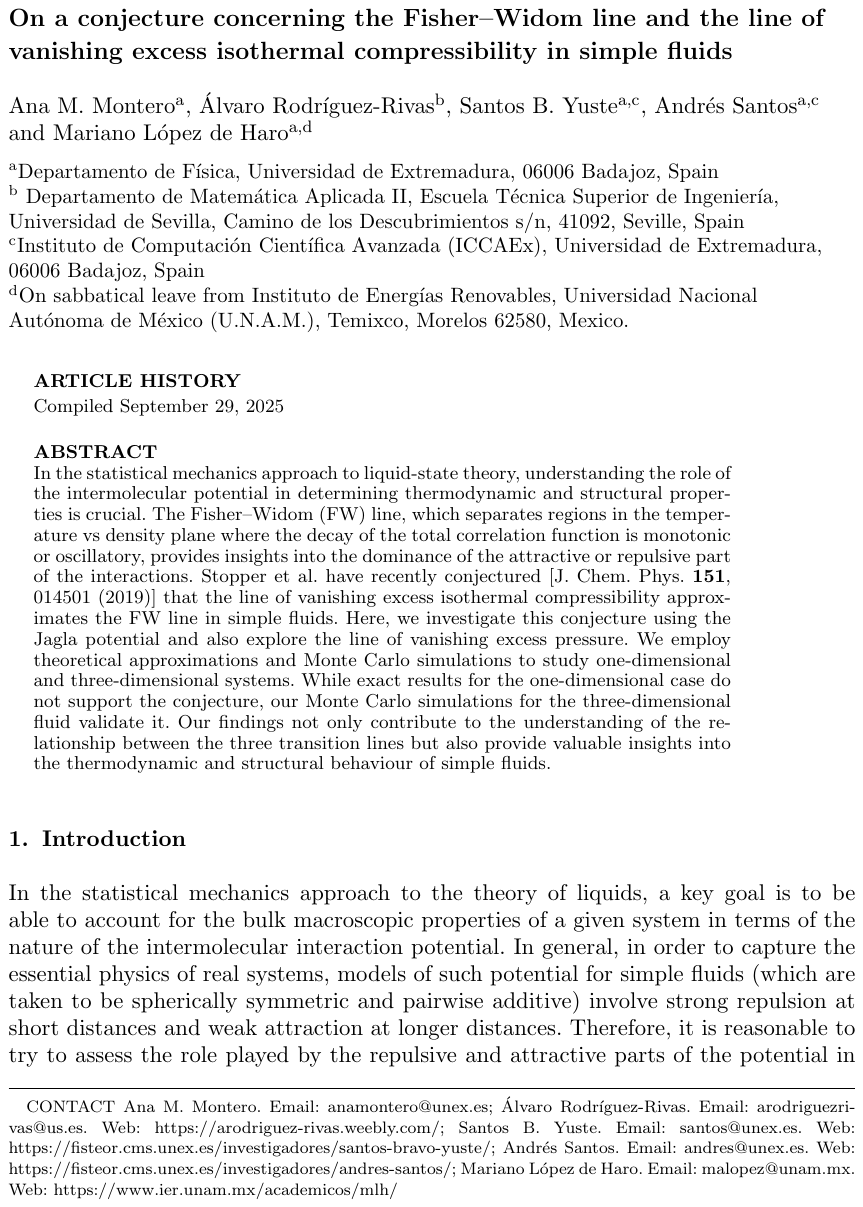}

\newpage
\section{Article 2}\label{a2}
\underline{\textbf{Title:}}\\ Discontinuous Structural Transitions in Fluids with Competing Interactions. \\
\underline{\textbf{Authors:}}\\ Ana M. Montero$^1$, Santos B. Yuste$^{1,2}$, Andr\'es Santos$^{1,2}$ and Mariano L\'opez de Haro$^{3}$.\\
\noindent \underline{\textbf{Affiliations:}}\\
$^1$ Departamento de Física, Universidad de Extremadura, E-06006 Badajoz,
Spain\\
$^2$  Instituto de Computación Científica Avanzada(ICCAEx), Universidad de Extremadura, E-06006 Badajoz, Spain\\\vspace{-0.6cm} \\
$^3$ Instituto de Energías Renovables, Universidad Nacional Autónoma de México (UNAM), Temixco 62580, Mexico\\\vspace{-1.1cm}
\begin{flushleft}
	\begin{tabular}{@{}ll@{}}
		\underline{\textbf{Journal:}} & Entropy \\[5pt]  
		\underline{\textbf{Volume:}}  & 27 \\[5pt] 
		\underline{\textbf{Number:}}  & 95 \\[5pt]      
		\underline{\textbf{Year:}}    & 2025 \\[5pt]   
		\underline{\textbf{DOI:}}     & \href{https://doi.org/10.3390/e27010095}{10.3390/e27010095} \\[5pt]   
	\end{tabular}
\end{flushleft}

\includepdf[pages=-, fitpaper=false,pagecommand={}, width=\textwidth, frame, offset=3.8mm -3mm]{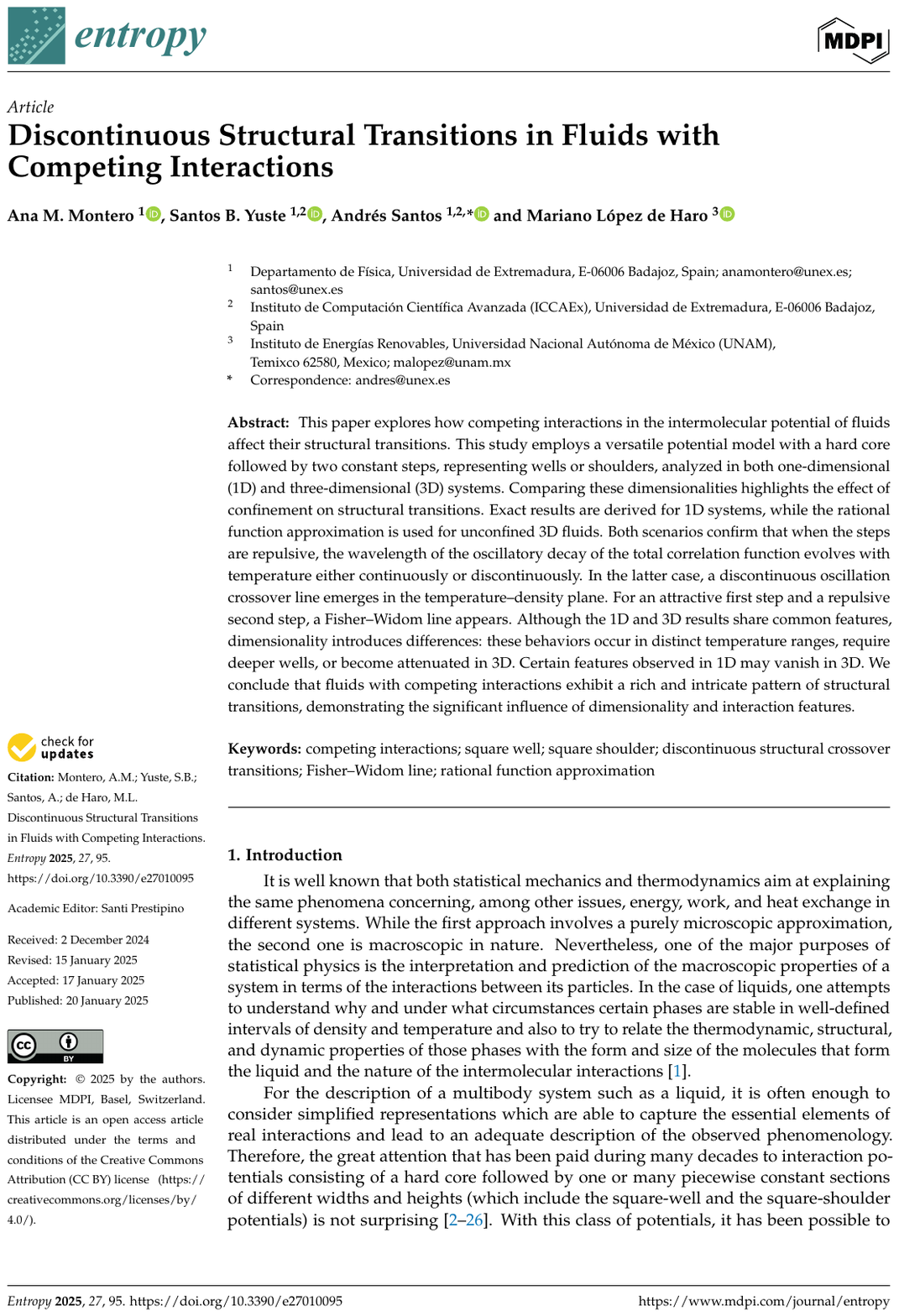}

%% file: chapters/C4_Q1DHardDisks.tex
\thispagestyle{empty}

\chapter{Quasi one-dimensional hard disks}\label{Q1D_Disks}

\section{Summary}

This chapter addresses the study and characterization of the thermodynamics and structure of a \acrshort{Q1D} system of hard disks with only \acrshort{NN} interactions. These systems correspond to those illustrated in Fig.~\ref{fig:c2_q1dDisk}(a). To ensure that only first \acrshort{NN} interactions occur, and assuming disks of diameter 1, which defines the unit of length in this chapter, the maximum width available for the center of the disks is set to $\epsilon=\sqrt{3}/2$. Across this series of three articles, this system is examined in depth from multiple perspectives, and a comprehensive and novel theoretical framework that enables the exact calculation of both thermodynamic and structural properties is developed.

\nameref{a3} deals with the longitudinal thermodynamic properties of the system. Although the exact solution for the equation of state for the \acrshort{Q1D} system of hard disks with only \acrshort{NN} interactions was already known via the \acrshort{TM} approach~\cite{KP93,VBG11}, ultimately this solution requires numerically solving an eigenvalue equation, and no analytical solution has yet been found. Here, we revisit the \acrshort{TM} solution and perform perturbation analysis to compute the exact third and fourth virial coefficients (the second one already having an analytical expression~\cite{KMP04,}). The distinctive properties of confined fluids are evident in the discrepancy between these coefficients and those obtained when incorrectly applying the standard diagrammatic method~\cite{M14b,M15,M20}, showing that the textbook cancellation of the so-called reducible diagrams does not hold in confined geometries. In the high-pressure regime, the analytical asymptotic behavior of the compressibility factor is derived. This result once again highlights the peculiar features of confined fluids. Notably, the high-pressure limit of the compressibility factor differs by a factor of 2 from that of the Tonks gas. This change arises due to the additional transverse degree of freedom in the \acrshort{Q1D} system.

Two alternative approximations for the equation of state are proposed, both of which eliminate the need to solve the eigenvalue equation derived from the \acrshort{TM} formalism. The first approximation is constructed based on the low-density behavior: it reproduces the exact second virial coefficient and, in the high-pressure limit, correctly captures the system's close-packing density. The second approximation is based on the exact high-pressure asymptotic behavior. Remarkably, despite being rooted in the high-density regime, it also yields the exact second virial coefficient, indicating its accuracy across both low- and high-pressure limits.

The structural properties of the system are addressed in \nameref{a4}. Because, to the best of our knowledge, no previous exact solution for the structural properties of the system was known, most of the theoretical background from Sec.~\ref{th_mapping} related to the mapping approach that transforms the \acrshort{Q1D} model into an exactly solvable \acrshort{1D} mixture is developed here, and later on applied to the \acrshort{Q1D} system of hard disks. Quantities like the \acrshort{NN} probability distribution, the longitudinal \acrshort{RDF}, and the structure factor are calculated exactly, and comparison with simulation data from the literature shows excellent agreement, thus validating the theory. Spatial correlations between particles at different transverse positions are also calculated exactly. These results shed light on the disappearance of defects in the zigzag arrangement as the system approaches high density. In this context, the exact scaling form of the defect density as a function of pressure is derived, revealing that the number of defects vanishes exponentially with increasing pressure. This behavior highlights the increasing order of the system and the emergence of a nearly perfect zigzag structure in the high-pressure limit.

By applying pole analysis techniques to the longitudinal \acrshort{RDF}, both the correlation length and the asymptotic oscillation frequency are determined across the full pressure range. In the high-pressure regime, where the zigzag structure is fully established, the analysis reveals a marked contrast between the behavior of the total \acrshort{RDF} and that of the partial \acrshort{RDF}s corresponding to particles involved in the zigzag arrangement. This difference becomes particularly pronounced at large distances, highlighting the distinct long-range ordering patterns within specific subsets of particle pairs compared to the overall fluid structure.

The final step in completing the comprehensive study of the \acrshort{Q1D} \acrshort{HD} model involves incorporating transverse properties and analyzing the system’s anisotropic behavior. \nameref{a5} is devoted to this task. It begins by extending the previously developed mapping framework to account not only for longitudinal but also for transverse degrees of freedom, and then proceeds to compute some of the most relevant anisotropic properties. On the thermodynamic side, the transverse pressure component is obtained exactly, along with its second, third, and fourth virial coefficients. By examining both components of the compressibility factor in parallel, an interesting feature emerges: for narrow channels, the longitudinal pressure is consistently greater than the transverse one. However, for high enough channel widths, there exists a unique pressure at which the two components cross and the transverse pressure overtakes the longitudinal one thereafter. Original \acrshort{MC} simulations for both pressure components separately confirm the theoretical predictions.

The anisotropic nature of the system is further emphasized through the analysis of the pair distribution function, also addressed in \nameref{a5}. In contrast to bulk systems, where translational and rotational invariance allow for a straightforward definition of the \acrshort{RDF}, the confined geometry in \acrshort{Q1D} systems breaks these symmetries. As a result, the standard definition of the \acrshort{RDF} becomes ambiguous. To address this, an \acrshort{RDF}-like quantity is introduced, one that explicitly incorporates the system's anisotropy and measures the average number of particles located at a given distance from a reference particle, regardless of orientation. This definition captures the directional dependence introduced by the confinement. Comparison with \acrshort{MC} simulations confirms the accuracy of the theoretical predictions, validating the approach and providing further insight into the spatial correlations in such anisotropic environments.

The complete source code used to perform the computations and generate the results presented in~\nameref{a3} and~\nameref{a4} of this chapter has been made publicly available online. The complete implementation corresponding to the analysis in~\nameref{a3} is provided in~\textcite{SingleFile}, while the source code associated with~\nameref{a4} can be found in~\textcite{SingleFile2}.

Before concluding the summary, we highlight that the notation for the longitudinal pressure differs slightly across the three articles in this series. For clarity, we summarize the notation conventions adopted in each article below:
\begin{itemize}
	\item \textbf{Article 3}: The unit energy was taken as $\beta = 1$ and therefore the longitudinal pressure is denoted by $\bp = p$.
	\item \textbf{Article 4:} The factor $\bp$ related to the longitudinal pressure is kept as it is, representing the pressure measured in thermal energy units.
	\item \textbf{Article 5:} The longitudinal pressure in thermal energy units is now represented by $\bpp$ to differentiate this component from the transverse pressure component $\beta p_\perp$.
\end{itemize}
This distinction in notation reflects the evolving complexity of the theory considered and, finally, it ensures that pressure components in different spatial directions are clearly identified.

\newpage
\section{Article 3}\label{a3}
\underline{\textbf{Title:}}\\ Equation of state of hard-disk fluids under single-file confinement. \\
\underline{\textbf{Authors:}}\\ Ana M. Montero$^1$ and Andr\'es Santos$^{1,2}$.\\
\noindent \underline{\textbf{Affiliations:}}\\
$^1$ Departamento de Física, Universidad de Extremadura, E-06006 Badajoz,
Spain\\
$^2$  Instituto de Computación Científica Avanzada(ICCAEx), Universidad de Extremadura, E-06006 Badajoz, Spain\\\vspace{-1.1cm}
\begin{flushleft}
	\begin{tabular}{@{}ll@{}}
		\underline{\textbf{Journal:}} & The Journal of Chemical Physics \\[5pt]  
		\underline{\textbf{Volume:}}  & 158 \\[5pt] 
		\underline{\textbf{Number:}}  & 15 \\[5pt]   
		\underline{\textbf{Pages:}}   & 154501 \\[5pt]   
		\underline{\textbf{Year:}}    & 2023 \\[5pt]   
		\underline{\textbf{DOI:}}     & \href{https://doi.org/10.1063/5.0139116}{10.1063/5.0139116} \\[5pt]   
	\end{tabular}
\end{flushleft}

\includepdf[pages=-, fitpaper=false,pagecommand={}, width=\textwidth, frame, offset=3.8mm -3mm]{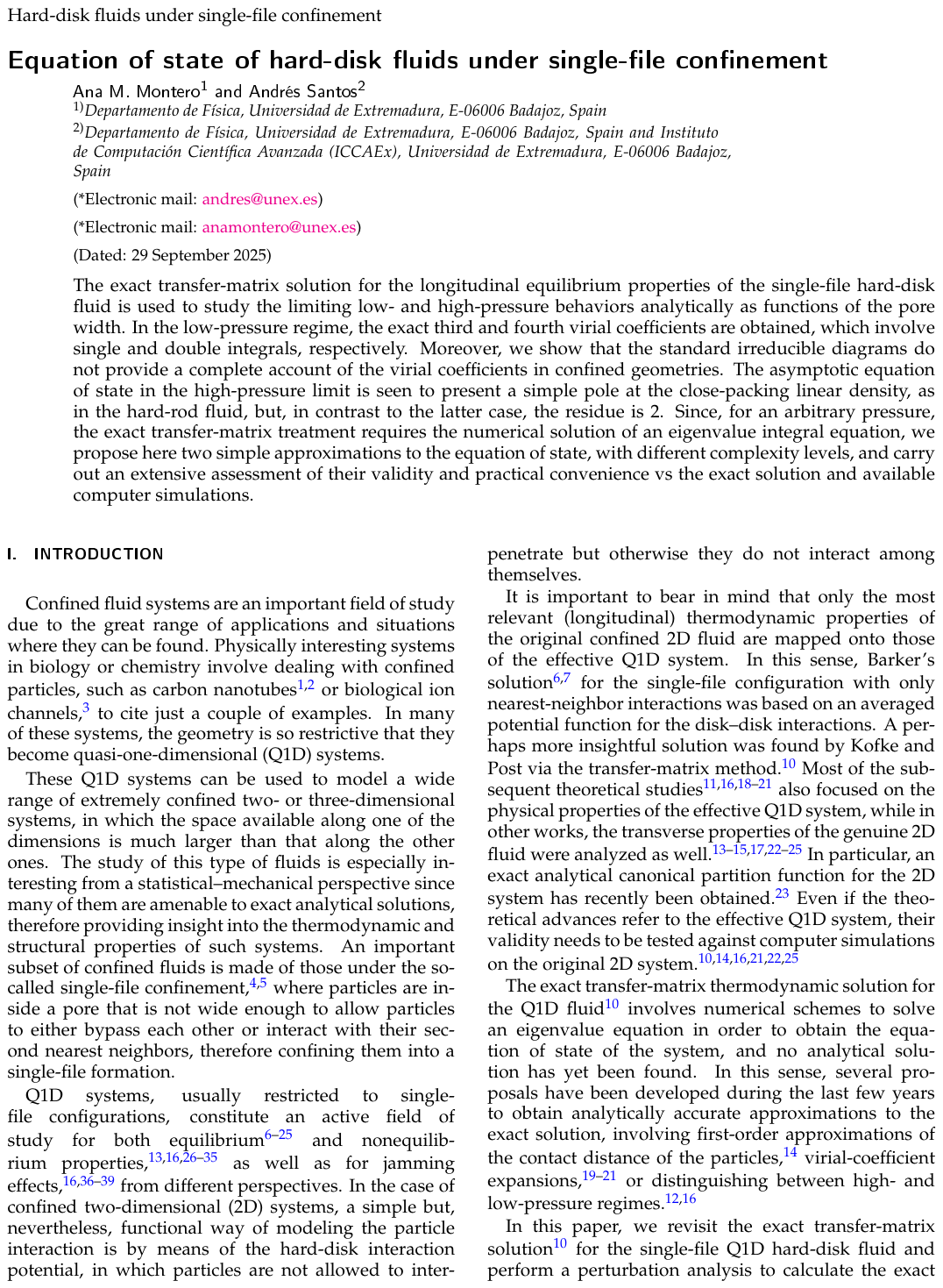}

\newpage
\section{Article 4}\label{a4}
\underline{\textbf{Title:}} Structural properties of hard-disk fluids under single-file confinement. \\
\underline{\textbf{Authors:}} Ana M. Montero$^1$ and Andr\'es Santos$^{1,2}$.\\
\noindent \underline{\textbf{Affiliations:}}\\
$^1$ Departamento de Física, Universidad de Extremadura, E-06006 Badajoz,
Spain\\
$^2$  Instituto de Computación Científica Avanzada(ICCAEx), Universidad de Extremadura, E-06006 Badajoz, Spain\\\vspace{-1.1cm}

\begin{flushleft}
	\begin{minipage}{0.5\textwidth}\raggedright
		\underline{\textbf{Journal:}} The Journal of Chemical Physics\vspace{0.2cm}
		
		\underline{\textbf{Volume:}} 159\vspace{0.2cm}
		
		\underline{\textbf{Number:}} 3\vspace{0.2cm}
		
		\underline{\textbf{Pages:}} 034503\vspace{0.2cm}
		
		\underline{\textbf{Year:}} 2023\vspace{0.2cm}
		
		\underline{\textbf{DOI:}} \href{https://doi.org/10.1063/5.0156228}{10.1063/5.0156228}
	\end{minipage}
\end{flushleft}

\includepdf[pages=-, fitpaper=false,pagecommand={}, width=\textwidth, frame, offset=3.8mm -3mm]{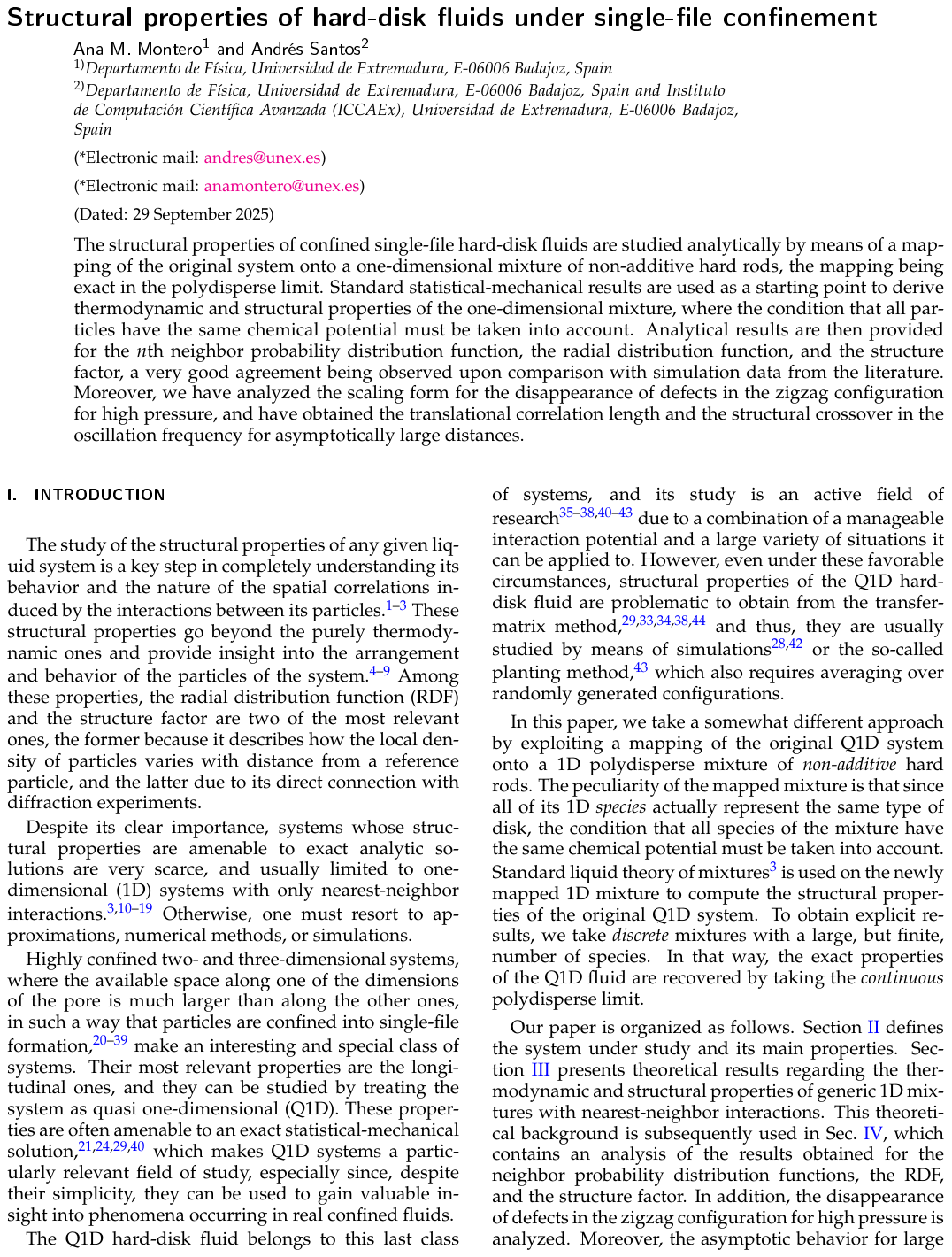}

\newpage
\section{Article 5}\label{a5}
\underline{\textbf{Title:}}\\ Exploring anisotropic pressure and spatial correlations
in strongly confined hard-disk fluids: Exact results. \\
\underline{\textbf{Authors:}}\\ Ana M. Montero,$^1$ and Andr\'es Santos$^{1,2}$.\\
\noindent \underline{\textbf{Affiliations:}}\\
$^1$ Departamento de Física, Universidad de Extremadura, E-06006 Badajoz,
Spain\\
$^2$  Instituto de Computación Científica Avanzada(ICCAEx), Universidad de Extremadura, E-06006 Badajoz, Spain\\\vspace{-1cm}

\begin{flushleft}
	\begin{minipage}{0.5\textwidth}\raggedright
		\underline{\textbf{Journal:}} Physical Review E\vspace{0.2cm}
		
		\underline{\textbf{Volume:}} 110\vspace{0.2cm}
		
		\underline{\textbf{Pages:}} L022601\vspace{0.2cm}
		
		\underline{\textbf{Year:}} 2024\vspace{0.2cm}
		
		\underline{\textbf{DOI:}} \href{https://doi.org/10.1103/PhysRevE.110.L022601}{10.1103/PhysRevE.110.L022601}
	\end{minipage}
\end{flushleft}

\includepdf[pages=-, fitpaper=false,pagecommand={}, width=\textwidth, frame, offset=3.8mm -3mm]{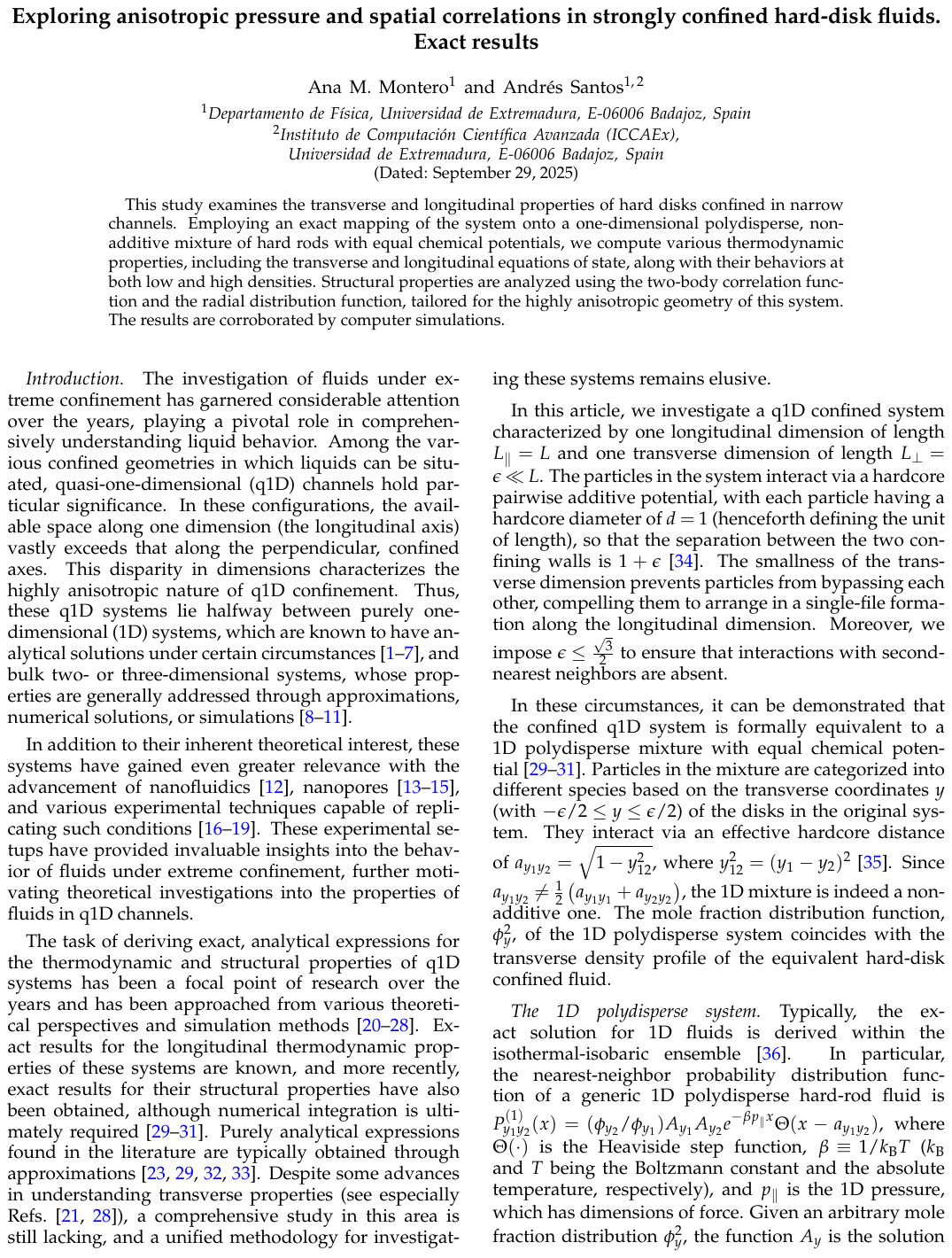}

%% file: chapters/C5_Q1DSWSS.tex
\thispagestyle{empty}

\chapter{Quasi one-dimensional square-well and square-shoulder disks}\label{1DSWSS}

\section{Summary}

This chapter summarizes the results of \nameref{a6}, where the focus remains on \acrshort{Q1D} disks but the simple hard-core interaction is replaced by two more intricate pair potentials: the \acrshort{SW} and~\acrshort{SS} models. Both potentials are characterized by an impenetrable hard core paired with either an attractive well or a repulsive step, respectively. The \acrshort{SS} potential is purely repulsive and belongs to the family of the so-called core-softened potentials, whereas the \acrshort{SW} potential combines the hard core with an attractive well, thereby creating competing attractive and repulsive forces.

\nameref{a6} revisits the mapping framework introduced in Chapter \ref{Q1D_Disks} for hard disks and generalizes it to \acrshort{Q1D} fluids whose particles interact through a hard core paired with an arbitrary attractive or repulsive tail. Although the analysis focuses on the \acrshort{SW} and \acrshort{SS} models, the formalism is entirely general and can be applied to any pair potential with only \acrshort{NN} interactions, regardless of the specific form taken by the additional interaction beyond the hard core.

Within this extended framework, we derive exact expressions for all relevant thermodynamic and structural quantities.  The equation of state and the excess internal energy are obtained at several temperatures, and the results for both potentials smoothly approach that of the \acrshort{HD} case in the high-temperature limit. The partial and total \acrshort{RDF}s are also calculated across the full temperature range. \acrshort{MC} simulations performed in both the canonical and isothermal--isobaric ensembles validate these results, exhibiting excellent agreement with the theoretical predictions.

We then examine the asymptotic decay of correlations and the associated correlation length across the entire density range for several temperatures. Although the correlation length is always continuous, it develops distinct kinks that coincide with discontinuous jumps in the oscillation frequency of the long-range correlations. For the \acrshort{SW} fluid, the attractive well generates a \acrshort{FW} line: below this line---at sufficiently low densities and temperatures---the longitudinal \acrshort{RDF} displays a purely monotonic decay instead of the more usual damped oscillatory decay. The \acrshort{SS} fluid lacks such a line, and its \acrshort{RDF} retains an oscillatory tail in the entire temperature--density plane.

Finally, all the code used to carry out the computations and produce the results presented in this chapter is available online in~\textcite{SingleFile3}.

\newpage
\section{Article 6}\label{a6}
\underline{\textbf{Title:}}\\ Exact equilibrium properties of square-well and square-shoulder disks in single-file confinement. \\
\underline{\textbf{Authors:}}\\ Ana M. Montero,$^1$ and Andr\'es Santos$^{1,2}$.\\
\noindent \underline{\textbf{Affiliations:}}\\
$^1$ Departamento de Física, Universidad de Extremadura, E-06006 Badajoz,
Spain\\
$^2$  Instituto de Computación Científica Avanzada(ICCAEx), Universidad de Extremadura, E-06006 Badajoz, Spain\\\vspace{-1.1cm}

\begin{flushleft}
	\begin{minipage}{0.5\textwidth}\raggedright
		\underline{\textbf{Journal:}} Physical Review E\vspace{0.2cm}
		
		\underline{\textbf{Volume:}} 110\vspace{0.2cm}
		
		\underline{\textbf{Pages:}} 024601\vspace{0.2cm}
		
		\underline{\textbf{Year:}} 2024\vspace{0.2cm}
		
		\underline{\textbf{DOI:}} \href{https://doi.org/10.1103/PhysRevE.110.024601}{10.1103/PhysRevE.110.024601}
	\end{minipage}
\end{flushleft}

\includepdf[pages=-, fitpaper=false,pagecommand={}, width=\textwidth, frame, offset=3.8mm -3mm]{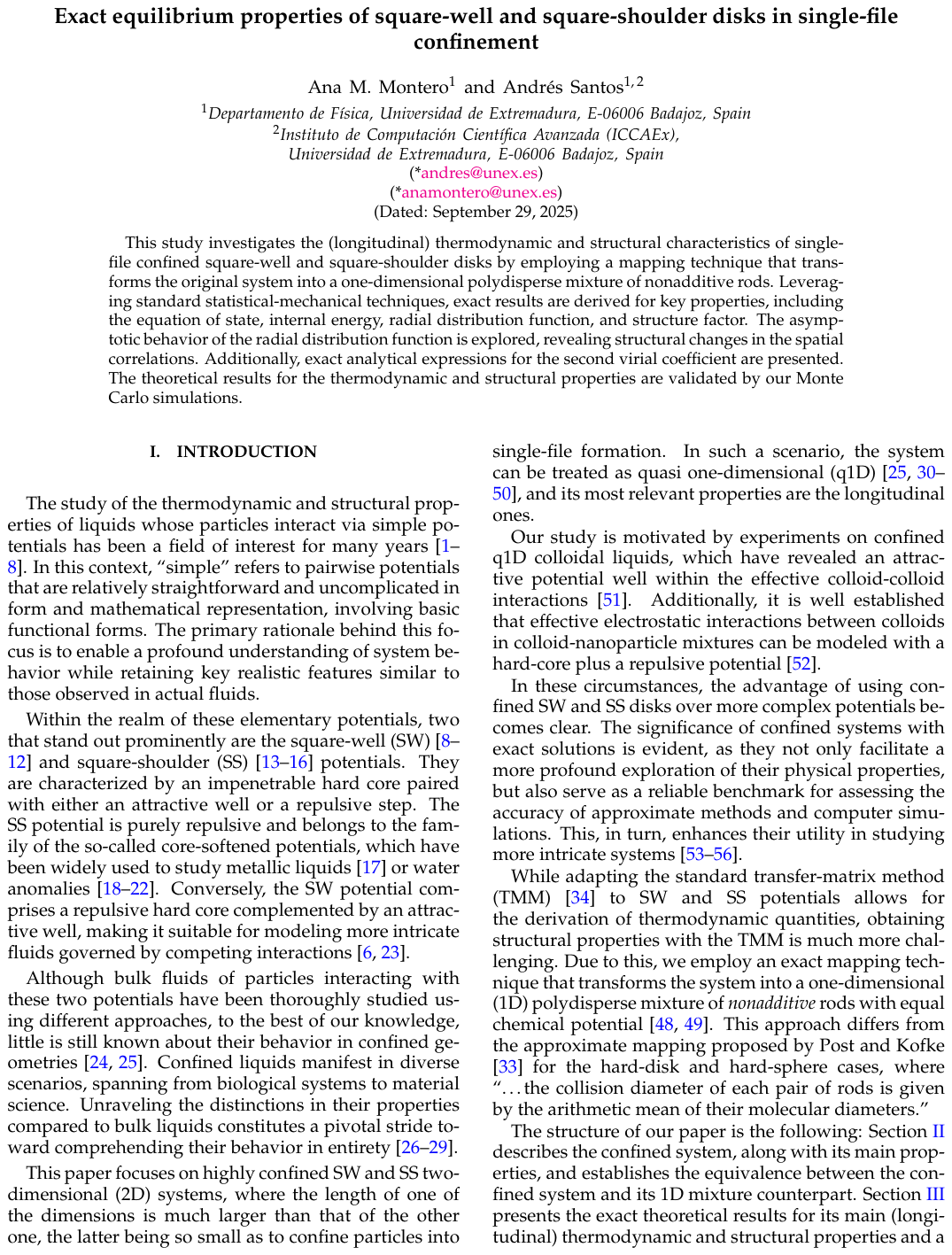}

%% file: chapters/C6_Q1DHardSpheres.tex
\thispagestyle{empty}

\chapter{Quasi one-dimensional hard spheres}\label{Q1D_HardSpheres}

\section{Summary}

In this chapter, we move forward to a \acrshort{Q1D} model of hard spheres, where particles are now allowed to move along two different confined transverse directions, in addition to the free longitudinal axis. Employing the same mapping framework as in previous chapters, we derive the exact solution for this geometry and evaluate the longitudinal and transverse properties. The full analysis, presented in \nameref{a7}, yields a comprehensive description of the system’s thermodynamic and structural behavior under geometrical cylindrical confinement.

\nameref{a7} extends the mapping technique to incorporate the additional transverse degree of freedom available to the hard spheres in a cylindrical pore. Although the conceptual strategy mirrors the one used for \acrshort{Q1D} hard disks, the algebra becomes more intricate. Each \enquote{species} in the mapped \acrshort{1D} mixture now corresponds to a specific location within the pore's cross section, and must be labeled by a pair of transverse coordinates rather than a single one. Careful treatment of this \acrshort{2D} spatial structure and cylindrical symmetry of the pore is essential for correctly constructing the mapping and ultimately obtaining exact expressions.

For this system we calculate both the longitudinal and transverse pressure components and find a trend analogous to that observed for \acrshort{Q1D} hard disks. When the pore is narrow, the longitudinal pressure is always higher than the transverse pressure at every density. However, as the pore widens, there exists a density at which the two components cross. Beyond this point the transverse pressure becomes the largest one.

We also analyze the limiting forms of the equation of state. In the low-pressure regime, exact expressions for the second and third virial coefficients are derived. In the high-pressure limit, an analytical expression for the compressibility factor is obtained, which allows for a detailed description of the differences in the asymptotic behavior of both components.

Regarding structural properties, we calculate both the longitudinal \acrshort{RDF} along the unconfined axis and a \acrshort{3D} \acrshort{RDF}-like function that tracks how particle ordering within the pore evolves as density increases, as derived in Sec.~\ref{th_rdf3D}. As previously noted, we refer to it as a \acrshort{RDF}-like function, since the conventional bulk \acrshort{RDF} relies on spatial isotropy—an assumption that no longer holds under the confinement considered here. The longitudinal \acrshort{RDF} at fixed transverse positions is also obtained, providing an exact measure of the rate at which defects vanish as close-packing is approached.

The code used for the simulations and calculations related to the hard-sphere system discussed in this chapter is openly available in~\textcite{SingleFileHardSpheres}, allowing for full reproducibility and further investigation of the results.

\newpage
\section{Article 7}\label{a7}
\underline{\textbf{Title:}}\\ Exact anisotropic properties of hard spheres in narrow cylindrical confinement. \\
\underline{\textbf{Authors:}}\\ Ana M. Montero,$^1$ and Andr\'es Santos$^{1,2}$\\
\noindent \underline{\textbf{Affiliations:}}\\
$^1$ Departamento de Física, Universidad de Extremadura, E-06006 Badajoz,
Spain\\
$^2$  Instituto de Computación Científica Avanzada(ICCAEx), Universidad de Extremadura, E-06006 Badajoz, Spain\\\vspace{-1.1cm}

\begin{flushleft}
	\begin{minipage}{0.5\textwidth}\raggedright
		\underline{\textbf{Journal:}} The Journal of Chemical Physics\vspace{0.2cm}
		
		\underline{\textbf{Volume:}} 163\vspace{0.2cm}

		\underline{\textbf{Number:}} 2\vspace{0.2cm}

		\underline{\textbf{Pages:}} 024506\vspace{0.2cm}

		\underline{\textbf{Year:}} 2025\vspace{0.2cm}

		\underline{\textbf{DOI:}} \href{https://doi.org/10.1063/5.0273930}{10.1063/5.0273930}  
	\end{minipage}
\end{flushleft}

\includepdf[pages=-, fitpaper=false,pagecommand={}, width=\textwidth, frame, offset=3.8mm -3mm]{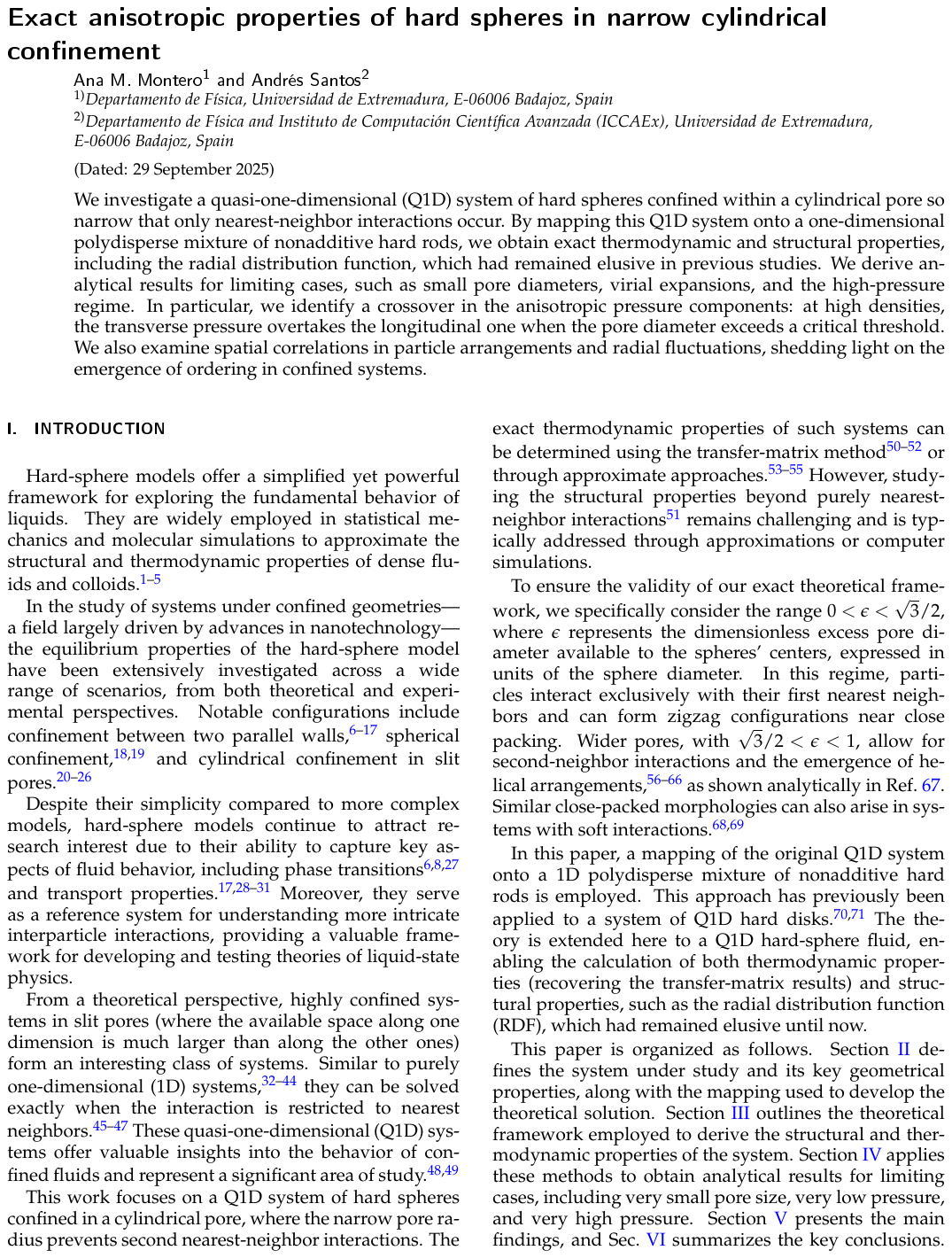}

%% file: chapters/C7_Dumbbells.tex
\thispagestyle{empty}

\chapter{One-dimensional anisotropic hard particles}\label{Dumbbells}

\section{Summary}

This chapter considers a different class of \acrshort{Q1D} systems, namely those composed of hard bodies whose centers are restricted to move along a straight line, while the particles themselves are anisotropic and may adopt several orientations. \nameref{a8} presents results for two representative shapes: (i) prisms that project three distinct longitudinal lengths, and (ii) dumbbells made of two tangent spheres. For each shape, we analyze models in which a particle can occupy either two or three allowed orientations with respect to the movement axis.

We extend the mapping technique---originally developed in \nameref{a3} and \nameref{a4} to account for spatially confined \acrshort{Q1D} systems---to be applicable to systems in which movement occurs along a single axis while particles retain orientational freedom. We then use this method along with the \acrshort{TM} method to compare exact results coming from both theories with the approximate Parsons--Lee~(\acrshort{PL}) theory, a well-established approximation that has proven to be very accurate for describing orientational ordering properties of hard nonspherical particles in \acrshort{2D} and \acrshort{3D} systems.

We study the exact properties of the systems and demonstrate that both models exhibit positional ordering with increasing density, where the phase is isotropic only in the limit of the dilute regime. The mapping technique gives access to the positional correlation length, which diverges in the high-density limit for both types of particles. The orientational correlation length studied via the \acrshort{TM} method, on the other hand, reveals that the hard prisms lack orientational correlation, whereas that of the hard dumbbells diverges at close packing.

By examining the equation of state, the \acrshort{RDF}, and the correlation lengths of the proposed models, we show that the \acrshort{PL} theory is exact when the hard bodies have additive interactions, as is the case of the prism model. However, the nonadditive interactions in the hard-dumbbell case cannot be fully accounted for by the \acrshort{PL} theory, which yields results that fail to capture the behavior of hard dumbbells in the high-pressure limit.

\newpage
\section{Article 8}\label{a8}
\underline{\textbf{Title:}}\\ Ordering properties of anisotropic hard bodies in one-dimensional channels. \\
\underline{\textbf{Authors:}}\\ Ana M. Montero,$^1$ Andr\'es Santos,$^{1,2}$ P\'eter Gurin,$^{3}$ and Szabolcs Varga$^{3}$\\
\noindent \underline{\textbf{Affiliations:}}\\
$^1$ Departamento de Física, Universidad de Extremadura, E-06006 Badajoz,
Spain\\
$^2$  Instituto de Computación Científica Avanzada(ICCAEx), Universidad de Extremadura, E-06006 Badajoz, Spain\\
$^3$ Physics Department, Centre for Natural Sciences, University of Pannonia, P.O. Box 158, Veszprém H-8201, Hungary\\\vspace{-1.1cm}

\begin{flushleft}
	\begin{tabular}{@{}ll@{}}
		\underline{\textbf{Journal:}} & The Journal of Chemical Physics \\[5pt]  
		\underline{\textbf{Volume:}}  & 159 \\[5pt] 
		\underline{\textbf{Pages:}}   & 154507 \\[5pt]   
		\underline{\textbf{Year:}}    & 2023 \\[5pt]   
		\underline{\textbf{DOI:}}     & \href{https://doi.org/10.1063/5.0169605}{10.1063/5.0169605} \\[5pt]   
	\end{tabular}
\end{flushleft}


\includepdf[pages=-, fitpaper=false,pagecommand={}, width=\textwidth, frame, offset=3.8mm -3mm]{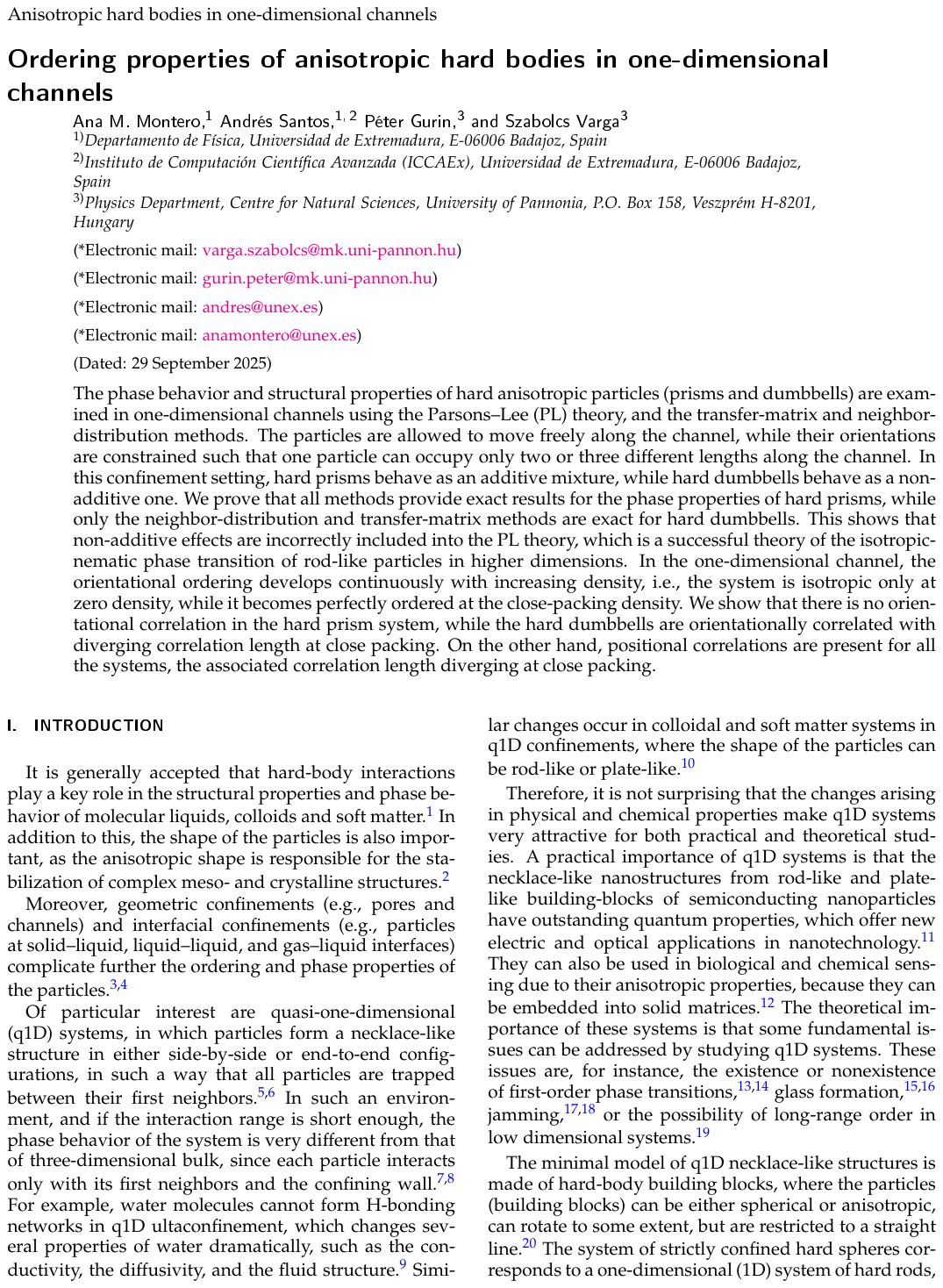}

%% file: chapters/C8_Results_and_Discussion.tex
\thispagestyle{empty}

\setcounter{chapter}{8}
\chapter{Results and discussion}\label{results}

This chapter reviews and synthesizes the principal results of the thesis, discussing their significance and their physical implications on the considered models. It is divided into three sections, which reflect the three main topics covered within this work.

\section{Competing interactions in 1D and 3D systems}

The study of interaction potentials with competing attractive and repulsive components in both \acrshort{1D} and \acrshort{3D} systems provides insight into how dimensionality influences the balance between these forces. This competition manifests itself in different ways depending on whether the system is analyzed from a thermodynamic perspective, through quantities such as response functions and the compressibility factor, or from a structural perspective, such as the observation of crossovers in the asymptotic oscillation frequency of the \acrshort{RDF}. These effects can differ significantly between \acrshort{1D} and \acrshort{3D} systems. In one dimension, exact results can be derived as long as only \acrshort{NN} interactions are present. In contrast, the analysis of the corresponding \acrshort{3D} systems must rely on approximate theoretical methods, such as the \acrshort{RFA}, or numerical techniques, such as \acrshort{MC} simulations.

An example of this type of analysis is the study of the conjecture by \textcite{SHRE19}, carried out in \nameref{a1} for the Jagla model [see Fig.~\ref{fig:c3_potentials}(a)]. This study illustrates how the same interaction potential can lead to different consequences depending on the dimensionality. Throughout this discussion, the density of both the \acrshort{1D} and \acrshort{3D} systems is denoted by $\rho$. Additionally, the dimensionless density $\rho^* = \rho \sigma^d$ is sometimes used, where $\sigma$ is the particle diameter and $d$ is the dimensionality. This standardization facilitates comparison between systems in different dimensions.

For the \acrshort{1D} Jagla model [see Fig.~1 of \nameref{a1}], in the low-density limit both the Zeno and Seno lines originate at the Boyle temperature, which is formally defined as the temperature at which the second virial coefficient vanishes. In contrast, the \acrshort{FW} line diverges at $\rho \to 0$, highlighting not only a quantitative difference but also a deeper qualitative distinction between the two sets of curves. All three curves terminate at the Boyle density $\rho_\mathrm{B} = \lambda_1^{-1}$, where $\lambda_1$ is the position of the minimum in the Jagla potential, but the overall shapes of the curves are significantly different, making it evident that the conjecture by \textcite{SHRE19} does not hold in the \acrshort{1D} case.

The case of the \acrshort{3D} system is also illustrated [see Fig.~4 of \nameref{a1}] and suggests that, in this case, the conjecture proposed in Ref.~\cite{SHRE19} is reasonably well satisfied in three dimensions, at least within the density range $0.20 \le \rho^* \le 0.40$. This outcome is not unexpected because the conjecture relies on estimating the \acrshort{FW} line from the ideal-gas-like isothermal compressibility, and this type of mean-field argument is generally more reliable in higher dimensions. Moreover, within this same density range, we also observe a notable proximity among the Zeno, Seno, and \acrshort{FW} lines, further reinforcing the apparent consistency of the conjecture in the \acrshort{3D} case. This analysis also shows that studying transition lines and structural crossovers using approximations and \acrshort{MC} simulation data requires great care. As an example, the \acrshort{RFA} and \acrshort{MC} approaches predict different \acrshort{FW}, Zeno, and Seno lines [see Fig.~4 of \nameref{a1}], even though the general form they predict for the \acrshort{RDF} is almost the same [see Fig.~5 of \nameref{a1}], which means that the study of these lines is very delicate.

\nameref{a2} is also focused on the study of competing interactions, this time focusing solely on structural properties, where again exact results are obtained for the \acrshort{1D} system and the \acrshort{RFA} is used for the \acrshort{3D} case. Results of the systematic study for different parameters of the two-step interaction potential [see Fig.~\ref{fig:c3_potentials}(b)] agree with the fact that both the \acrshort{1D} and the \acrshort{3D} models share many common features, particularly in terms of their qualitative structural behavior.

An example of these shared characteristics can be found in the analysis of interaction potentials with purely repulsive barriers. In both dimensions, the systems exhibit similar behavior in the limiting cases of high and low temperatures, aligning with theoretical expectations. Specifically, in the low-temperature limit, the asymptotic oscillation frequency is governed by the range of the repulsive barrier, while in the high-temperature limit, it is determined by the hard-core diameter. Despite these similarities, the behavior at intermediate temperatures can be very different. This is clearly illustrated by comparing results side by side [see Figs.~3(a–b) and~6(a) of \nameref{a2}]. Additionally, certain features observed in the \acrshort{1D} case at high densities and low temperatures are not present in the \acrshort{3D} case. While this may partly stem from limitations of the \acrshort{RFA} approach. It is also important to note that such features in three dimensions could occur beyond the system's freezing point, a phenomenon that does not exist in one dimension.

Further similarities emerge when one of the steps in the potential is attractive. In this case, a systematic study of the \acrshort{FW} line reveals a strong qualitative agreement between the \acrshort{1D} and \acrshort{3D} systems [compare Figs.~4 and~8 of \nameref{a2}].

Despite the similarities between both systems reported here, it must be noted that the study of these models, although they are made of simple potentials, show a remarkably complex pattern of structural transitions for which a full study would require inspecting a very broad parameter space. In this sense, although \acrshort{1D} and \acrshort{3D} systems share many common characteristics, the appearance or disappearance of these characteristics can be determined by certain thresholds in this parameter space, which can indeed look very different quantitatively in both the \acrshort{1D} and \acrshort{3D} cases.

\section{Spatially confined Q1D models}

The main result that should be highlighted from the chapters on spatially confined \acrshort{Q1D} systems with only \acrshort{NN} interactions is the development of an exact, unified theoretical framework to perform an in-depth study of all equilibrium properties of these systems. The central tool of this analysis is always the same: each confined \acrshort{2D} or \acrshort{3D} system is mapped onto a \acrshort{1D} polydisperse mixture of nonadditive rods whose species label reflects the transverse coordinate of the original system. Requiring a common chemical potential for all species closes the problem, as described in Sec. III of \nameref{a4}.

\subsection{Thermodynamics}

Among the quantities computed for these models, the equation of state provides the fundamental thermodynamic characterization. Here, we analyze it through both components of the global compressibility factor: the longitudinal $Z_\|$, and the transverse one, $Z_\perp$. Exact expressions for both of them are derived for the \acrshort{Q1D} \acrshort{HD} and \acrshort{HS} systems. The results show that, although they coincide in the dilute limit and both diverge as the system approaches close packing, their behavior across the rest of the density range is very different. This difference is shown explicitly in Fig.~1 of \nameref{a5} for the \acrshort{HD} case and in Fig. 2 of \nameref{a7} for the \acrshort{HS} case.

In the case of the \acrshort{Q1D}~\acrshort{SW} and~\acrshort{SS} models, both the equation of state and the internal energy acquire temperature dependence due to additional attractive or repulsive interactions beyond the hard-core repulsion. Results for these quantities clearly demonstrate that, in the high-temperature limit, both models converge to the behavior of the \acrshort{HD} system, as the influence of the extra well or shoulder becomes negligible. In contrast, at very low temperatures, the behavior of the two models diverges significantly. These contrasting temperature-dependent behaviors are illustrated in Fig.~5 of \nameref{a6}.

Regarding limiting behaviors, virial expansions stand out as one of the most common approaches for characterizing fluid behavior in the low-density regime. Although they are traditionally formulated as power series in the number density, it is shown---particularly in Fig.~2 of \nameref{a3}---that expansions in powers of the longitudinal pressure yield significantly better behavior for \acrshort{Q1D} systems. For this reason, all virial expansions throughout these articles are expressed in terms of the longitudinal pressure.

Exact expressions for the second and third virial coefficients are obtained for both the \acrshort{Q1D} \acrshort{HD} and \acrshort{HS} models and, additionally, exact results for the fourth virial coefficient are derived in the \acrshort{HD} case. While the results for the second virial coefficients match previously reported values, discrepancies arise at higher orders. We show that the origin of this discrepancy lies in the improper application of standard irreducible diagram techniques, which assume that reducible diagrams cancel out---a condition that does not hold due to the lack of translational invariance caused by confinement. This issue is analyzed in detail in Sec.~IIC of \nameref{a3}. In the \acrshort{Q1D} \acrshort{SW} and \acrshort{SS} fluid we obtain the exact second virial coefficient and show that its analytical form is directly related to that of the corresponding \acrshort{Q1D} \acrshort{HD} system [see Eqs.~(3.24) and~(3.25) of \nameref{a6}].

In the high-density regime, particles tend to arrange in a zigzag configuration because the minimum contact distance, $a_\mathrm{min}$, is achieved by particles sitting on opposite ends of the pore. This zigzag configuration marks the close-packing density of the system, $\lambda_\mathrm{cp}=1/a_\mathrm{min}$, where all particles are at the same distance apart from their left and right neighbors. While it is tempting to assume that in this limit the \acrshort{Q1D} system will behave like a \acrshort{1D} Tonks gas of hard rods of diameter $a_\mathrm{min}$, we derived the analytic asymptotic behavior of the compressibility factor and showed that this is not the case, since the contribution from the higher-dimensional nature of the system cannot be neglected. Table~\ref{tab:SummaryZHighPressure} summarizes these findings. While the high-pressure form always involves the same denominator, the prefactor multiplying it depends on the particular system under study.

\begin{table}[htb]
	\centering
	\caption{Summary of the high-pressure behavior of both components of the compressibility factor for different hard-particle systems under single-file confinement.}
	\label{tab:SummaryZHighPressure}
	\renewcommand{\arraystretch}{2}
	\begin{tabular}{@{} p{3.5cm}  C{2.5cm}  C{2.5cm} @{}}
		\toprule
		\textbf{System}           & $Z_{\parallel}$ & $Z_{\perp}$ \\
		\midrule
		1D hard rods                 & $\displaystyle\frac{1}{1-\lambda/\lambda_\mathrm{cp}}$  & ---\\
		Q1D hard disks            & $\displaystyle\frac{2}{1-\lambda/\lambda_\mathrm{cp}}$  & $\displaystyle\frac{2(\lambda_\mathrm{cp}^2-1)}{1-\lambda/\lambda_\mathrm{cp}}$\\
		Q1D hard spheres (cylindrical pore)          & $\displaystyle\frac{5/2}{1-\lambda/\lambda_\mathrm{cp}}$  & $\displaystyle\frac{\frac{5}{4}(\lambda_\mathrm{cp}^2-1)}{1-\lambda/\lambda_\mathrm{cp}}$\\
		\bottomrule
	\end{tabular}
\end{table}

Table~\ref{tab:SummaryZHighPressure} also reveals an interesting and qualitatively similar behavior in both the \acrshort{Q1D} \acrshort{HD} and \acrshort{HS} systems. When the available pore width $\epsilon$ is lower than a certain threshold, $Z_{\parallel}$ remains consistently larger than $Z_{\perp}$ across all densities because the channel is not wide enough to allow significant transverse structuring. However, beyond this threshold value of $\epsilon$ this changes, and a density appears beyond which $Z_{\perp}$ surpasses $Z_{\parallel}$, reflecting a shift in the dominant direction of particle interactions due to increased transverse accessibility. This threshold value differs depending on the number of confined directions. For the \acrshort{Q1D} \acrshort{HD} system its value is $\epsilon_{\mathrm{th}} = 1/\sqrt{2}$, while for the \acrshort{Q1D} \acrshort{HS} case it is $\epsilon_{\mathrm{th}} = \sqrt{2/3}$.

\subsection{Structure}

One of the key strengths of the mapping method developed in this thesis is its ability to provide exact results for the \acrshort{RDF}, both for the system as a whole and for specific pairs of particles at particular transverse coordinates. This capability enables not only a detailed understanding of the global structural organization, but also the resolution of spatial correlations between particles at distinct transverse positions. Both quantities are calculated for all considered systems and are shown to agree very well with \acrshort{MC} simulation results in \nameref{a6} and with simulations from the literature in \nameref{a4}, which further validates the accuracy of the theoretical framework. The evolution of structural ordering in the system is clearly reflected in the increasingly pronounced peaks of the longitudinal \acrshort{RDF}. However, the development of the zigzag structure can also be analyzed through the correlations between particles at specific transverse positions. In particular, the strongest signature of zigzag ordering arises from the correlations between particles on opposite sides of the pore, which form the alternating structure. These targeted correlations offer a more direct and sensitive probe of the onset and growth of zigzag arrangements than the global \acrshort{RDF} alone.

Of particular interest is the analysis of the \acrshort{RDF} at contact for particles located on the same side of the pore. The value of this contact peak decreases with increasing pressure, signaling the progressive disappearance of defects as the zigzag structure becomes more pronounced. An exact analytical expression is obtained for both the \acrshort{HD} and \acrshort{HS} cases, where defects disappear quasi-exponentially as $\bpp^\gamma e^{-\bpp (1-a_{\mathrm{min}})}$, with $\gamma=1$ for disks and $\gamma=3/2$ for spheres. The exact expression can be found in Eq.~(4.3) of \nameref{a4} and Eq.~(5.3) of \nameref{a7}.

The asymptotic behavior of spatial correlations is also analyzed, providing direct access to the correlation length and the asymptotic oscillation frequency of the partial \acrshort{RDF}s. In the case of the \acrshort{HD} model, the correlation length increases smoothly with pressure, consistent with the absence of any true phase transition. However, a distinct kink appears at a specific pressure value [see Fig.~8 of \nameref{a4}], indicating a discontinuous change in the asymptotic oscillation frequency. This sharp transition marks a structural crossover between two different wavelengths in the asymptotic long-range correlations and signals the onset of zigzag ordering in this context. Additionally, the ability of the model to compute the exact (albeit numerical) \acrshort{RDF} at very large distances [see especially Fig.~9(a) of \nameref{a4}] allows us to confirm the predictions for the correlation length and oscillation frequency.

The situation for the \acrshort{SW} and \acrshort{SS} models is similar but considerably more complex. The correlation length now depends on both pressure and temperature, and the presence of an interaction beyond the hard core introduces a richer structural landscape, particularly in the case of the \acrshort{SW} fluid. Here, the asymptotic oscillation frequency undergoes not just one, but two distinct discontinuous transitions as pressure increases. At a given temperature, if the density is sufficiently low, the long-range decay of the \acrshort{RDF} transitions from damped oscillatory to a purely monotonic exponential decay. This behavior defines a \acrshort{FW} line in the \acrshort{SW} model, a structural boundary that marks a fundamental shift in the nature of particle correlations. The phase diagram of the asymptotic oscillation frequency is presented in Fig.~13 of \nameref{a6}, and illustrative examples of the \acrshort{RDF} at different state points are shown in Fig.~14 of \nameref{a6}. In particular, Fig.~14(a) of \nameref{a6} clearly displays the monotonic exponential decay characteristic of the region below the \acrshort{FW} line, in contrast to the damped oscillatory behavior observed in Figs.~14(b,c) of \nameref{a6}. This extra complexity tends to disappear in the high-temperature limit, where the system again recovers the \acrshort{HD} limiting behavior.

So far, the discussion has focused exclusively on the \emph{longitudinal} \acrshort{RDF}s and correlation lengths. This is primarily due to the nature of confined \acrshort{Q1D} geometries, where translational invariance is broken in the transverse directions. As a result, defining a global \acrshort{RDF} that depends solely on the distance between two particles is no longer as straightforward as in the bulk case. This, once again, highlights that special care is required when extending bulk definitions to such highly constrained systems. In \nameref{a5} we study this difficulty in the \acrshort{HD} scenario, and describe main differences between the bulk and the confined geometries. To address this challenge, we analytically compute spatial correlations in a confined geometry for the ideal-gas case, where the density profile along the transverse direction is uniform. Interestingly, even in the absence of interparticle interactions, the resulting two-body distribution function is not constant, which means that this deviation arises solely from the geometric confinement, not from any intrinsic interaction between particles. Something similar occurs in cases where the density profile is nonuniform: even in the absence of interparticle correlations, the two-body distribution function deviates even more strongly from constancy, particularly at short distances. These observations highlight the difficulty of defining a simple, universal \acrshort{RDF} in \acrshort{Q1D} confined geometries. Nevertheless, it remains possible to define a meaningful quantity that captures spatial correlations: the average number of particles located between a distance $r$ and $r + \dd r$ from a reference particle. In \nameref{a5}, this function is denoted by $\hat{n}(r)/2\lambda$ [see Fig.~4 of \nameref{a4}], and its excellent agreement with \acrshort{MC} simulations further validates its use as a robust measure.

A similar approach is applied in the \acrshort{Q1D} \acrshort{HS} model. The resulting correlation functions, when expressed as functions of the interparticle distance, exhibit greater complexity in the positioning of local maxima compared to the correlations measured strictly along the longitudinal direction. The difference between the two types of correlations becomes more pronounced as the pore width increases, while for smaller pore widths, where transverse motion is more restricted, the two functions progressively converge.

\section{Orientationally free Q1D models}

This section discusses the results on the ordering properties of hard anisotropic particles, which are allowed to move freely along a single \acrshort{1D} direction while also possessing a discrete set of possible orientations. A key theoretical outcome is the successful extension of the mapping methodology---originally developed for spatially confined \acrshort{Q1D} systems---to systems where particles exhibit orientational degrees of freedom. Although the mapping framework is general enough to accommodate continuous orientational motion, the models studied here focus on particles restricted to two or three discrete orientations, called the 2-state and 3-state models, respectively.

Two different hard-body shapes were selected for this study because they are similar---but also different enough---as to offer meaningful insight into the orientational and spatial behaviors. On the one hand, we considered hard prisms with three distinct side lengths ($\sigma_1, \sigma_2, \sigma_3$), and on the other hand, hard dumbbells made up of two tangent spheres, which feature only two different effective lengths along the movement axis ($\sigma, 2\sigma$). These shapes highlight the contrast between additive and nonadditive interactions. In the case of prisms, the system is additive: for any pair of orientations ($i,j=1,2,3$), the contact distance is given by $\sigma_{ij} = (\sigma_i + \sigma_j)/2$, which is not true for dumbbells. This setup allows us to investigate how additivity and the number of distinct longitudinal sizes influence structural and thermodynamic behavior. A visual representation of both configurations is provided in Fig.~1 of \nameref{a8}. The selection of different orientations for the 2-state and 3-state models analyzed in \nameref{a8} is summarized, for convenience, in Table~\ref{tab:Summary2and3state}.

\begin{table}[htb]
	\centering
	\caption{Summary of the length along the movement axis of each hard-body shape for the orientations considered in \nameref{a8}.}
	\label{tab:Summary2and3state}
	\renewcommand{\arraystretch}{2}
	\begin{tabular}{@{} p{3.5cm}  C{2.5cm}  C{2.5cm} @{}}
		\toprule
		\textbf{System}           & $2$-state & $3$-state \\
		\midrule
		Prisms                 & $\sigma_1\leq \sigma_2$  & $\sigma_1\leq \sigma_2\leq\sigma_3$\\
		Dumbbells            & $\sigma, 2\sigma$  & $\sigma, \sigma, 2\sigma$\\
		\bottomrule
	\end{tabular}
\end{table}

All models are analyzed using three approaches: the approximate \acrshort{PL} theory, the \acrshort{TM} method, and the mapping approach, referred to in this article as the neighbor distribution (\acrshort{ND}) method for convenience. General findings show that both the \acrshort{TM} and \acrshort{ND} methods yield exact results for all systems, regardless of additivity. In contrast, the \acrshort{PL} theory provides exact results only in the additive case and fails to fully capture the features associated with nonadditive interactions in the dumbbell model.

\subsubsection{Prisms}

The analysis of the bulk properties of the hard prisms shows how the structure of the fluid transitions from isotropic to perfect nematic with increasing density. In the low-density limit, the system approaches the behavior of an ideal gas, with equal number fractions for all orientations. At high density, particles get closer to each other and reduce the available space. This forces particles to align along the direction of minimal contact distance [see Fig.~2 of \nameref{a8}].

Regarding spatial correlations, we show that the prisms obey the shift property of additive \acrshort{1D} hard-body mixtures: all pair distributions coincide if the origin of each curve $g_{ij}(z)$ is shifted to $z=\sigma_{ij}$~\cite{S07b,GDER04,GDER05}. At high densities, oscillatory features in the spatial correlations become more pronounced as the perfect nematic order starts developing, and correlations become identical to those of hard rods with diameter $\sigma_1$. The extent of positional and orientational order can be studied with the help of the correlation lengths. The \acrshort{ND} method provides access to the spatial correlation length, $\xi_p$, which presents a nonmonotonic behavior as density increases---signaling a competition between different ordering structures---and diverges in the high-density limit. The \acrshort{TM} analysis of orientational correlation length, $\xi_o$, shows that the prisms lack long-range orientational order.

\subsubsection{Dumbbells}

A similar analysis is carried out for the dumbbells, where now differences between the $2$-state and the $3$-state models are more pronounced due to particle nonadditivity. The $3$-state case is the most interesting one because, if we define orientations such that $\sigma_1=\sigma$, $\sigma_2=\sigma$, and $\sigma_3=2\sigma$, the shape of the dumbbells implies that $\sigma_{12} \le (\sigma_1+\sigma_2)/2=\sigma$. This means that, instead of random orientational ordering of states 1 and 2 in the high-density limit, particles form clusters where neighboring particles are perpendicular to each other, alternating particles in states 1 and 2. At close packing, cluster length diverges to maximize density. This structure does not form in the 2-state case, where particles form a perfect nematic order in state 1 to maximize density.

The pair distribution function, which captures spatial correlations, no longer satisfies the shift property present in the prism model. By analyzing the trends in the heights and positions of the peaks in the various correlation functions $g_{ij}(z)$, additional insight is gained into the system's orientational ordering. Specifically, the results show that first-neighbor pairs preferentially adopt perpendicular orientations, while second-neighbor pairs tend to be parallel, indicating a preference for alternating patterns in the local structure.

In this context, both the spatial and orientational correlation lengths, $\xi_p$ and $\xi_o$, are nonzero throughout the entire pressure range. However, their behaviors differ significantly, particularly near the close-packing limit. For both the 2- and 3-state systems, the spatial correlation length $\xi_p$ diverges, reflecting long-range positional order. In contrast, the orientational correlation length $\xi_o$ exhibits different trends: in the 3-state system, $\xi_o$ also diverges as particles form alternating, cross-oriented clusters, indicating the development of complex orientational ordering. In the 2-state system, however, $\xi_o$ vanishes in the high-pressure limit as the system approaches perfect nematic order, where all particles align in a single orientation.

%% file: chapters/C9_Conclusions_and_Outlooks.tex
\thispagestyle{empty}

\chapter{Conclusions and outlook}\label{conclusions}

This chapter summarizes the key conclusions drawn from the work presented, highlighting the main contributions to the field. An outline of planned and potential future directions that could extend the research line initiated here is also presented.

\section{Conclusions}

\subsection{The effect of dimensionality}

An important part of the thesis is focused on analyzing the effect of dimensionality on systems interacting with simple pairwise potentials with competing interactions (potentials with attractive and repulsive parts). A general conclusion extracted from this analysis is that \acrshort{1D} and \acrshort{3D} systems often share many qualitative features. However, dimensionality introduces significant differences that can lead to the disappearance of certain phenomena or require stronger interaction parameters in \acrshort{3D} to observe features that are more easily present in \acrshort{1D}. Our results indicate that there is no clear or reliable criterion to determine \emph{a priori} whether a specific feature observed in \acrshort{1D} will also appear in \acrshort{3D}, or vice versa, highlighting the subtle and nontrivial role of dimensionality.

A detailed analysis of the Seno, Zeno, and \acrshort{FW} lines---each representing distinct signatures of the effects of competing interactions---was carried out for \acrshort{1D} and \acrshort{3D} models. The results reveal that these three curves are very similar in \acrshort{3D}, at least in the range of intermediate densities, suggesting a coherent structural and thermodynamic response to competing interactions. However, this similarity breaks down in the \acrshort{1D} case, where the lines differ significantly across the entire density range. This discrepancy highlights the dimensional dependence of the relationship between these quantities and suggests that the conjecture by~\textcite{SHRE19} about a close correspondence between the \acrshort{FW} line and the line of vanishing excess isothermal compressibility (Seno line) appears to hold primarily in three dimensions. In lower-dimensional systems, such as the \acrshort{1D} case studied here, the conjecture loses its validity.

\subsection{Exact solution of Q1D models}

A major focus of this work was the development of an exact solution for \acrshort{Q1D} models via a mapping from the \acrshort{Q1D} system to a \acrshort{1D} mixture, which is exact and has been successfully applied to a variety of models. From these applications, several very general features and limitations of the method should be highlighted:
\begin{itemize}
	\item The method is exact only when considering \acrshort{Q1D} systems in which particles are forced to stay in single-file formation and interactions are restricted to nearest-neighbors.
	\item All the information of the original higher-dimensional \acrshort{Q1D} system---including both longitudinal and transverse properties---is fully encoded in the corresponding \acrshort{1D} mixture.
	\item The mapping approach is not limited to \acrshort{Q1D} systems with purely positional degrees of freedom; it can also be applied to systems with orientational freedom, enabling the exact treatment of anisotropic particles.
	\item The methodology is equally valid for systems with either discrete or continuous additional degrees of freedom, providing a flexible framework that can adapt to a wide range of physical situations.
	\item Despite its general applicability, the numerical complexity of the method increases rapidly with the number of mapped variables. As a result, the approach becomes computationally demanding---and in many cases impractical---for \acrshort{Q1D} systems with more than two or three additional degrees of freedom, at least when using standard numerical techniques.
\end{itemize}

This mapping has then been applied to a variety of systems for which the exact solution for their thermodynamic and structural quantities was obtained. The main results for both systems---\acrshort{Q1D} hard disks and hard spheres---demonstrate a qualitatively similar behavior, even though their quantitative features differ. Below we summarize the key conclusions regarding their properties:
\begin{itemize}
	\item The exact transverse and longitudinal components of the equation of state are obtained. For wide enough channels, the transverse pressure component becomes larger than the longitudinal at high densities, indicating a progressive transverse development as the channel widens.
	\item In the low-density limit, the correct third and fourth virial coefficients were obtained, showing that the standard irreducible diagram methods miss essential \enquote{reducible} contributions that survive under confinement.
	\item In the high-density limit, particles arrange in a zigzag configuration to optimize packing. Under these conditions, the compressibility factor diverges differently than in the Tonks gas, due to additional contributions arising from transverse fluctuations in the zigzag structure.
	\item The exact \acrshort{RDF} was computed and the results were validated against \acrshort{MC} simulation data from the literature, showing excellent agreement.
	\item Additionally, for the \acrshort{HD} model, we identified a structural crossover in the asymptotic oscillation frequency of the \acrshort{RDF}, which also corresponds to a kink in the spatial correlation length. This transition, occurring at a specific pressure, marks the onset of zigzag ordering and serves as a precursor to long-range structural rearrangements.
\end{itemize}

All the previously discussed features also hold for the \acrshort{Q1D} models with \acrshort{SW} and \acrshort{SS} interactions, although the inclusion of temperature as a variable adds complexity to the parameter space. The \acrshort{SS} model, being purely repulsive, shows behavior that remains quantitatively closer to the \acrshort{HD} case. In contrast, the \acrshort{SW} model, due to its attractive well, exhibits a richer structural behavior, including two distinct structural crossovers instead of only one in the asymptotic oscillation frequency and the appearance of a \acrshort{FW} line. These features, absent in the \acrshort{HD} and \acrshort{SS} models, highlight the significant influence of attractive interactions.

\subsection{Q1D hard-bodies with orientational freedom}

The last model analyzed was that of hard-bodies constrained to move along a single axis but that can occupy different orientations. We examined the effects of additivity and nonadditivity on phase behavior and correlation lengths.

Our results show that the \acrshort{PL} theory provides exact thermodynamic properties for additive \acrshort{Q1D} fluids, but the theory is no longer valid when nonadditive interactions are introduced. In contrast, the \acrshort{TM} method and the mapping approach remain exact for both additive and nonadditive cases. Because particles possess both orientational and positional degrees of freedom, ordering can arise from two different mechanisms. In additive fluids, orientational correlations are absent and structural evolution is driven purely by positional packing. Nonadditive interactions, however, couple both the orientation and position: the system then develops simultaneous orientational and positional order, and its behavior depends sensitively on the specific type of interparticle interaction.

\section{Outlook}

One of the objectives of the present work has been to study the role of dimensionality in systems where the interaction potential presents competing interactions. Although the conjecture by~\textcite{SHRE19}, which suggests similarities between the \acrshort{FW} line and the Seno line, has been shown not to hold for the \acrshort{1D} Jagla potential studied in \nameref{a1}, this conclusion does not necessarily generalize to all \acrshort{1D} systems. Given that \acrshort{1D} models allow for an exact analysis, it would be valuable to further test this conjecture by using other interaction potentials, such as the two-step potential with competing interactions analyzed in \nameref{a1}, and to systematically test if the conjecture holds throughout the entire parameter space for each potential. Whether there are some characteristics of the interaction potential for which the~\textcite{SHRE19} conjecture does or does not hold can provide interesting insight into the mechanisms that govern the relationship between structural and thermodynamic properties. Such a study could also be complemented by comparing the exact \acrshort{1D} results with the approximate findings for the corresponding \acrshort{3D} systems, assessing whether the conjecture also holds for the entire range of the parameter space.

Within the field of particles in \acrshort{Q1D} confined geometries, the main idea of the theoretical framework developed during this thesis could be extended to account for other systems. Among others, we highlight
\begin{itemize}
	\item \underline{\acrshort{Q1D} mixture of hard disks/spheres with different sizes:} By considering a binary mixture confined inside a pore, it becomes possible to generalize the theoretical framework. However, this extension introduces additional mathematical complexity. In particular, the condition of equal chemical potential applied in Sec.~\ref{th_q1dEQC} for the monodisperse case must now be enforced separately for each species in the binary mixture, a condition that adds a layer of difficulty to the analysis, both conceptually and computationally. Some work in this direction has already been carried out.
	\item \underline{\acrshort{Q1D} system with interacting walls:} The theoretical framework can also be extended to systems in which particles interact with the confining walls through more than just hard-core exclusion, and experience attraction or repulsion from the walls. Such systems are particularly relevant for experimental setups, as wall-particle interactions play a crucial role in real confined environments. In this context, the relationship between the chemical potential and the largest eigenvalue discussed in Sec.~\ref{th_q1dEQC} must be modified to incorporate the effects of wall interactions, which introduces additional complexity into the theoretical formulation, but also broadens its applicability. Some preliminary work in this regard has also been done.
	\item \underline{\acrshort{Q1D} model of hard particles with continuous rotation:} For systems of hard particles with orientational degrees of freedom, as in the setup of \nameref{a8}, the obvious extension is to consider continuous orientations and not only a discrete set. This generalization shifts the mapping framework from the discrete mixture regime to a fully polydisperse one, significantly increasing the numerical complexity of the problem. This extension may also impact the physical behavior of the system: with a continuous range of orientations, discrete orientation jumps are replaced by smooth changes, and the role of orientational entropy becomes more prominent. Fluctuations around specific preferred orientations can now lead to subtle ordering effects that were absent or suppressed in the discrete models. The study of this extended system is already well underway and we expect to publish the results in the near future.
	\item \underline{\acrshort{Q1D} model of hard particles with different interactions:} Beyond the distinction between discrete and continuous orientations, systems with hard anisotropic particles can be further generalized to explore a broader range of hard-body shapes, potentially representing more realistic or experimentally relevant particle geometries. Additionally, introducing interaction potentials beyond pure hard-core exclusion, such as extra soft repulsions or directional attractions, would greatly enrich the model and bring it closer to experimental systems.
	\item \underline{\acrshort{Q1D} system with orientational and spatial freedom}: Finally, another interesting extension of the theoretical framework would involve studying a \acrshort{Q1D} system in which particles are allowed to move along one or two spatially confined directions while also possessing orientational degrees of freedom. This hybrid system would combine the geometric complexity of spatial confinement with the richness introduced by particle anisotropy and orientational freedom. Such a model would be particularly relevant for approaching realistic experimental setups---such as colloidal rods, ellipsoids, or Janus particles confined in narrow channels, where both position and orientation play critical roles in determining the behavior of the system.
\end{itemize}

Ongoing work is focusing on an extension of our work about hard spheres confined inside cylindrical pores, examining correlation lengths for both the overall fluid and for specific pairs of particles located at distinct transverse positions. Preliminary findings reveal a rich interplay among these correlation lengths, shaped by the development of zigzag ordering and the system’s cylindrical symmetry. We expect to publish the results of this work soon.

For all the systems described above, it is essential to emphasize that the theoretical framework developed in this thesis is exact only under the condition that interparticle interactions are restricted to nearest neighbors. This constraint must always be taken into account when designing the geometry of the confining walls, as well as the shape and interaction potential of the particles. While this may seem like a strong limitation, we believe that the insights obtained within this framework remain highly valuable. Even under these restrictive conditions, the results provide a solid foundation for understanding the fundamental mechanisms at play in \acrshort{Q1D} confined systems and can serve as a useful guide for interpreting and modeling more general, and potentially more complex, cases.

%% file: chapters/A1_ProofOfMaxA2.tex
\thispagestyle{empty}

\chapter{Physical argument about the need for the largest eigenvalue in Eq.~\eqref{eq:q1d_eig}}\label{ProofOfMaxA2}

The aim of this Appendix is to provide a physical reasoning for why only the largest eigenvalue in Eq.~\eqref{eq:q1d_eig} needs to be retained. We begin by revisiting the eigenvalue equation presented in Eq.~\eqref{eq:q1d_eig}
\begin{equation}\label{eq:app_eig1}
	\sum_j \hat{\Omega}_{ij}(\bpp,\beta)\phi_j  = \frac{1}{A^2(\bpp,\beta)} \phi_i,
\end{equation}
where we have explicitly represented the dependency of $A^2$ on the temperature and the pressure. Equation~\eqref{eq:app_eig1} can be recast in matrix form as an eigenvalue problem,
\begin{equation}\label{eq:app_eig2}
	\hat{\mathsf{\Omega}}(\bpp,\beta) \cdot \boldsymbol{\phi}  = \mu(\bpp,\beta) \boldsymbol{\phi}, \qquad \mu(\bpp,\beta) = \frac{1}{A^2(\bpp,\beta)}.
\end{equation}
This equation yields as many eigenvalues as the number $M$ of species in the mixture, each one associated with an eigenvector $\boldsymbol{\phi}$. Among all possible pairs $(\mu, \boldsymbol{\phi})$, only the one corresponding to the largest eigenvalue, $\mu_{\mathrm{max}} = \max({\mu})$, is physically relevant to describe the system. This selection is justified by examining the behavior of the solution in the low-pressure limit.

We begin by noting that the short-range nature of the potential imposes the condition $\lim_{x \to 0} \psi_{ij}(x) = 0$. Applying the final-value theorem for the Laplace transform to this condition yields $\lim_{\bpp \to 0} \bpp \hat{\Omega}_{ij}(\bpp) = 1$. Therefore, in the low-pressure limit, Eq.~\eqref{eq:app_eig2} reduces to
\begin{equation}\label{eq:app_eig3}
	 \boldsymbol{1} \cdot \boldsymbol{\phi}  = \nu(\beta) \boldsymbol{\phi}, \qquad \nu(\beta) =\lim_{\bpp\to 0}\bpp\mu(\bpp,\beta),
\end{equation}
where $\boldsymbol{1}$ denotes the all-ones matrix of size $M \times M$.

The matrix $\boldsymbol{1}$ has a well-known eigenspectrum in which only the maximum eigenvalue is nonzero, $\nu_{\mathrm{max}} = M$, with its normalized eigenvector $\boldsymbol{v}_{\mathrm{max}} = \frac{1}{\sqrt{M}} [1,1,\cdots,1]^T$~\cite{HJ85}. Furthermore, this eigenpair is the one that maximizes Rayleigh quotient~\cite{S06b}, defined as
\begin{equation}
	R_{\mathrm{Q}}(\boldsymbol{v}) = \frac{\boldsymbol{v}^T\cdot \boldsymbol{1} \cdot \boldsymbol{v}}{\boldsymbol{v}^T \cdot \boldsymbol{v}},
\end{equation}
for every nonzero vector $\boldsymbol{v}$. The maximum value of $R_{\mathrm{Q}}(\boldsymbol{v})$, coincides with the maximum eigenvalue $\nu_{\mathrm{max}}$, when $\boldsymbol{v} =\boldsymbol{v}_{\mathrm{max}} $.

The expected physical behavior of the mixture at low pressure is that it recovers the ideal-gas behavior, therefore $\lim_{\bpp \to 0}\boldsymbol{\phi} = \frac{1}{\sqrt{M}} [1,1,\cdots,1]^T$, where the constant prefactor is determined by the normalization condition $\sum_i \phi_i^2 = 1$. This means that the physically relevant solution of Eq.~\eqref{eq:app_eig3} is the eigenvalue-eigenvector pair corresponding to a constant eigenvector, i.e., $\mathbf{\phi}=\boldsymbol{v}_\mathrm{max}$. From this analysis, we conclude that the physically meaningful solution to Eq.~\eqref{eq:app_eig3} is given by the largest eigenvalue, which, in the low-pressure limit, satisfies
 \begin{equation}
 	\mu = \frac{M}{\bpp}.
 	\end{equation}

Although this argument is formally valid only in the limit $\bpp \to 0$, an analytical continuation to finite pressure supports the conclusion that the physically meaningful solution of Eq.~\eqref{eq:q1d_eig} corresponds to the maximum eigenvalues, that is, the minimum value of $A$.